%% file: paper.tex
\PassOptionsToPackage{prologue,dvipsnames}{xcolor}
\documentclass[acmsmall,screen]{acmart}
\usepackage{breqn}

\newcommand\grumbler[3]{\textcolor{#1}{#2 says: #3}}
\newcommand\mike[1]{\grumbler{blue}{Mike}{#1}}
\newcommand\chujun[1]{\grumbler{orange}{Chujun}{#1}}
\newcommand\yang[1]{\grumbler{red}{Yang}{#1}}

\newtoggle{extended-version}
\toggletrue{extended-version}
\usepackage{amsmath}
\allowdisplaybreaks
\usepackage{graphicx}
\usepackage{xspace}
\usepackage{svg}
\usepackage{caption}
\usepackage{subcaption}
\usepackage{empheq}
\usepackage{multirow}
\usepackage[normalem]{ulem}
\usepackage{algorithm}
\usepackage[noend]{algpseudocode}
\usepackage{listings}
\usepackage{url}
\usepackage{enumitem}
\usepackage{placeins}
\usepackage{siunitx}
\usepackage{minibox}

\newcommand{\readevent}[1]{\textsf{read(\ensuremath{#1})}}
\newcommand{\writeevent}[1]{\textsf{write(\ensuremath{#1})}}
\newcommand{\commitevent}{\textsf{commit}\xspace}
\newcommand{\tx}{\ensuremath{\mathit{T}}\xspace}
\newcommand{\so}{\ensuremath{\mathit{so}}\xspace}
\newcommand{\po}{\ensuremath{\mathit{po}}\xspace}
\newcommand{\writeread}{\ensuremath{\mathit{wr}}\xspace}
\newcommand{\pos}{\ensuremath{\mathit{rdpos}_\ast\xspace}}
\newcommand{\posk}{\ensuremath{\mathit{rdpos_k}\xspace}}
\newcommand{\ogwr}{\ensuremath{\mathit{wr_{obs}}}\xspace}
\newcommand{\hb}{\ensuremath{\mathit{hb}}\xspace}
\newcommand{\co}{\ensuremath{\mathit{co}}\xspace}
\newcommand{\cocausal}{\ensuremath{\mathit{co_{causal}}}\xspace}
\newcommand{\corc}{\ensuremath{\mathit{co_{rc}}}\xspace}
\newcommand{\comin}{\ensuremath{\mathit{pco}}\xspace}
\newcommand{\pco}{\comin}
\newcommand{\arcausal}{\ensuremath{\mathit{ww}_{\mathit{causal}}}\xspace}
\newcommand{\arrc}{\ensuremath{\mathit{ww}_{\mathit{rc}}}\xspace}
\newcommand{\arserial}{\ensuremath{\mathit{ww}}\xspace}
\newcommand{\antidependency}{\ensuremath{\mathit{rw}}\xspace}
\newcommand{\rank}{\ensuremath{\mathit{rank}}\xspace}
\newcommand{\boundary}{\ensuremath{\mathit{boundary}\xspace}}

\newcommand{\smtso}{\ensuremath{\phi_{\so}}\xspace}
\newcommand{\smtco}{\ensuremath{\phi_{\co}}\xspace}
\newcommand{\smtarserial}{\ensuremath{\phi_{\arserial}}\xspace}
\newcommand{\smtarcausal}{\ensuremath{\phi_{\arcausal}}\xspace}
\newcommand{\smtarrc}{\ensuremath{\phi_{\arrc}}\xspace}
\newcommand{\smthb}{\ensuremath{\phi_{\hb}}\xspace}
\newcommand{\smtwritereadk}{\ensuremath{\phi_{\writeread_k}}\xspace}
\newcommand{\smtwriteread}{\ensuremath{\phi_{\writeread}}\xspace}
\newcommand{\smtchoice}{\ensuremath{\phi_{\mathit{choice}}}\xspace}
\newcommand{\smtcomin}{\ensuremath{\phi_{\comin}}\xspace}
\newcommand{\smtpco}{\smtcomin}
\newcommand{\smtantidependency}{\ensuremath{\phi_{\antidependency}}\xspace}
\newcommand{\smtcocausal}{\ensuremath{\phi_{\cocausal}}\xspace}
\newcommand{\smtcorc}{\ensuremath{\phi_{\corc}}\xspace}
\newcommand{\smtboundary}{\ensuremath{\phi_{\boundary}}\xspace}
\newcommand{\smtobs}{\ensuremath{\phi_{\mathit{obs}}}\xspace}

\newcommand{\bench}[1]{\textsf{#1}\xspace}
\newcommand\mc[3]{\multicolumn{#1}{#2}{#3}}

\newcommand{\causal}{\textsc{causal}\xspace}
\newcommand{\Causal}{\causal}
\newcommand{\rc}{\textsc{rc}\xspace}

\newcommand{\serializable}{\textsc{serializable}\xspace}

\newcommand{\unserializable}{\textsc{unserializable}\xspace}
\newcommand{\Unserializable}{\textsc{Unserializable}\xspace}

\newcommand{\isopredict}{IsoPredict\xspace}
\newcommand{\Isopredict}{\isopredict}
\newcommand{\IsoPredict}{\Isopredict}
\newcommand{\configFull}{\textsf{Exact-Strict}\xspace}
\newcommand{\configExpress}{\textsf{Approx-Strict}\xspace}
\newcommand{\configRelaxed}{\textsf{Approx-Relaxed}\xspace}

\definecolor{GrayCodeBlock}{RGB}{241,241,241}
\definecolor{BlackText}{RGB}{110,107,94}
\definecolor{RedTypename}{RGB}{182,86,17}
\definecolor{GreenString}{RGB}{96,172,57}
\definecolor{PurpleKeyword}{RGB}{184,84,212}
\definecolor{GrayComment}{RGB}{120,120,120}
\definecolor{GoldDocumentation}{RGB}{180,165,45}
\lstdefinelanguage{rust}
{
    columns=fullflexible,
    keepspaces=true,
    showstringspaces=false,
    frame=single,
    xleftmargin=4pt,
    xrightmargin=4pt,
    backgroundcolor=\color{GrayCodeBlock},
    basicstyle=\ttfamily\color{BlackText},
    keywords={
        true,false,
        unsafe,async,await,move,
        use,pub,crate,super,self,mod,
        struct,enum,fn,const,static,let,mut,ref,type,impl,dyn,trait,where,as,
        break,continue,if,else,while,for,loop,match,return,yield,in
    },
    keywordstyle=\color{PurpleKeyword},
    ndkeywords={
        bool,u8,u16,u32,u64,u128,i8,i16,i32,i64,i128,char,str,
        Self,Option,Some,None,Result,Ok,Err,String,Box,Vec,Rc,Arc,Mutex,Cell,RefCell,HashMap,BTreeMap,
        macro_rules
    },
    ndkeywordstyle=\color{RedTypename},
    comment=[l][\color{GrayComment}\slshape]{//},
    morecomment=[s][\color{GrayComment}\slshape]{/*}{*/},
    morecomment=[l][\color{GoldDocumentation}\slshape]{///},
    morecomment=[s][\color{GoldDocumentation}\slshape]{/*!}{*/},
    morecomment=[l][\color{GoldDocumentation}\slshape]{//!},
    stringstyle=\color{GreenString},
    string=[b]",
    escapeinside={(*@}{@*)}
}

\lstnewenvironment{Rust}[1][]
    {\lstset{language=Rust,
      basicstyle=\footnotesize,
      literate={-}{-}1,   
      columns=fullflexible,
      numberstyle=\tiny\color{gray},
      stepnumber=1,
      numbers=left,
      numbersep=5pt,
      backgroundcolor=\color{white},
      showspaces=false,
      showstringspaces=false,
      showtabs=false,
      frame=single,
      rulecolor=\color{black},
      tabsize=2,
      captionpos=b,
      breaklines=true,  
      breakatwhitespace=false, 
      title=\lstname,
      keywordstyle=\color{blue},
      commentstyle=\color{green},
      escapeinside={\%*}{*)},
      morekeywords={*,...},
      #1
    }
  }
  {}
\lstnewenvironment{Java}[1][]
  {
  \lstset{language=Java,
      basicstyle=\footnotesize,
      literate={-}{-}1,   
      columns=fullflexible,
      numberstyle=\tiny\color{gray},
      stepnumber=1,
      numbers=left,
      numbersep=5pt,
      backgroundcolor=\color{white},
      showspaces=false,
      showstringspaces=false,
      showtabs=false, 
      frame=single,
      rulecolor=\color{black},
      tabsize=2,
      captionpos=b,
      breaklines=true,  
      breakatwhitespace=false, 
      title=\lstname,
      keywordstyle=\color{blue},
      commentstyle=\color{green},
      stringstyle=\color{mauve},
      escapeinside={\%*}{*)},
      morekeywords={*,...},
      #1
      }
  }
  {}

\begin{document}

\iffalse
%\onecolumn
\pagestyle{empty}
\input{cover}
%\twocolumn
\clearpage
\pagestyle{standardpagestyle}
\setcounter{page}{1}
\fi

\newcommand\mytitle{\IsoPredict: Dynamic Predictive Analysis for Detecting Unserializable Behaviors in Weakly Isolated Data Store Applications}
\newcommand\myshorttitle{\IsoPredict: Dynamic Predictive Analysis for Detecting Unserializable Behaviors in Data Store Applications}
\title{\mytitle\iftoggle{extended-version}{\smallskip\\\minibox[frame]{\normalsize \textsf{\textnormal{This extended version of a PLDI 2024 paper adds an appendix with additional material}}}}{}\title[\myshorttitle]{\mytitle}}
% \title{Predicting Unserializable Behavior under Weak Isolation}
% \title{Predicting Unserializable Behavior on Weakly Isolated Data Stores}


\author{Chujun Geng}
\orcid{0009-0000-6149-0208}
\affiliation{%
  \institution{Ohio State University}
  \city{Columbus}
  \country{USA}
}
\email{geng.195@osu.edu}

\author{Spyros Blanas}
\orcid{0009-0004-2703-7177}
\affiliation{%
  \institution{Ohio State University}
  \city{Columbus}
  \country{USA}
}
\email{blanas.2@osu.edu}

\author{Michael D. Bond}
\orcid{0000-0002-8971-4944}
\affiliation{%
  \institution{Ohio State University}
  \city{Columbus}
  \country{USA}
}
\email{mikebond@cse.ohio-state.edu}

\author{Yang Wang}
\orcid{0000-0002-9721-4923}
\affiliation{%
  \institution{Ohio State University}
  \city{Columbus}
  \country{USA}
}
\email{wang.7564@osu.edu}


\begin{abstract}
\input{0.abs.tex}
\end{abstract}






%%% The following is specific to PLDI '24 and the paper
%%% 'IsoPredict: Dynamic Predictive Analysis for Detecting Unserializable Behaviors in Weakly Isolated Data Store Applications'
%%% by Chujun Geng, Spyros Blanas, Michael D. Bond, and Yang Wang.
%%%
\setcopyright{rightsretained}
\acmDOI{10.1145/3656391}
\acmYear{2024}
\copyrightyear{2024}
\acmSubmissionID{pldi24main-p59-p}
\acmJournal{PACMPL}
\acmVolume{8}
\acmNumber{PLDI}
\acmArticle{161}
\acmMonth{6}
\received{2023-11-03}
\received[accepted]{2024-03-31}

\maketitle

% \yang{The title feels a bit too broad to me. Could we put the idea of dynamic analysis in the title?}
% \mike{Revised. How's that?}

% \mike{Writing/\LaTeX stuff:
% \begin{itemize}
% \item Use lowercase for terms like write--read, causal consistency, event, events, etc.
% \item Use an en dash (--) instead of a hyphen (-) for write--read and key--value
% \item Use macros \causal, \rc, \serializable, when referring to them as formal consistency models, \emph{not} for these terms in general.
% \item Use \textbackslash mathit for multi-letter names in math mode
% \item Use multline* (or align*) for equations that aren't actually referenced by number in the text.
% \item Capitalize sections and subsections like titles. Capitalize subsubsections and paragraphs like sentences.
% \end{itemize}}
% 
% \mike{Name ideas: IsoPredict and Peregrine}

\input{1.intro.tex}

\input{2.background.tex}

\input{3.design.tex}

\input{4.implementation}

\input{5.experiment.tex}
\input{6.related}

\input{7.conclusion.tex}

\section*{Data-Availability Statement}
An artifact reproducing this paper's results is publicly available~\cite{isopredict-artifact}.

\begin{acks}
We thank the MonkeyDB authors~\cite{monkeydb} for making their implementation publicly available and answering our questions about it;
% the ECRacer authors~\cite{ecracer} for sharing their artifact with us although we ended up going in a different direction;
Vincent Beardsley and Noah Charlton for helpful discussions;
and the anonymous reviewers for valuable feedback.
This material is based in part upon work supported by the National Science Foundation under Grant Numbers NSF CCF-2118745, CSR-2106117, and OAC-2112606, and by Oracle America, Inc.
\end{acks}

\bibliographystyle{ACM-Reference-Format}
\bibliography{paper}

\iftoggle{extended-version}{
% \clearpage
\appendix
\input{proof}

\input{smt}

\input{predictions}

%\input{monkeydb_bugs}
}{}

\end{document}

%% file: 0.abs.tex
Distributed
% transactional \chujun{I don't think our approach is limited to transactional data stores. Cassandra for example doesn't support transactions, and technically it's possible to apply \isopredict on Cassandra-based applications.}
data stores typically provide weak isolation levels, which are efficient but can lead to unserializable behaviors, which are hard for programmers to understand and often result in errors.
% do not provide sstrong guarantees of \textcolor{red}{transaction} isolation such as a common order of all transactions executed across different nodes.
% This could lead to unexpected behaviors in applications such as timing related bugs that are hard to detect and hard to reproduce.
This paper presents the first dynamic predictive analysis for data store applications under weak isolation levels, called \emph{\isopredict}. Given an observed \emph{serializable} execution of a data store application, \isopredict generates and solves SMT constraints to find an \emph{unserializable} execution that is a feasible execution of the application.
\Isopredict introduces novel techniques that handle divergent application behavior; solve mutually recursive sets of constraints; and balance coverage, precision, and performance. An evaluation on four transactional data store benchmarks shows that \isopredict often predicts unserializable behaviors, 99\% of which are feasible.

% For distributed transactional data stores, causal consistency is the strongest isolation level that is achievable under network partition.
% \mike{Don't just mention \causal; include \rc.}
% It is hard to detect abnormal behaviors such as violation of serializability on causally consistent databases especially when the applications are able to handle some of the anomalies from weak isolation levels.
% We develop a sound dynamic analysis tool that predicts unserializable behaviors from execution traces using SMT solvers, and we further improve its performance by adding graph-based constraints with minimal sacrifice of completeness.
% \mike{Revise to say something about results.}

%% file: 1.intro.tex
\section{Introduction}

Distributed data stores are the foundation of today's service infrastructure, due to
their scalability, fault tolerance, and ease of use~\cite{Corbett2012Spanner,Elhemali2022DynamoDB,snowflake,mysqlcluster}.
% \yang{Two citations cannot show properly. Not sure what is going wrong.}
% \mike{They needed a year}
% Client applications issue groups of operations called \emph{transactions}, and the data store
% guarantees these operations are executed in an atomic and consistent manner.
Many real-world data stores only support \emph{weak isolation} levels, such as
% also called \emph{weak consistency}.
\emph{causal consistency} (\causal)~\cite{ahamad95causalmemory}, which is the strongest level that achieves availability under network partitions~\cite{burckhardt2014,lynchCAP}. Another weak isolation level is \emph{read committed} (\rc)~\cite{Berenson1995Critique}, which is commonly used by database applications to balance performance and correctness~\cite{Crooks2017SBC,pavlo2017sigmod,Cheng2023Responsibility,Tang2022AdHoc}.
Under weak isolation, an execution may be \emph{unserializable}, producing an outcome that is impossible for any serial execution.
Unserializable behaviors are poorly understood by most programmers, and often lead to errors and failures in real-world systems~\cite{Cheng2023Responsibility,Tang2022AdHoc,Warszawski2017ACIDRain}.
% Unserializable behaviors are difficult to detect, debug, and fix~\cite{Cheng2023Responsibility,Tang2022AdHoc,Warszawski2017ACIDRain}.
% \yang{The above two citation places look similar to me. Are we trying to convey something different here?}
% They are elusive, occurring intermittently in production runs.

Prior work has introduced techniques to find unserializable behaviors in data store applications under weak isolation, but has scalability or accuracy limitations.
Static analysis can find unserializable behaviors, but its precision scales poorly with program complexity, leading to many false positives (infeasible unserializable behaviors)~\cite{serializability-for-causal-consistency-2018,nagar2018automated,clotho}.
Dynamic analysis can avoid false positives by analyzing only the observed execution~\cite{monkeydb,serializability-for-eventual-consistency-2017},
or it can extrapolate from an observed execution but report numerous false positives~\cite{isodiff,Warszawski2017ACIDRain}.
% Hybrid static--dynamic analysis can analyze an execution in an order-oblivious way but reports false positives~\cite{isodiff}.
% \mike{This sentence is kinda vague. I removed it since we cover it in \S\ref{sec:related}.}
% \mike{Is that a reasonable framing of IsoDiff?} \yang{Yes to me.}
% \mike{Is there a recent model checking arXiv paper that we should mention? Also a PLDI'23 paper (maybe the same paper?}
% \chujun{They're the same paper~\cite{Bouajjani2023}. I've cited it somewhere in the Design section.}
% \mike{Shall we mention model checking and \cite{Bouajjani2023}? I think of this paragraph in Intro as a summary of Related Work, in which case it'd make sense to mention everything that's covered there (very briefly).}
\S\ref{sec:related} discusses prior work in more detail.

\subsubsection*{Motivating example}
Algorithm~\ref{alg:credit} shows code of a transactional data store application. The $\mathit{DataStore}$ provides a key--value interface. Our execution model requires that every $\mathit{get}$ (read) or $\mathit{put}$ (write) operation to execute in a transaction, so an operation starts a new transaction if the current session (i.e., client) is not in a transaction. A $\mathit{commit}$ operation ends the session's ongoing transaction.

\begin{figure}[t]
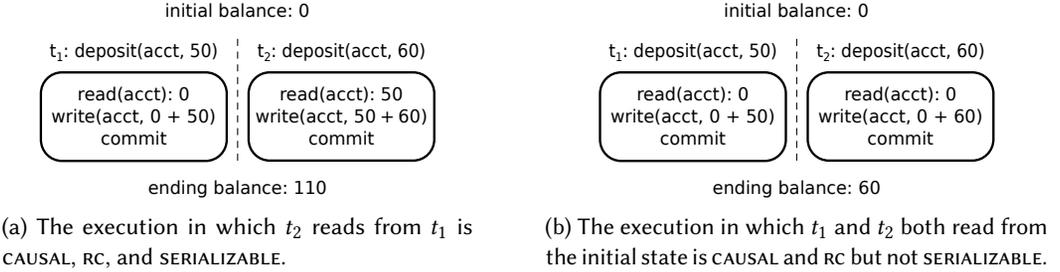

     \centering
     \begin{subfigure}[h]{0.45\textwidth}
         \centering
         \includesvg[inkscapelatex=false,scale=1.1]{img/motivate.svg}
         \caption{The execution in which $t_2$ reads from $t_1$ is \causal, \rc, and \serializable.}
         \label{fig:motivating_serial}
     \end{subfigure}
     \hfill
     \begin{subfigure}[h]{0.48\textwidth}
         \centering
         \includesvg[inkscapelatex=false,scale=1.1]{img/motivate_alt.svg}
         \caption{The execution in which $t_1$ and $t_2$ both read from the initial state is \causal and \rc but not \serializable.}
         \label{fig:motivating_causal}
     \end{subfigure}

     % \mike{``Execution'' seems too close to ``execution history.'' How about adding the word ``informal'' or something like that?}
     \caption{Different executions of two sessions (clients) concurrently on the same account.}
     \label{fig:motivating_example}
\end{figure}

Figure~\ref{fig:motivating_example} shows two different executions of the application. In each execution, two sessions (i.e., clients) call \textsc{deposit} concurrently on the same empty account to deposit 50 and 60, respectively.
Developers would expect that the ending balance will be 110, which is the only serializable outcome.
% and is likely to be the case in practice (depending on the data store implementation and exact timing).
However, under weak isolation levels \causal and \rc, the ending balance may be 110, 50, or 60.
% \chujun{Is it safe to say that under most circumstances the ending balance would more likely to be 110? Then it would make sense to use our technique to predict the other outcomes such as 50 or 60. Our technique also avoids predicting 0 ending balance because that violates causal consistency.}
% \mike{Yes. Revised.}

\begin{algorithm}[t]
\small
\caption{A procedure in a data store application that deposits money in an account.}\label{alg:credit}
\begin{algorithmic}
\Procedure{deposit}{$\mathit{account}$, $amount$}
\State $\mathit{balance} \gets \mathit{DataStore}.\mathit{get}(\mathit{account})$ \Comment{Read balance; implicitly starts transaction if not in one}
%\State $balance \gets balance + amount$
\State $\mathit{DataStore}.\mathit{put}(\mathit{account}, \mathit{balance} + \mathit{amount})$ \Comment{Update balance}
\State $\mathit{DataStore}.\mathit{commit}()$ \Comment{Commits transaction}
\EndProcedure
\end{algorithmic}
% \mike{Use mathit for multi-letter words in math mode (note the difference above).}
\end{algorithm}

\subsubsection*{Contributions}
This paper introduces \isopredict, the first predictive analysis for transactional data store applications, and shows that the approach is effective at finding unserializable behaviors.
Given a serializable execution such as Figure~\ref{fig:motivating_serial} as input, \isopredict finds an unserializable execution such as Figure~\ref{fig:motivating_causal}.
% \footnote{If our analysis is given an \emph{unserializable} execution, it will report the same or a different unserializable execution.}
% with an ending balance of 110 as input, can find and report an unserializable execution with an ending balance of 50 or 60.
% \yang{Do we have to start from a serializable execution?} \mike{Yes and no; see footnote I added but then commented out. Removed ``serializable'' above.}
% \subsubsection*{Contributions}
\Isopredict uses
% \emph{sound\footnote{Following the predictive analysis literature~\cite{said-nfm-2011,wcp,smarttrack,rvpredict-pldi-2014,predict-deadlocks},
% a predictive analysis is \emph{sound} if it reports no false positives (infeasible executions).}
\emph{dynamic predictive analysis}, which analyzes an observed execution of a program
and detects alternative \emph{feasible, unserializable} executions of the program.

Predictive analysis is powerful because, in essence, it explores many executions at once.
% and detects only feasible behaviors.
% In contrast, traditional dynamic analysis and testing explore a single execution at a time.
% The advantage of predictive analysis over traditional dynamic analysis and testing is that users do not need to repetitively test the application in the hope of finding any anomalies.
% Given the trace of a single execution, from predictive analysis predicts unserializable behaviors that they are known to be feasible they are feasible.
% In the context of shared-memory programs, much work has introduced predictive analysis to predict data races, atomicity violations, and deadlocks (e.g.,~\cite{said-nfm-2011,wcp,smarttrack,rvpredict-pldi-2014,sinha2012,predict-deadlocks}), but the techniques are not directly applicable to predicting unserializable behaviors in data store applications.
% We are not aware of any sound predictive analysis for transactional data store applications.
% which is not that surprising considering the challenges associated with encoding predicted unserializable behaviors under weak consistency, as we show.
% This paper introduces, to the best of our knowledge, the first predictive analysis for transactional data store applications.
% Our analysis, called \emph{\isopredict}, takes an observed serializable execution under weak consistency as input, and outputs a feasible unserializable execution
% if one is knowable from the observed execution. (Given an \emph{unserializable} execution, \isopredict outputs the same or a different unserializable execution.)
To predict an unserializable execution from an observed serializable execution, \isopredict generates SMT constraints
% based on the observed execution
that encode execution feasibility, unserializability, and weak isolation level (\causal or \rc), and uses an off-the-shelf SMT solver to solve them.
% \Isopredict has the potential to report infeasible unserializable behaviors, since transactions that committed in the observed execution may abort in the predicted execution (or vice versa).
%To explore the space, 
We introduce analysis variants that trade coverage for performance, and precision for coverage.
To account for the possibility of predicting infeasible executions, \isopredict can optionally \emph{validate} a predicted unserializable execution. % by attempting to execute it.
An evaluation on transactional data store benchmarks shows that \isopredict is effective at predicting unserializable executions from observed executions under \causal and \rc.
More than 99\%
% \chujun{Changed from 95\% to 99\%}
of predictions are validated as feasible executions.

While prior work introduces predictive analysis for shared-memory programs~\cite{said-nfm-2011,wcp,smarttrack,rvpredict-pldi-2014,sinha2012,predict-deadlocks},
to our knowledge \isopredict is the first predictive analysis approach for transactional data store applications, which present unique challenges (\S\ref{sec:related}).
Compared to prior work MonkeyDB~\cite{monkeydb}, \isopredict is comparably effective at finding unserializable executions of the evaluated programs (\S\ref{subsec:cmp_monkeydb}).
% of the evaluated benchmarks.
However, \isopredict and MonkeyDB use completely different approaches to find erroneous executions. While MonkeyDB uses random exploration to produce an erroneous execution, \isopredict uses predictive analysis to evaluate an equivalence class of many executions at once.
Furthermore, MonkeyDB requires applications to run on its specialized data store, while \isopredict's predictive analysis approach is in principle suitable for analyzing executions from any data store, although demonstrating so is outside the scope of this paper.
%We discuss the differences between MonkeyDB and \isopredict in depth in \S\ref{subsec:cmp_monkeydb}.
% \yang{In system community, the introduction should present a few key challenges. Maybe PL papers do not require that?
% When I read ``While prior work introduces predictive analysis
% for shared-memory programs [ 28, 30, 37--39 , 42] (§8), to our knowledge IsoPredict is the first pre-
% dictive analysis for transactional data store application'', my natural expectation is that the next sentence should say how transactional data stores are different from shared-memory and thus introduce new challenges that cannot be addressed by existing techniques
% and I think we have good answers for that.}
% \mike{Revised that sentence, admittedly with a forward reference.}
%Overall, this paper introduces the first predictive analysis for transactional data store applications and shows that the approach is effective at finding unserializable behaviors.
% that is competitive with alternative approaches.

%% file: 2.background.tex
\section{Background}

This section introduces this paper's formalisms for weakly isolated executions of transactional data store applications, which are closely based on the axiomatic framework of Biswas and Enea~\cite{biswas2019}.
% Biswas and Enea define weak isolation levels using an axiomatic framework leveraging that places constraints on a ``commit order'' over an ``execution history''~\cite{biswas2019}.
We use this framework
because it supports a variety of isolation levels, is well suited to encoding as constraints, and has been employed by recent work~\cite{monkeydb,Bouajjani2023}.

\subsection{Weakly Isolated Execution Histories}
\label{subsec:background-execution-history}

% and is well suited to encoding in a constraint solver.
% \mike{Looking at Enea's work since 2019, it looks like Enea et al.\ may have several more papers that build on \cite{biswas2019}?}
% \mike{Most is quite theoretical and not very related.}

% Following Biswas and Enea's framework,
% database isolation levels are defined by a series of restrictions on commit orders over an abstract execution history that includes transactions, session orders and write-read relationships.
A transactional data store is modeled as a distributed store of key--value pairs.
% where the exact values are ignored since they do not affect the result of our predictive analysis. \yang{Do you plan to introduce predictive analysis before this? Do any analysis methods use exact values?}
A data store application performs read (get) and write (put) operations on keys, all executed in transactions. Non-transactional applications can be handled by treating each read and write operation as a separate transaction.
% Every event is in exactly one transaction.
An execution consists of \emph{events} in committed transactions (aborted transactions are not part of an execution). Each event is either \readevent{k}, or \writeevent{k} or \commitevent, where $k$ is a key.
% \[
% \mathit{Event} \in \bigcup_{k \textnormal{ is a key}} \{\mathit{read}_{k}, \mathit{write}_{k}\}
% \]
% \mike{The formalisms $\mathit{Event}$, $\mathit{read}_k$, and $\mathit{write}_k$ are (almost) never used again. What's used instead, i.e., how to connect this definition to the rest of the paper?}
Other operations, such as insertion into and deletion from a set, can be modeled in terms of reads and writes.
Multiple clients may open connections, or \emph{sessions}, to the data store.
If a session is not in a transaction, its next event implicitly starts a new transaction, ensuring every event is in a transaction.
The \commitevent event ends the current transaction.
Within a session, transactions are ordered by the strict partial
% \spyros{``totally ordered by a partial order'' --- huh?!}
order \emph{session order} (\so):
\begin{align*}
%\label{eq:session_order}
\so(t_1, t_2) \coloneq \textnormal{$t_1$ precedes $t_2$ in the same session}
\end{align*}

An important property of an execution is \emph{which} write each read reads from.
The strict partial order $\writeread_k$ (write--read on key $k$) orders transactions if one reads from the other:
\begin{align*}
\writeread_k(t_1, t_2) &\coloneq \textnormal{$t_2$ reads the write of $t_1$ on $k$}
\end{align*}
If a read reads from a write in the same transaction, the read is not included as an event in the transaction (and thus this write--read ordering is not included in $\writeread_k$).
% If a \writeevent{k} is followed by a \readevent{k} in the \emph{same} transaction, the read always reads from its own transaction, but this read is \emph{not} included in the history.
If a transaction writes $k$ multiple times, only the last write is included as an event in the transaction.
% If a transaction issues multiple writes on the same key $k$, only the last \writeevent{k} is included in the history.
Thus a \readevent{k} event always reads from a \writeevent{k} in another transaction, which is the transaction's last write to $k$.
If a transaction $t$ reads $k$ from the data store's initial state,
then $\writeread_k(t_0,t)$, where $t_0$ is a special transaction representing the initial state.
% \spyros{Maybe we should say $t_0$ everywhere instead of $t_0$? In the remainder of the text we seem to be using $t_i$ to indicate transactions and $T$ to indicate a set of transactions. (Except the figures, it would be nice to have uniform presentation.) We can leave this for the camera ready.}
% \chujun{Updated all text and figures to use lower case.}
% Every read event has exactly one \emph{last writer}
% transaction, unless the read is preceded in the same transaction by a write to the same key (such reads have no last-writer transaction). A read from
% The strict partial order $\writeread_k$ orders
% Then there exists a \emph{write--read} order, \writeread, that orders reads after their last-writer transactions, defined in terms of write--read orders on each key, $\writeread_k$:
The union of $\writeread_k$ over all keys is \writeread, i.e.,
$\writeread \coloneq \bigcup_{k \textnormal{ is a key}} \writeread_k$.
% \begin{align*}
% \writeread(t_1, t_2) &\coloneq \exists k, \writeread_k(t_1, t_2)
% \end{align*}
%\yang{Does it require $t_2$ to happen after $t_1$? If the value $t_2$ reads is different from the value $t_1$ writes, does it still form
%the write-read relation? Actually even with values, there are still confusions since a $t_3$ may write the same value to the key. That's why
%Adya's work introduces the concepts of versions.}
%\chujun{I think in our model it's the other way around: when $t_2$ reads from $t_1$, we consider $t_2$ happens after $t_1$. MonkeyDB said it is impossible to use Adya's version in a distributed setting because we don't have a definitive ordering between conflicting accesses from different nodes. \co is the alternative to version in our axiomatic framework}
% \subsubsection*{Happens-before}
The transitive closure of \so and \writeread is \emph{happens-before} order, i.e.,
% \begin{align*}
% \hb(t_1,t_2) \coloneq so(t_1,t_2) \lor wr(t_1,t_2) \lor \exists t, (hb(t_1,t) \land hb(t,t_2))
% \end{align*}
% Or more simply,
$\hb \coloneq (\so \cup \writeread)^+$.
An execution \emph{history} of a data store application is the set of all committed transactions (\tx),
%\yang{Actually under very weak isolation levels like read uncommitted, an aborted transaction may be visible to other transactions. Not sure about causal though: does a causal consistent database ensure
%read will only get committed data?}
%\chujun{MonkeyDB's prior work~\cite{biswas2019} ignored aborted transactions in their models (including Causal), so I guess it's safe to assume only committed data is read}
session order (\so), and write--read order (\writeread), i.e.,
$History \coloneq \langle T, so, wr \rangle$.
Every history includes the special transaction $t_0$ mentioned above that represents the initial state. $t_0$ implicitly writes the initial value to every key, and $t_0$ is \so-ordered before all other transactions.

\subsubsection*{Example}

Figures~\ref{fig:causal_serial_history} and \ref{fig:causal_unserial_history} each show an execution history as a graph.
% corresponding to Figures~\ref{fig:motivating_serial} and \ref{fig:motivating_causal}, respectively.
Transactions are boxes containing read and write events implicitly concluded by a commit event.
$t_1$ and $t_2$ execute in different sessions, and $t_0$ is the initial state transaction.
The $\writeread_k$ edges indicate each read's writer.
% In each history, the transactions $t_1$ and $t_2$ execute in two different sessions, so there are no \so edges between them.
% The special transaction $t_0$ representing the initial state of the data store writes a value to every key accessed in the history and is \so ordered before all other transactions.
% neither figure shows these \hb edges explicitly because in each figure, $t_1$ and $t_2$ are already (transitively) \hb ordered before $t_0$ by \writeread edges.

\subsection{Serializablility}
\label{subsec:background-serializablity}

An execution history $\langle T, so, wr \rangle$ is \emph{\serializable} if and only if
it could have been produced by a serial execution of the transactions in $T$.
(In a serial execution, transactions execute one at a time, and every read to $k$ reads from the most recent write to $k$.)
Equivalently,
% the axiomatic framework says that
an execution is \serializable if and only if there exists a \emph{commit order}, \co, with the following constraints:
% There are two constraints that the axiomatic framework places on \co to ensure
% that it represents a serial execution with execution history $\langle T, so, wr \rangle$.
(1) \co must be consistent with happens-before (\hb) order.
% i.e., $\hb \subseteq \co$.
% ($\hb \coloneq (\so \cup \writeread)^+$).
% In order to reason about the consistency level of the execution history, we need to consider another property called commit order (\co).
% \co is a strict total order of all committed transactions and it directly affects the final state of the database.
% \mike{\co is just a formalism used to define whether an execution is \causal, \serializable, etc. In any case, this paragraph won't make sense to readers as is. Here I think we can just omit mention of \co here, or mention briefly that the next two subsections will both use the same kind of formalism, called \co.}
% The axiomatic framework contains a set of properties for different consistency levels (e.g., causal consistency and serializability), and a history satisfies certain consistency level if there exists a \co that satisfies the corresponding axiom.
% For the purpose of our work, we focus on Causal and Serializable.
(2)
% for \co to represent a valid serial execution with history $\langle T, so, wr \rangle$,
% so every read reads from the latest write,
a transaction that writes to $k$ cannot be \co-ordered between two transactions ordered by $\writeread_k$. The second constraint's ordering is called \emph{arbitration order} and represented by the strict partial order \arserial, which is defined as follows:
%  that must be ordered in order for \co to be consistent with $\writeread_k$:
% \mike{More intuition for why it's necessary and sufficient as \serializable's arbitration order?}
% \yang{Yes, I find it hard to understand why defining arbitration order.}
% \mike{Revised and added intuition.}
% Serializable Axiom~\cite{biswas2019} resolves the arbitration of conflicting writes in a similar way to Causal Consistency, except that the condition where the conflicting write being a Happens Before predecessor of the read transaction is replaced with another one where conflicting write is a commit order ($co$) predecessor of the read:
% \begin{multline} \label{eq:serial_ar}
% \forall t_1, t_2 \in T, t_1 \ne t_2, \forall k, \\
% \co(t_1, t_2) \Leftarrow \textnormal{$t_1$ and $t_2$ write to $k$} \\
% \land\:\exists t_3 \in T\setminus\{t_1, t_2\}, \writeread_k(t_2, t_3) \land \co(t_1, t_3)
% \end{multline}
% \mike{Also need equation stating that happens-before ordering (which hasn't been defined yet) implies \co ordering, right?}
% \chujun{That is true, I'll add those implications once I'm done modifying the explanation on \co}
% \mike{On second thought, I understand why it's unnecessary to explicitly state $\hb \implies \co$.}
% To distinguish \co implied by the arbitration between conflicting writes (Equation~\ref{eq:serial_ar}) from the rest, we call them \arserial just for illustration purposes.
\begin{align} \label{eq:serial_ar}
\arserial(t_1, t_2) \coloneq \exists k, \textnormal{$t_1$ and $t_2$ write to $k$}
\land\:\exists t_3 \in \tx, wr_k(t_2, t_3) \land \co(t_1, t_3)
\end{align}
% \mike{What's the purpose of the above equation and \arserial? Isn't it subsumed by Equation~\ref{eq:serial_ar}?}
% \chujun{\arserial might makes it easier to explain what \comin is consisted of (Equation~\ref{eq:comin}). Might have to move it to the design section}
% \mike{I think it's fine and good to define and use \arserial here.}
Note the circular dependency between \arserial and \co: Commit ordering may imply additional arbitration ordering, which in turn may imply additional commit ordering.
% it is insufficient to define \serializable in terms of whether $(\hb \cup \arserial)^+$ is acyclic.
% $(\hb \cup \arserial)^+$ is a partial order, while \co is a total order, and the additional edges in \co
% may require additional edges in $\arserial$ according to Equation~(\ref{eq:serial_co}), imperiling acyclicity.
This property leads to challenges in encoding SMT constraints that \S\ref{sec:predictive-analysis} explains and addresses.
% \noindent
% Arbitration constraints with rank:
% \begin{multline}
% \forall t_1, t_2 \in T, t_1 \ne t_2, \forall k, \\
% \co(t_1, t_2) \Leftarrow \textnormal{$t_1$ and $t_2$ write to $k$} \\
% \land\:\exists t_3 \in T\setminus\{t_1, t_2\},\:wr_k(t_2, t_3) \land co(t_1, t_3) \\
% \land \: rank(t_1, t_2) > rank(t_1, t_3) \\
% \end{multline}
%
% \noindent
% \begin{minipage}{1\linewidth}
% Transitivity constraints with rank:
% \begin{multline}
% \forall t_1, t_2 \in T, \\
% \co(t_1, t_2) \Leftarrow hb(t_1, t_2) \lor \\
%                         (\exists t', co(t_1, t') \land co(t', t_2) \: \land \\
%                                      rank(t_1, t_2) > rank(t_1, t') \land rank(t_1, t_2) > rank(t', t_2)) \\
% \end{multline}
% \end{minipage}
% As mentioned above,
% a history is serializable if and only if there exists a total order \co that respects happens-before (\hb) and % arbitration orders (\arserial). More formally, \yang{So \co has nothing to do with the real timing of transaction commits?}
% \chujun{That is correct. \co is not part of the execution. It's a potential order in which transactions would commit to a "global memory", but such a global memory doesn't exist in weakly isolated distributed databases.}
Thus a history is \serializable if and only if there exists a \co that is consistent with \hb and \arserial:
\begin{align*}
%\label{eq:serial_co}
\langle T, so, wr \rangle \textnormal{ is } \serializable \iff
\exists \co, \hb \cup \arserial \subseteq \co
% where: \co is a strict total order of committed transactions}\\
\end{align*}
Equivalently, the history is \serializable if and only if there exists \co such that $(\hb \cup \arserial \cup \co)^+$ is acyclic.
An execution is \unserializable if and only if it is \emph{not} \serializable.

% \mike{In, $\writeread \cup \so \cup \arserial \cup \co$ above, can \writeread and/or \arserial and/or \co be removed?}
% \chujun{I don't think \writeread, \arserial can be removed because they are all partial orders while \co is a total order. The purpose of that equation is that \co does not contradict \writeread, \arserial, etc}
% \mike{Agreed. Revised.}
% \yang{By reading this, my feeling is that if $(\hb \cup \arserial)^+$ is acyclic, then we can always find a \co? If so, why introducing \co? }
% \chujun{In Causal Consistency we don't really need \co, but in Serializability we will need \co for the recursive implication. The reason \co is included in Causal Consistency is because MonkeyDB's axiomatic framework uses \co as a "template" that fits all consistency models}
% \mike{Addressed this issue}

\subsubsection*{Example}

\newcommand\ltco{\ensuremath{<_{\co}}}

Figure~\ref{fig:causal_serial_history}'s history is \serializable because there exists a commit order ($t_0 \ltco t_1 \ltco t_2$), shown in Figure~\ref{fig:causal_serial_co}, that is consistent with the \serializable axioms.
% (\S\ref{subsec:background-serializablity}).
Note that the arbitration rule (Equation~\ref{eq:serial_ar}) never applies in Figure~\ref{fig:causal_serial_history}, and so Figure~\ref{fig:causal_serial_co} shows no \arserial edges.

The history in Figure~\ref{fig:causal_unserial_history} is \unserializable because there does \emph{not} exist a commit order that satifies the \serializable axioms. For example, as Figure~\ref{fig:causal_unserial_co} shows, if $\co(t_1, t_2)$, then $\arserial(t_1, t_0)$ by Equation~\ref{eq:serial_ar}, which implies $\co(t_1, t_0)$ and thus \co is cyclic.
Alternatively, if $\co(t_2, t_1)$, then $\arserial(t_2, t_0)$ and thus $\co(t_2, t_0)$, and again \co is cyclic.
% Since no total commit order exists satisfying the \serializable axioms, the execution history is \unserializable.

% Clearly $t_0 \ltco t_1$ and $t_0 \ltco t_2$ because \hb order implies commit order.

% \begin{figure}[ht]
%     \centering
%     % \includesvg[width=0.35\textwidth]{img/causal_not_serial.svg}
%     \includegraphics[width=0.25\textwidth]{img/causal_and_unserial}
%     \begin{minipage}{0.45\textwidth} % choose width suitably
% {\centering\footnotesize
% $Session_0 t_0$: write(x), write(y) \\
% $Session_1 t_1$: write(x), read(y) \\
% $Session_2 t_2$: write(y), read(x) \par}
% \end{minipage}
%     \caption{A \causal and \unserializable Execution}

%     \label{fig:causal_unserial}
% \end{figure}

\begin{figure}
\centering
    \begin{subfigure}[h]{0.47\textwidth}
         \centering
         \includesvg[inkscapelatex=false,scale=1]{img/causal_and_serial.svg}
         \caption{Execution history}
         \label{fig:causal_serial_history}
     \end{subfigure}
     \hfill
     \begin{subfigure}[h]{0.47\textwidth}
         \centering
         \includesvg[inkscapelatex=false,scale=1]{img/causal_and_serial_co.svg}
         \caption{A \co (dashed arrows) consistent with the \serializable axioms.}
         \label{fig:causal_serial_co}
     \end{subfigure}

% \mike{\antidependency doesn't exist in the axiomatic framework according to this section. Examples in this section should omit \antidependency.}
% \chujun{Changed it to \co edges instead.}

% \mike{Need \hb edges from $t_0$.}
% \chujun{Isn't that implied by the two \writeread edges?}
% \mike{Explained that in the text.}
\vspace*{-0.5em}
\caption{A \causal, \textbf{\serializable} history corresponding to Figure~\ref{fig:motivating_serial}.}

% \yang{Since the text uses $t_0$, please make it consistent here, though $t_0$ is more frequently used in DB papers. $t_0$ feels like a timestamp to me.}
% \chujun{Changed it.}
% \chujun{Not sure how to display the two figures side by side.}
\label{fig:causal_serial_history_both}
% \end{figure}
\bigskip
% \begin{figure}
\centering
    \begin{subfigure}[h]{0.47\textwidth}
         \centering
         \includesvg[inkscapelatex=false,scale=1]{img/causal_and_unserial.svg}
         \caption{Execution history}
         \label{fig:causal_unserial_history}
     \end{subfigure}
     \hfill
     \begin{subfigure}[h]{0.47\textwidth}
         \centering
         \includesvg[inkscapelatex=false,scale=1]{img/causal_and_unserial_co.svg}
         \caption{A \co (dashed arrows) inconsistent with the \serializable axioms
         (contradiction shown in red).}
         % A \serializable \co (dashed arrows) because it leads to an arbitration edge from $t_1$ to $t_0$ that contradicts $t_0 \ltco t_1$ (shown in red)}
         \label{fig:causal_unserial_co}
     \end{subfigure}
\vspace*{-0.5em}
\caption{A \causal, \textbf{\unserializable} history corresponding to Figure~\ref{fig:motivating_causal}.}
\label{fig:causal_unserial_both}
\end{figure}

\subsection{Causal Consistency}
\label{subsec:background-causal}

Causal consistency (\causal) is a weak isolation level that
preserves the order of operations that are causally related~\cite{ahamad95causalmemory}.
\Causal is of theoretical and practical interest
because it is the strongest isolation level achievable when a data store requires availability under network partitions~\cite{burckhardt2014,lynchCAP,mahajan11cacTR}.
% \mike{Give intuitive explanation of causal consistency?}

% Following Biswas and Enea,
Similar to \serializable, \causal is defined in terms of whether there exists a commit order that
is consistent with happens-before (\hb) and an arbitration order, which we call \arcausal to distinguish it from the arbitration order for \serializable (\arserial).
% ~\cite{biswas2019}.
% \causal's arbitration order (\arcausal) is a partial order on transactions that write to the same key.
Two transactions $t_1$ and $t_2$ are ordered by $\arcausal$ if they write the same key and if there is a third transaction $t_3$ that happens-after $t_1$ ($\hb(t_1,t_3)$) and reads from $t_2$'s write to the same key ($\writeread(t_2,t_3)$).
More formally,
% The Causal Axiom~\cite{biswas2019} states that for any write-read pair where $t_2$ writes $k$ and $t_3$ reads $k$, any conflicting transaction $t_1$ that writes to $k$ must precede $t_2$ in the commit order $co$ as long as $t_1$ is a $(wr \cup so)^+$ predecessor of $t_3$.
% For simplicity, we call $(wr \cup so)^+$ Happens-Before relation, which is the transitive closure of write-read relation and session order:
% \begin{equation} \label{eq:happens_before}
%     hb \coloneq (wr \cup so)^+
% \end{equation}
\begin{align} \label{eq:causal_ar}
\arcausal(t_1, t_2) \coloneq \exists k, \textnormal{$t_1$ and $t_2$ write to $k$}
\land \exists t_3 \in \tx, wr_k(t_2, t_3) \land hb(t_1, t_3)
\end{align}
% %
A history is \causal if and only if there exists a commit order consistent with \hb and \arcausal:
\begin{align} \label{eq:causal_history}
\langle T, \so, \writeread \rangle \textnormal{ is } \causal \iff
\exists \co, \hb \cup \arcausal \subseteq \co
\end{align}
Equivalently, a history is \causal if and only if $(\hb \cup \arcausal)^+$ is acyclic.%
% \mike{Changed $\hb \cup \arcausal \cup co)^+$ to $(\hb \cup \arcausal)^+$. Isn't that correct?}
% \chujun{That's correct.}
% The definitions of \serializable and \causal are structurally quite similar. The difference between them is their arbitration orders (\arserial and \arcausal).
\footnote{Unlike \serializable, \causal can be defined in terms of whether $(\hb \cup \arcausal)^+$ is acyclic, which implies that a total commit order must exist.
In contrast, \serializable's arbitration order (\arserial) is dependent on the commit order, so \serializable must be defined in terms of whether $(\hb \cup \arserial \cup \co)^+$ is acyclic.}
% \mike{(Although I may have been the one to write that paragraph) it doesn't make sense to me now why what needs to be acyclic is different between the two models.}

\subsubsection*{Example}

The history in Figure~\ref{fig:causal_serial_history} is \causal because
there exists a commit order $t_0 \ltco t_1 \ltco t_2$ that is consistent with the \causal axioms.
(Or, since the history is \serializable, which is strictly stronger than \causal, the history must be \causal.)
The history in Figure~\ref{fig:causal_unserial_history} is \causal because there exists a commit order, $t_0 \ltco t_1 \ltco t_2$ (or $t_0 \ltco t_2 \ltco t_1$),
that is consistent with the \causal axioms.

\subsection{Read Committed}
\label{subsec:background-rc}

Read committed (\rc) is a popular weak isolation level because of the balance between performance and consistency it provides~\cite{Berenson1995Critique}.
Whereas \causal requires transactions ordered by happens-before (\hb) to be viewed by other transactions in the same order, \rc's arbitration order, \arrc, only applies to write transactions that are read by multiple read events from the same transaction.
% that is strictly weaker than \causal.
% In \causal, transactions that are ordered by the happens-before relations must be viewed by other transactions in the same order.
% This restriction does not exist for \rc.
% As suggested by its name, \rc only requires read events to read from committed transactions.
% \rc's arbitration order among writer transactions only applies when they are being read by multiple read events from the same transaction, where the arbitration order needs to be consistent with these read events' program order.
\newcommand{\eventwriteread}{\ensuremath{\overline{\mathit{wr}}}\xspace}%
More formally, \rc is defined based on whether there exists a commit order that is consistent with \hb and \arrc, which is defined as follows:
% Intuitively, \rc's arbitration order is weaker than causal consistency's because transactions ordered by session order no longer implies arbitration orders.
% \mike{Why mention this? In any case, first show and explain the rule.}
\begin{align} \label{eq:rc_ar}
\arrc(t_1, t_2) \coloneq \exists k, \textnormal{$t_1$ and $t_2$ write to $k$}
\land
\exists\:\!\alpha, \beta, \po(\beta,\alpha) \land \eventwriteread_k(t_2,\alpha) \land \exists k', \eventwriteread_{k'}(t_1,\beta)
% \exists t_3 \in \tx, \exists\:\!\alpha, \beta \in t_3, \po(\beta,\alpha)
\end{align}
where \po is \emph{program order}, a strict partial order that orders events within a transaction; and $\eventwriteread_k(t,e)$ is true if and only if $e$ is a read \emph{event} that reads from a write in transaction $t$ (and thus $e \ne t$).
Thus $\alpha$ and $\beta$ must be events in the same transaction such that $\alpha$ is a \readevent{k} event that reads from \writeevent{k} in $t_2$, and $\beta$ is a read event that reads from any write in $t_1$.
% \mike{Tried to crisp up the definition above. Please check.}
% \chujun{Looks good.}
% \mike{We have to express this as text since we don't have formalisms ordering individual events. How do the generated constraints handle that?}
% \chujun{It's part of the Python code where each event has an attribute representing its program order. For all $\beta$ that come before $\alpha$ in the program order, we generate rc constraints for them based on formula 4. For all the other events we simply set their rc constraints as false.}
An execution history is \rc if and only if there exists a commit order that is consistent with \hb and \arrc:
\begin{align} \label{eq:rc_history}
\langle T, \so, \writeread \rangle \textnormal{ is } \rc \iff
\exists \co, \hb \cup \arrc \subset \co
\end{align}

\subsubsection*{Example}

The execution histories in Figures~\ref{fig:causal_serial_history} and \ref{fig:causal_unserial_history} are \rc because there exist commit orders (in fact, the same commit orders used to establish \causal) satisfying the above condition. Or, the histories are \rc because they are \causal, which is strictly stronger than \rc.

%% file: 3.design.tex
\section{\IsoPredict Overview}
\label{sec:design-overview}

% \emph{\Isopredict} is an analysis for predicting \unserializable executions from an observed \serializable execution on a weakly isolated data store.
% Given an execution trace from a data store providing a weak isolation model $M$ (\causal or \rc), \isopredict either (1) produces a \emph{predicted execution} that is feasible, that is valid under $M$, and is unserializable; or (2) it reports that no such execution exists.
% \Isopredict can optionally validate the feasibility of the predicted execution in the face of potential divergence by attempting to execute the predicted execution under $M$ and checking unserializability.

% \subsection{Overview}

\begin{figure}[tp]
  \centering
  \includesvg[inkscapelatex=false,width=0.80\textwidth]{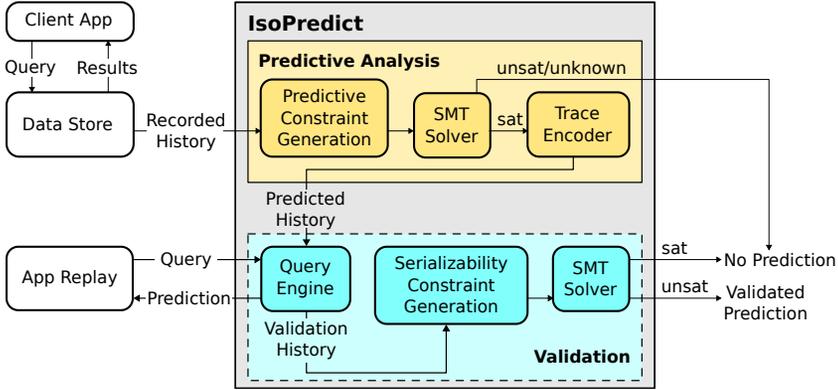}

 % \mike{Label the arrows between application and data stores as ``database connection'' or similar, and/or use a different kind of arrow?}

 % \mike{I'm not sure whether this belongs in Design or Eval (or Intro), but it doesn't seem like an implementation detail. Putting it at the beginning of Design would make it possible to reference the figure at relevant places in Design.}
% \mike{Change ``check'' to ``checker.'' Make numbers bigger. \url{https://tex.stackexchange.com/questions/7032/good-way-to-make-textcircled-numbers}}
  % \mike{Change constraint generation to predictive constraint generation?}
  \caption{\Isopredict's components and workflow.}
  \label{fig:impl}
\end{figure}

\Isopredict consists of two main components, as shown in Figure~\ref{fig:impl}: \emph{predictive analysis} and \emph{validation}.

% \paragraph{Predictive analysis}

The predictive analysis component takes as input an \emph{observed} execution history that is recorded at the client application's backend data store, generates SMT constraints, and uses an SMT solver to find a predicted \unserializable execution if one exists.
% If a predictable \unserializable execution exists then, depending on how \isopredict is configured, the execution can either be reported to the user,
% or it can be sent to the validation component (the figure shows only the latter option).
\S\ref{sec:predictive-analysis} describes \isopredict's predictive analysis.

% \paragraph{Validation}

The validation component tries to execute the predicted execution history to determine if it is feasible, and it generates and solves constraints to determine if the resulting execution is \unserializable.
% \Isopredict optionally validates that a predicted execution is feasible by feeding the predicted history to a query engine that tries to ``play back'' the execution.
% (\S\ref{subsec:design-validation}).
%$ The query engine records its execution history (called the validation history) during the playback.
% After the playback process is finished, \Isopredict checks whether the validation history is \serializable.
If so,
% \isopredict has successfully predicted an abnormal execution that may lead to undesirable outcomes, and
\isopredict outputs the validated history alongside a visualization of the validated \unserializable execution.
\S\ref{subsec:design-validation} describes \isopredict's validation component.

Validation is optional; developers may choose to skip it for two reasons. First, it may be overkill---in our experiments, over 99\% of predicted \unserializable executions are successfully validated. Second, validation may be impractical if the application cannot be replayed easily.
% (for example, when the database is too big).
%, or if a query engine for validation that can handle the application is unavailable.
Validation is, however, useful to our evaluation to measure how many predicted executions are feasible.

% \mike{There's probably more to overview here, like the prediction strategies and validation.}
% \chujun{Added the validation part.}

\section{Predictive Analysis}
\label{sec:predictive-analysis}

\Isopredict's predictive analysis component takes as input an \emph{observed} execution history of a data store application.
% is recorded from the client application's backend data store.
The observed history $\mathit{History} = \langle \tx, \so, \ogwr \rangle$ consists of a set of transactions \tx, session order \so between transactions, and observed write--read ordering \ogwr. % and it is valid under a weak isolation model $M$ (\causal or \rc).
% \mike{What exactly is \ogwr? It's ordering between events (not transactions, right?). What do pairs in \ogwr (and \writeread) look like?}
% \chujun{\ogwr doesn't really contain any ordering, it is all the pairs of write--read transactions from the observed execution. The pairs look like ($t_{write}$, $t_{read}$).}
The goal of \isopredict is to find a feasible, \unserializable execution that is valid under a weak isolation model $M$ (i.e., \causal or \rc).
% (1) is feasible, i.e., the execution \emph{could} have occurred, based on the observed execution;
% (2) is unserializable; and
% (3) is valid under $M$.
To find such an execution,
\isopredict encodes and solves the following necessary and sufficient constraints for a \emph{predicted} execution history, $\mathit{History}' = \langle T', \mathit{so}, \writeread \rangle$:
% \mike{Changed $T$ to $T'$ because $T'$ does not need to be equal to $T$---it might be a subset of $T$, right?}
% \chujun{That makes sense}
\begin{enumerate}
    \item $\mathit{History}'$ must be a feasible execution prefix\footnote{We allow $T'$ to be a subset of $T$ to exclude transactions that may diverge from the observed execution (\S\ref{subsec:divergent}). An execution prefix is sufficient: If $\mathit{History}'$ exists, a full execution history exists that has $\mathit{History}'$ as a prefix and meets the criteria above.} of the program that produced $\mathit{History}$ (\S\ref{subsec:encode_execution}).
    \item $\mathit{History}'$ must be \unserializable (\S\ref{subsec:encode_unserial}).
    \item $\mathit{History}'$ must be valid under $M$ (\S\ref{subsec:encode_weak}).
\end{enumerate}
As an example, Figure~\ref{fig:causal_serial_history} shows a \serializable execution history that contains two deposit transactions (Algorithm~\ref{alg:credit}) running concurrently.
\Isopredict generates and solves the constraints sketched above, in order to predict the \causal and \rc but \unserializable execution from Figure~\ref{fig:causal_unserial_history}.
% by letting transaction $t_2$ read from the initial state $t_0$.
% \chujun{Added a short example on predictive analysis.}

\subsection{Encoding of Feasible Execution} \label{subsec:encode_execution}

This section describes the constraints that \isopredict generates to ensure that
$\mathit{History}' = \langle T', \so, \writeread \rangle$ is a feasible execution of the application that produced $\mathit{History} = \langle T, \so, \ogwr \rangle$.
% Intuitively, $\writeread \ne \ogwr$ because some reads in the predicted execution must have different last writers than in the observed execution in order for $\mathit{History}'$ to be \unserializable (since $\mathit{History}$ is \serializable).
% At the same time, the prediction must account for the potential downstream effects of a read reading a different value than in the observed execution, affecting subsequent control-dependent events.

% \Isopredict encodes the predicted execution as constraints on symbolic transactions.
% Symbolic transactions are defined as a custom data type in the SMT solver where its possible values are the universally unique transaction identifiers from the observed execution ($\mathit{\langle T, \so, \ogwr \rangle}$).
% \mike{Could be explained better.}

\subsubsection*{Session order}

The predicted execution must preserve the observed execution's session order (\so).
\Isopredict generates constraints over a Boolean SMT function $\smtso(t_1,t_2)$ that takes two transactions as input; a transaction is an SMT data type representing the set of all executed transactions $T$.
The analysis generates the following constraints to preserve the observed execution's \so:
% Session Order (\so) is the ordering between consecutive transactions from the same session.
% It is encoded as an SMT function that takes in 2 symbolic transactions and produces a Boolean outcome.
\begin{align*}
\multirow{2}{*}{$\forall t_1, t_2 \in T, t_1 \ne t_2, \quad$}
%\begin{cases}
\quad \boxed{\smtso(t_1,t_2)} & \quad \textnormal{if } \so(t_1,t_2) \\
\boxed{\neg\smtso(t_1,t_2)} & \quad \textnormal{otherwise}
%\end{cases}
\end{align*}
For clarity, SMT constraints generated by \isopredict are \boxed{\textnormal{boxed}} throughput the paper. The way to understand the above is that, for every $t_1,t_2 \in T$ such that $t_1 \ne t_2$,\footnote{%
Although the partial and total orders throughout the paper are irreflexive,
the analysis never needs to generate irreflexivity constraints (e.g., $\forall{t}, \boxed{\neg \phi_r(t,t)}$ for relation $r$) because it never generates any constraints that \emph{use} $\phi_r(t,t)$.}
the analysis generates a constraint---either $\smtso(t_1,t_2)$ or $\neg\smtso(t_1,t_2)$ depending on whether the transactions are ordered by \so.

% \chujun{trying out a different approach}
% \begin{align*}
% SO_{set} = \{(t_1, t_2) \vert (t_1.session = t_2.session) \\
% \land \textnormal{$t_1$ precedes $t_2$, where $t_1, t_2 \in T$}\}
% \end{align*}
% \begin{align*}
% %\begin{cases}
% \quad \boxed{\so(t_1,t_2)} & \quad \textnormal{if } (t_1,t_2) \in SO_{set}\\
% \boxed{\neg\so(t_1,t_2)} & \quad \textnormal{otherwise}
% %\end{cases}
% \end{align*}

\subsubsection*{Write--read order}

% Unlike session order, the write--read relation in the observed execution (\ogwr) is different in the predicted execution (\writeread).
Each read in the predicted execution can potentially read from any transaction that writes the same key.\footnote{Recall that a read to $k$ can only read from another transaction's \emph{last} write to $k$ (\S\ref{subsec:background-execution-history}).}
To help reason about multiple reads in a transaction to the same key that have different writer transactions (and to help exclude potentially divergent events; \S\ref{subsec:divergent}), we introduce the notion of an event's \emph{position}: In each session, events are numbered with monotonically increasing integers.
To ensure each read has exactly one writer transaction in the predicted execution, \isopredict introduces an SMT function $\smtchoice(s,i)$ that takes as input a session and the position of a read event in the session, and returns the writer transaction that the read reads from.
Like transactions,
sessions are a finite SMT data type representing the set of all sessions.
% \footnote{Since event positions are with respect to sessions rather than transactions, \smtchoice could take a session instead of a transaction as input. However, \smtchoice takes a transaction as input for simplicity since transactions are already SMT types.}
(Note that $\smtchoice(s,i)$ is left undefined if $i$ is not the position of a read event in $s$.) \Isopredict generates the following constraints to ensure that $\smtchoice(s, i)$ is equal to some transaction that writes the same key:
% %
% \begin{align*}
% & \forall k \textnormal{ is a key},
% \forall t_2 \textnormal{ reads } k,
% \forall i \in \posk(t_2), \\
% & \boxed{\bigvee_{t_1 \textnormal{ writes } k} \smtchoice(t_2, i) = t_1}
% \end{align*}
% where $\posk(t_2)$ is the set of positions of reads to $k$ in the in $t_2$.
\begin{align*}
& \forall k \textnormal{ is a key},
\forall t_2 \textnormal{ reads } k,
\forall i \in \posk(t_2),
\quad \boxed{\bigvee_{t_1 \ne t_2 \textnormal{ writes } k} \smtchoice(s_2, i) = t_1}
\end{align*}
where $s_2$ is $t_2$'s session, and $\posk(t)$ is the set of positions of reads to $k$ in transaction $t$.
% \mike{Should $t_1 \textnormal{ writes } k$ be changed to $t_1 \ne t_2 \textnormal{ writes } k$?}
% \chujun{Changed it.}
% \begin{align*}
% & \forall k \textnormal{ is a key},
% \forall r \textnormal{ reads } k,
% \quad \boxed{\bigvee_{e \textnormal{ writes } k} \smtchoice(r) = e}
% \end{align*}

% \chujun{We represent each read event as $read_k(t, i)$ where $k$ is the data key it reads, $t$ is the transaction this event belongs to and $i$ is the event's position in $t$'s program order.
% Similarly, we define write events as $write_k(t, i)$.
% We define a symbolic integer for every read event indicating which write event it reads from in the prediction.
% }

% \begin{align*}
% & \forall k \textnormal{ is a key},
% \forall t_2 \textnormal{ s.t.\ $t_2$ reads } k, \\
% & \boxed{\smtwritereadk(t_1,t_2) = \bigvee_{t_1 \in \tx, i, j \in po} choice(read_k(t_2, i)) = write_k(t_1, j)}
% \end{align*}
% \chujun{How about this?}

\Isopredict encodes $\writeread_k$ by generating constraints on Boolean SMT functions $\smtwritereadk(t_1,t_2)$:
% for each key $k$:
% %
\begin{align*}
& \forall k \textnormal{ is a key},
\forall t_1 \textnormal{ writes } k,
\forall t_2 \textnormal{ reads } k,
t_1 \ne t_2,
\quad \boxed{\smtwritereadk(t_1,t_2) = \bigvee_{i \in \posk(t_2)} \smtchoice(s_2, i) = t_1}
\end{align*}
where $s_2$ is $t_2$'s session.

To encode $\writeread(t_1,t_2)$, the analysis generates constraints on a Boolean SMT function $\smtwriteread(t_1,t_2)$ that represents the union of all $\smtwritereadk(t_1,t_2)$:
% %
\begin{align*}
& \forall t_1,t_2 \in \tx, t_1 \ne t_2,
\quad \boxed{\smtwriteread(t_1, t_2) = \bigvee_{k \textnormal{ is a key}} \smtwritereadk(t_1, t_2)} \\
\end{align*}

\subsection{Encoding Unserializability}
\label{subsec:encode_unserial}

This section describes how the analysis encodes constraints for the predicted execution to be \unserializable. The constraints must ensure that \emph{all possible commit orders
% (\co; \S\ref{subsec:background-serializablity})
are cyclic}.
%  (Equation~(\ref{eq:serial_co})).
\S\ref{subsubsec:exact-unser-constraints} presents an approach that encodes the needed constraints exactly,
% (technically, the negation of an existential quantifier),
resulting in long solving times. \S\ref{subsubsec:sufficient-unser-constraints} presents an alternative approach that encodes a sufficient
% but unnecessary
condition for unserializability,
which has lower solving time than the first approach, but still has high coverage in our experiments.

\subsubsection{Constraints that encode an exact condition}
\label{subsubsec:exact-unser-constraints}

To encode that no acyclic \co exists for the predicted execution history, \isopredict generates the following constraint:
% \isopredict must either generate an exponential number of constraints in the size of the execution (defeating the purpose of using the solver), or it must generate constraints with universal quantification---which SMT solvers are typically slow at solving~\cite{Leino2016}.
% We choose the lesser of two evils: \Isopredict generates the following constraint to find an execution history for which a total order \co does \emph{not} exist:
% %
\begin{align*}
\boxed{\forall \smtco, \neg\mathit{IsSerializable}(\smtco)}
\end{align*}
where $\mathit{IsSerializable}$ is defined as shown below.
Note that in the constraint above, $\smtco(t)$, which takes a transaction $t$ as input and evaluates to an integer indicating $t$'s position in the \co total order, is not an SMT function---it is a bound variable of the quantifier.
Function $\mathit{IsSerializable}$ is defined as follows:
\begin{empheq}[box=\fbox]{align*}
\mathit{IsSerializable}(\smtco) \coloneq \; & \mathit{Distinct}(\smtco(t_1),\dots,\smtco(t_n)) \; \land \\
& \bigwedge_{\forall t_1, t_2 \in T, t_1 \ne t_2}
(\smtwriteread(t_1, t_2) \lor \smtso(t_1, t_2) \lor \mathit{Arbitration}(t_1, t_2)) \Rightarrow \smtco(t_1) < \smtco(t_2)
% \land & \bigwedge_{\forall t_1, t_2 \in T, t_1 \ne t_2} \smtwriteread(t_1, t_2) \Rightarrow \smtco(t_1) < \smtco(t_2)\\
% \land & \bigwedge_{\forall t_1, t_2 \in T, t_1 \ne t_2} \smtso(t_1, t_2) \Rightarrow \smtco(t_1) < \smtco(t_2)\\
% \land & \bigwedge_{\forall t_1, t_2 \in T, t_1 \ne t_2} \mathit{Arbitration}(t_1, t_2) \Rightarrow \smtco(t_1) < \smtco(t_2)
\end{empheq}
where $t_1, \dots, t_n$ are all transactions in \tx,
and $\mathit{Distinct}(v_1,\dots,v_k)$ is a built-in SMT function that requires all input values to be distinct from each other.
By mapping \smtco(t) to a unique integer for each $t$, the first line of the equation above ensures that \co is a total order.

The second line of the equation ensures that \co is consistent with \writeread, \so, and \arserial, respectively.
For simplicity and to reduce the size of the constraints, arbitration constraints are factored out into the $\mathit{Arbitration}$ function, which is defined as follows:
\begin{align*}
% \label{eq:smt_wwquant}
\boxed{\mathit{Arbitration}(t_1, t_2) \coloneq \bigvee_{\substack{\forall k, t_1 \textnormal{ and } t_2 \textnormal{ write } k \\ \forall t_3 \in \tx \setminus \{t_1,t_2\}, t_3 \textnormal{ reads } k}} \smtwritereadk(t_2, t_3) \land \bigl(\smtco(t_1) < \smtco(t_3) \bigr)}
\end{align*}
which is a straightforward encoding of the \serializable arbitration constraints in Equation~\ref{eq:serial_ar}.

By using this approach we are pushing all the heavy lifting to the SMT solver.
% of finding a predicted execution such that all compatible commit orders (\co) are cyclic.
% While the solver may be able to avoid exploring an exponential number of possibilities exponentially,
However,
SMT solvers are known to be inefficient at solving constraints with universal quantifiers~\cite{Leino2016}---an issue confirmed by our performance results (\S\ref{subsec:isopredict-results}).

\subsubsection{Constraints encoding a sufficient but unnecessary condition}
\label{subsubsec:sufficient-unser-constraints}

% \mike{Aren't directed graphs and binary relations the same thing? It seems to me that what distinguishes this approach is that it computes a relation whose cyclicity is a sufficient but unnecessary condition for \co cyclicity. (I removed the language about graphs and changed the subsubsection title.)}

% It is computationally expensive to find an execution such that all of its possible commit orders (\co) are cyclic.
Alternatively, the analysis can encode a sufficient, but unnecessary, condition for predicting an \unserializable execution.
% While it is not very efficient to encode every possible \serializable(\co), we solve this issue by switching to a graph-based approach at the cost of potentially higher false negative rate.
% We model transactions as nodes and the relations between transactions as directed edges in a graph.
% The problem of finding a \serializable(\co) is equivalent to finding a fully connected directed acyclic graph by adding new edges implied by the commit order (\co).
We introduce a partial order, \comin, that is a \emph{subset of every commit order for every valid predicted execution}.
% \chujun{added \serializable here because \causal \co may not follow \comin. Maybe we should explain the differences between \serializable $\mathit{History}$ and \serializable \co}
% \mike{What is ``\serializable \co''---I think you might mean \co for the \serializable definition, e.g., $\co_{\mathit{ser}}$?. I don't think we need to distinguish that; \co is for the serializability definition, and \cocausal is for the causal consistency definition, and $\co_{\mathit{rc}}$ can be for the read committed definition. Also, see my comment before Intro about writing \serializable versus serializable.}
If there exists a predicted execution for which \comin is cyclic, then there cannot exist an acyclic \co for the predicted execution, meaning it is \unserializable.
% If \comin is cyclic, then it means finding an acyclic \co is impossible, hence the predicted history is \unserializable.
In theory, this approach has the potential for missing \unserializable executions that \S\ref{subsubsec:exact-unser-constraints}'s approach finds. But in our experiments, the \pco-based approach predicts all \unserializable executions that \S\ref{subsubsec:exact-unser-constraints}'s approach finds (\S\ref{subsec:isopredict-results}).

% \chujun{Do we also add \serializable to the \co in this paragraph?} \mike{No, \co is unique to the serializability definition.}
We define \pco to include all orders that must be in \co: session (\so), write--read (\writeread), and arbitration (\arserial) orders. We also introduce an \emph{anti-dependency order} (\antidependency) that must be in every \co, which allows adding more edges to \pco and thus finding more \unserializable executions.
A challenge with encoding \pco is that the arbitration and anti-dependency orders are both defined in terms of commit order, creating a circular dependency that leads to erroneous self-justifying edges in \pco. We break both circular dependencies by introducing the notion of \emph{rank} in the generated constraints.
Next we describe anti-dependency order (\antidependency), the circular dependency problem and our rank-based solution to it, and finally the constraints that the analysis generates.

\paragraph{Adding anti-dependency order (\antidependency) to \pco}
\label{subsec:anti-dependency}

% In a weakly ordered execution history, conflicting accesses
% are unordered in general.
% The axiomatic framework orders conflicting accesses if they have a write--read relationship (\writeread) or if they are ordered by the arbitration rule (Equation~\ref{eq:serial_ar}).
% While these orders on conflicting accesses are sufficient to ensure that acyclicity implies serializability,
% we would like
To make \pco as large as possible while still being consistent with % (i.e., a subset of)
every valid \co,
% in order to predict more \unserializable executions.
% To accomplish this, we can add
% In an effort to add as many required orders to \pco as possible, we introduce
we add
an \emph{anti-dependency} (\antidependency) order
% \footnote{Prior work uses the concept of anti-dependency to order events in \serializable executions~\cite{adya2000, serializability-for-eventual-consistency-2017, serializability-for-causal-consistency-2018}, but\ldots \mike{Explain why what they're doing is different from what we're doing.} \chujun{they assume there is a total order among all transactions while we don't?} \mike{Isn't it that they assume conflicting pairs of events are ordered while we don't?}}
to \pco. \antidependency must be part of any valid \co, as we prove in \iftoggle{extended-version}{Appendix~\ref{sec:proof}}{the extended version of this paper~\cite{isopredict-extended-arxiv}}. Intuitively, for any write--read relation $\writeread_k(t_1, t_2)$, anti-dependency prevents future transactions that also write $k$ from being ordered between $t_1$ and $t_2$ in the commit order.
% \spyros{first and only mention of ``conflicting''. we should define before use, or rewrite.}
% \chujun{Revised the sentence above.}
More formally, we define $\antidependency(t_1,t_2)$ as follows:
\iffalse
\begin{definition}[Anti-dependency]
% In any write--read relation $(write_k(t_3), read_k(t_1))$, the read transaction $t_1$ anti-depends (\antidependency) on any conflicting write $t_2$ if $t_2$ is a successor of $t_3$ in the commit order (\co).
For any two transactions $t_1$ and $t_2$ from a $\mathit{History}$, $t_1$ anti-depends (\antidependency) on $t_2$ in a given commit order (\co) if:
\begin{itemize}
    \item $\exists k$, $t_1$ reads $k$ from transaction $t_3$ that writes $k$
    \item $t_2$ writes $k$ and $t_2 \ne t_3$
    \item $t_3$ precedes $t_2$ in the commit order \co
\end{itemize}
\end{definition}
\fi
\begin{align*}
\antidependency(t_1, t_2) \coloneq
\exists k, t_2 \textnormal{ writes } k \land
\exists t_w, \writeread_k(t_w, t_1) \land
\pco(t_w, t_2)
\end{align*}
% Note that the rule for \antidependency is analogous to the \serializable arbitration rule (Equation~\ref{eq:serial_ar}), but for read--write instead of write--write ordering.

Figure~\ref{fig:pco_with_rw} shows an example in which \pco is cyclic
% (meaning the history is \unserializable)
only if \antidependency is included.
%Figure~\ref{fig:pco_without_rw} shows \pco edges \emph{excluding} \antidependency edges,
%while Figure~\ref{fig:pco_with_rw} shows \pco  \emph{including} \antidependency.
% Only by including \antidependency is \pco cyclic, indicating an \unserializable history.
% Without anti-dependency, \pco is unable to describe the \unserializable behavior shown in Figure~\ref{fig:pco_without_rw} since there will be no cycles unless we try to reason about the order between $t_1$ and $t_2$ (Figure~\ref{fig:causal_unserial_co}).
% After adding the anti-dependency orders, it is obvious that $t_1$ and $t_2$ form a cycle, and we no longer have to search for conflicts by potentially iterating all possible orderings of transactions.
% \chujun{Added an example of anti-dependency above.}

\begin{figure}[t]
\begin{minipage}{0.47\linewidth}
     \centering
%     \begin{subfigure}[h]{0.45\textwidth}
%         \centering
%         \includesvg[inkscapelatex=false,scale=1]{img/pco_without_rw.svg}
%         \caption{An execution history with its \pco edges shown \underline{\emph{excluding} \antidependency}. By excluding \antidependency, \pco is acyclic.}
%         \label{fig:pco_without_rw}
%     \end{subfigure}
%     \hfill
%     \begin{subfigure}[h]{0.45\textwidth}
%         \centering
         \includesvg[inkscapelatex=false,scale=1]{img/pco_with_rw.svg}
         \caption{Including anti-dependency ordering (\antidependency; dashed arrows) in \pco makes \pco cyclic.}
         \label{fig:pco_with_rw}
%     \end{subfigure}
%     \label{fig:pco_rw}
\end{minipage}
% \end{figure}
\hfill
% \begin{figure}
\begin{minipage}{0.47\linewidth}
    \centering
    \includesvg[inkscapelatex=false,scale=1.1]{img/circular.svg}
    \caption{An example of circular dependency: $\arserial(t_1, t_2)$ depends on $\comin(t_1, t_3)$, which in turn depends on $\arserial(t_1, t_2)$.}
    \label{fig:circular}
\end{minipage}
\end{figure}

% \paragraph{Partial commit order}

The partial order \pco can now be defined as the union of all orders that must be part of \co:
\[
\pco = (\so \cup \writeread \cup \arserial \cup \antidependency)^+
\]
Adapting Equation~\ref{eq:serial_ar} to use \pco instead of \co, we define arbitration order, \arserial, as follows:
\begin{align*}
\arserial(t_1, t_2) \coloneq \exists k, \textnormal{$t_1$ and $t_2$ write to $k$}
\land \exists t_3 \in \tx, wr_k(t_2, t_3) \land \pco(t_1, t_3) \\
\end{align*}

\paragraph{Circular dependency and rank}

In the definitions above,
note the circular dependencies between \pco and \arserial and between \pco and \antidependency, which seem to permit ``self-justifying'' edges. As an example, consider Figure~\ref{fig:circular}. According to the definitions,
$\comin(t_1, t_3) \Rightarrow \arserial(t_1, t_2)$, and
$\arserial(t_1, t_2) \Rightarrow \comin(t_1, t_3)$, allowing us to \emph{wrongly} conclude
$\arserial(t_1, t_2)$ and $\comin(t_1, t_3)$.
To avoid such self-justifying edges, \pco, \arserial, and \antidependency in fact must be defined as the \emph{minimal} relations that satisfy the above definitions.

How can we encode this ``minimal relation'' property in the SMT constraints? If \isopredict simply encodes the above definitions as SMT constraints, the constraint solver will find self-justifying edges, resulting in spurious cycles and reporting  executions that are not actually \unserializable.
For example, for Figure~\ref{fig:circular}, the SMT solver would choose both $\arserial(t_1, t_2)$ and $\comin(t_1, t_3)$ to be true, finding a cycle and wrongly reporting a predicted execution that is actually \serializable.

We address this problem by introducing the notion of \emph{rank}, which orders \pco edges that depend on each other. \isopredict relies on an integer SMT function $\rank(t_1,t_2)$ to enforce the following rule:
\begin{quote}
    \em For any relations $r$ and $r'$, if $r(t_1, t_2)$ depends on $r'(t'_1,t'_2)$, then $\rank(t_1, t_2) > \rank(t'_1, t'_2)$.
    % \item For any transaction $t_3$ that is on the transitive closure path of $\comin(t_1, t_2)$, $\rank(t_1, t_2)$ must be greater than both $\rank(t_1, t_3)$ and $\rank(t_3, t_2)$
\end{quote}
Note that the rule does not require $t_1 \ne t'_1$ or $t_2 \ne t'_2$.
For Figure~\ref{fig:circular}, rank constraints disallow $\arserial(t_1, t_2)$ and $\comin(t_1, t_3)$, which would require both $\rank(t_1, t_2) > \rank(t_1, t_3)$ and $\rank(t_1, t_3) > \rank(t_1, t_2)$.
% By using rank constraints, the SMT solver will assign false to both $\arserial(t_1, t_2)$ and $\comin(t_1, t_3)$.

\paragraph{Generated constraints}

% To encode a sufficient condition for an \unserializable execution, \isopredict generates the following constraints.

% We aim to find an \unserializable predicted execution whose \comin is cyclic.
% \begin{multline}\label{eq:comin}
% \exists t_1, t_2 \in T, t_1 \ne t_2, \\
% \comin(t_1, t_2) \land \comin(t_2, t_1) \textnormal{ where,} \\
% \comin(t_1, t_2) = \so(t_1, t_2) \lor \writeread(t_1, t_2) \lor \arserial(t_1, t_2) \lor \antidependency(t_1, t_2) \\
% \end{multline}
% \mike{Does the implementation really use an $\exists$? Or does it use a disjunction ($\bigwedge$)? Or something else?}
% \chujun{The implementation uses a big Or over all pairs of transactions, which is equivalent to $\exists$?}
% \mike{Doesn't transitivity need to be encoded??}
% \chujun{Yes, transitive closure should be part of this equation}

\Isopredict generates arbitration and anti-dependency constraints on Boolean SMT functions $\smtarserial(t_1,t_2)$ and $\smtantidependency(t_1,t_2)$:
\begin{align*}
& \forall t_1, t_2 \in T, t_1 \ne t_2,
\end{align*}
\begin{align*}
%\label{eq:smt_wwcomin}
& \boxed{\smtarserial(t_1, t_2) =
\bigvee_{\substack{\forall k, t_1 \textnormal{ and } t_2 \textnormal{ write } k \\ \forall t_3 \in \tx \setminus \{t_1,t_2\}, t_3 \textnormal{ reads } k}} \smtwritereadk(t_2, t_3) \land \smtcomin(t_1, t_3)) \land \rank(t_1,t_2) > \rank(t_1,t_3)}
\end{align*}
% %
\begin{align*}
%\label{eq:smt_rwcomin}
\boxed{\smtantidependency(t_1, t_2) = \bigvee_{\substack{\forall k, 
 t_1 \textnormal{ reads } k \: \land \: t_2 \textnormal{ writes } k \\ \forall t_3 \in \tx \setminus \{t_1,t_2\}, t_3 \textnormal{ writes } k}} \smtwritereadk(t_3, t_1) \land \smtcomin(t_3, t_2) \land \rank(t_1,t_2) > \rank(t_3,t_2)}
\end{align*}

The following constraints ensure that \pco is a partial order implied by \so, \writeread, \arserial, and \antidependency:
% %
\begin{align*}
\forall t_1, t_2 \in T, t_1 \ne t_2,
\end{align*}
\begin{empheq}[box=\fbox]{align*}
\smtpco(t_1, t_2) =\;&
\smtso(t_1,t_2) \lor \smtwriteread(t_1,t_2) \lor \smtarserial(t_1,t_2) \lor \smtantidependency(t_1,t_2) \; \lor \\
&
% \Big(
\bigvee_{t \in \tx \setminus \{t_1,t_2\}} \!\! \smtpco(t_1,t) \land \smtpco(t,t_2) \land \rank(t_1,t_2) > \rank(t_1,t) \land \rank(t_1,t_2) > \rank(t,t_2)
% \Big)
\end{empheq}
% \mike{Is that right? Is rank correctly encoded / is anything missing?}
% \chujun{That's correct}

To ensure that \pco is cyclic, the analysis generates the following constraint:
% %
\begin{align*}
\boxed{\bigvee_{\forall t_1,t_2 \in \tx, t_1 \ne t_2} \smtpco(t_1,t_2) \land \smtpco(t_2,t_1)}
\end{align*}
% %
If the solver finds a satisfying solution, a predicted \unserializable execution exists. If the solver reports no satisfying solution, a predicted \unserializable execution may or may not exist. In our experiments, a predicted \unserializable execution never exists in this case.

% \paragraph{Exercise for the reader}

We have not been able to come up with an execution for which our \pco-based approach misses a predicted \unserializable execution.
We believe that such an execution should exist because otherwise it would imply a polynomial-time algorithm for deciding if an execution history is serializable---a problem that is NP-hard~\cite{biswas2019}.

\subsection{Encoding Weak Isolation}
\label{subsec:encode_weak}

This section describes the constraints that \isopredict generates to ensure that the execution conforms to the target weak isolation model (\causal or \rc).

Regardless of the model, \isopredict encodes \hb as the transitive closure of \so and \writeread (\S\ref{subsec:background-execution-history}), by generating constraints on a Boolean SMT function $\smthb(t_1,t_2)$:
\begin{align*}
& \forall t_1, t_2 \in T, t_1 \ne t_2,
\quad \boxed{\smthb(t_1, t_2) = \smtso(t_1, t_2)
\lor \smtwriteread(t_1, t_2)
\lor \bigvee_{\forall t \in T\setminus\{t_1, t_2\}}\smthb(t_1, t) \land \smthb(t, t_2)}
\end{align*}

\subsubsection{Causal consistency (\causal)}
\label{subsec:encode_causal}
To ensure that the predicted execution is \causal, \isopredict generates constraints that ensure that the transitive closure of \causal arbitration order (\arcausal) and happens-before (\hb) is acyclic (\S\ref{subsec:background-causal}).
% %
\Isopredict encodes the \causal axiom (Equation~\ref{eq:causal_ar}) by generating constraints on a Boolean SMT function $\smtarcausal(t_1, t_2)$ representing \arcausal:
\begin{align*}
%\label{eq:smt_wwcausal}
& \forall t_1, t_2 \in T, t_1 \ne t_2,
\quad \boxed{\smtarcausal(t_1, t_2) = \bigvee_{\substack{\forall k, t_1 \textnormal{ and } t_2 \textnormal{ write } k \\
\forall t_3 \in \tx \setminus \{t_1, t_2\}, t_3 \textnormal{ reads } k}} \smtwritereadk(t_2, t_3) \land \smthb(t_1, t_3)}
\end{align*}
% \mike{Revised above equations. Please check.}
% \chujun{Fixed typo in the above equation. Changed $t_3$ to $t_2$ in the last item $\smthb(t, t_3)$}
%
% \mike{General question: For relation $r$, do we want to always specify $t_1 \ne t_2$ before constraints involving $r(t_1,t_2)$, or do we want to have constraints $\forall t, \neg r(t,t)$? Based on the constraints in the text, it appears that it's the former. I think I may have removed some of the $t_1 \ne t_2$ instances throughout the text, but they probably need to be added back.}
% \chujun{In the code we have $\neg r(t, t)$, but I'm not sure if it's necessary to add them to the paper.}
% \mike{Addressed throughout paper}
%
To ensure the execution is \causal, there must exist a strict total order that is consistent with $(\hb \cup \arcausal)^+$ (Equation~\ref{eq:causal_history}).
\Isopredict generates the constraints on an \emph{integer} SMT function $\smtcocausal(t)$:
\begin{align*}
& \forall t, t_1, t_2 \in T, t_1 \ne t_2,
% 0 \leq \cocausal(t) < \textnormal{total number of transactions} \\
% \cocausal(t_1) \ne \cocausal(t_2) \\
\quad \boxed{\smthb(t_1, t_2) \lor \smtarcausal(t_1, t_2) \; \Rightarrow \; \smtcocausal(t_1) < \smtcocausal(t_2)}
\end{align*}

\subsubsection{Read committed (\rc)}
\label{subsec:encode_rc}

Similar to \causal, \Isopredict generates constraints so that the transitive closure of \rc arbitration order (\arrc) and happens-before (\hb) is acyclic (\S\ref{subsec:background-rc}).
\Isopredict encodes the \rc axiom (Equation~\ref{eq:rc_ar}) with the help of a Boolean SMT function $\smtarrc(t_1, t_2)$ that represents \arrc:
\begin{align*}
% \label{eq:smt_wwrc}
& \forall t_1, t_2 \in T, t_1 \ne t_2,
\;\;\; \boxed{\smtarrc(t_1, t_2) = \bigvee_{\substack{\forall k, \; t_1 \textnormal{ and } t_2 \textnormal{ write } k \\
\forall t_3 \in \tx \setminus \{t_1, t_2\}, \; t_3 \textnormal{ reads } k \\
\forall i \in \pos(t_3), \forall j \in \posk(t_3), \;  i < j}} \smtchoice(s_3, i) = t_1 \land \smtchoice(s_3, j) = t_2}
\end{align*}
where $\pos(t)$ is the set of positions of read events in transaction $t$, $\posk(t)$ is the set of positions of read to $k$ in transaction $t$, and $s_3$ is $t_3$'s transaction.
% %
To ensure there exists a strict total order that is consistent with $(\hb \cup \arrc)^+$ (Equation~\ref{eq:rc_history}), \isopredict generates constraints on an integer SMT function $\smtcorc(t)$:
\begin{align*}
& \forall t, t_1, t_2 \in T, t_1 \ne t_2,
% 0 \leq \corc(t) < \textnormal{total number of transactions} \\
% \cocausal(t_1) \ne \cocausal(t_2) \\
\quad \boxed{\smthb(t_1, t_2) \lor \smtarrc(t_1, t_2) \; \Rightarrow \; \smtcorc(t_1) < \smtcorc(t_2)}
\end{align*}

\subsection{Prediction Examples}
\label{subsec:prediction-examples}

This section shows \causal, \unserializable behaviors predicted by \isopredict on programs evaluated in \S\ref{sec:eval}. The actual executions consist of dozens of transactions and thousands of events; the figures show only the transactions and events relevant to predicting \unserializable behavior.

\begin{figure}
\centering
    \begin{subfigure}[h]{0.48\textwidth}
         \centering
         \includesvg[inkscapelatex=false,scale=1]{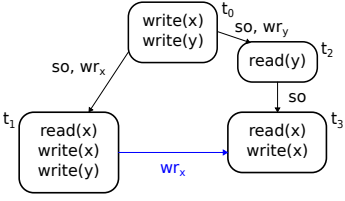}
         \caption{An observed execution of \bench{Wikipedia} for which a predicted \causal, \unserializable execution exists.}
         % Note that the only happens-before path from $t_1$ to $t_3$ is the $\writeread_x$ edge that will be removed by the prediction.}
         \label{fig:wiki_1_small_observed}
     \end{subfigure}
     \hfill
     \begin{subfigure}[h]{0.48\textwidth}
         \centering
         \includesvg[inkscapelatex=false,scale=1]{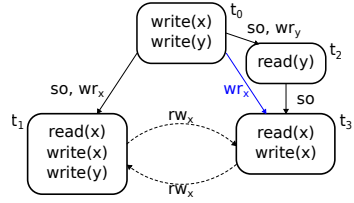}
         \caption{A \causal, \unserializable prediction; the \pco cycle (including \antidependency edges) shows it is \unserializable.}
         %by making $t_3$ read from the initial state $t_0$ instead of $t_1$. It is \unserializable because of the cycle made of \antidependency edges between $t_1$ and $t_3$.}
         \label{fig:wiki_1_small_prediction}
     \end{subfigure}
     \begin{subfigure}[h]{0.48\textwidth}
         \centering
         \includesvg[inkscapelatex=false,scale=1]{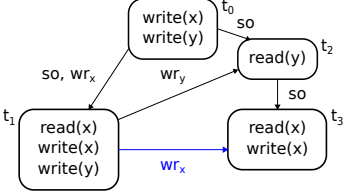}
         \caption{An observed execution of \bench{Wikipedia}, for which \emph{no} predicted \causal, \unserializable execution exists.}
         % Observed execution from a different trial of the same \bench{Wikipedia} program where a different interleaving of transactions caused $t_2$ to read from $t_1$, forming a second happens-before path from $t_1$ to $t_3$ that will make \causal and \unserializable predictions impossible.}
         \label{fig:wiki_1_large_observed}
     \end{subfigure}
     \hfill
     \begin{subfigure}[h]{0.48\textwidth}
         \centering
         \includesvg[inkscapelatex=false,scale=1]{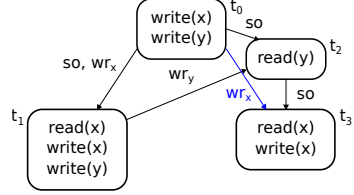}
         \caption{The non-\causal execution that results if we try to change (\subref{fig:wiki_1_large_observed}) so $t_3$ reads from $t_0$.}
         % (mimicking (\subref{fig:wiki_1_small_prediction})'s prediction).}
         % When there is an additional happens-before path from $t_1$ to $t_3$, making the same prediction as Figure~\ref{fig:wiki_1_small_prediction} will lead to an execution that is not \causal.}
         \label{fig:wiki_1_large_prediction}
     \end{subfigure}

\caption{Comparison of (relevant subsets of) executions from \bench{Wikipedia}. Blue edges highlight the differences between observed and predicted executions.}
\label{fig:wiki_1}
\end{figure}

Figure~\ref{fig:wiki_1_small_observed} shows an observed execution of the \bench{Wikipedia} benchmark, and Figure~\ref{fig:wiki_1_small_prediction} shows the \causal, \unserializable execution predicted by \isopredict.
In contrast, Figure~\ref{fig:wiki_1_large_observed} shows a different observed execution of \bench{Wikipedia}, from which no \causal, \unserializable execution can be predicted.
Figure~\ref{fig:wiki_1_large_prediction} serves to illustrate that changing $t_3$'s read of $x$ to read from $t_0$ would lead to a non-\emph{\causal} execution (and thus will not be reported by \isopredict).
%\yang{due to divergence?} \chujun{Not exactly. This example is to show that different timing of transactions in the observed trace will affect the prediction results. In Figure~\ref{fig:wiki_1_small_observed}, $t_1$ happens after $t_2$. In Figure~\ref{fig:wiki_1_large_observed}, $t_1$ happens before $t_2$, which makes it impossible to find an unserializable and causal prediction.} \mike{Revised for clarity.}).

\begin{figure}[t]
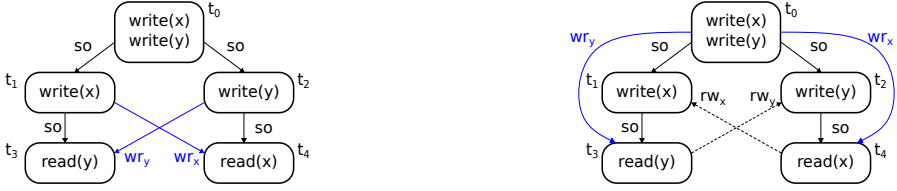

  \centering
  \begin{subfigure}[h]{0.45\textwidth}
      \centering
      \includesvg[inkscapelatex=false,scale=1]{img/smallbank_1_obs.svg}
      \caption{An observed execution for which a \causal, \unserializable predicted execution exists.}
      \label{fig:smallbank_1_observed}
  \end{subfigure}
  \hfill
  \begin{subfigure}[h]{0.45\textwidth}
      \centering
      \includesvg[inkscapelatex=false,scale=1]{img/smallbank_1_pred.svg}
      \caption{A \causal, \unserializable predicted execution as shown by the \pco cycles including \antidependency edges.}
      % by making $t_3$ and $t_4$ read from the initial state $t_0$.}
      \label{fig:smallbank_1_prediction}
  \end{subfigure}

  \caption{Observed and predicted executions of \bench{Smallbank}. For simplicity, each history shows a subset of the executed transactions, and each transaction shows a subset of the executed events.}
  \label{fig:smallbank_1}
\end{figure}

% Similar to Figure~\ref{fig:wiki_1},
% Figure~\ref{fig:smallbank_1} shows observed and predicted executions for the \bench{Smallbank} benchmark.
%\yang{It is a little weird to see one sentence here without further explanation. Are you showing Smallbank as another example in addition to Figure 7? Does it tell anything more we should pay attention to?}
%\chujun{Maybe we need to uncomment some of the explanations of this example?}
%\mike{Done. How's that?}
Figure~\ref{fig:smallbank_1_observed} shows an observed execution of the \bench{Smallbank} benchmark, and Figure~\ref{fig:smallbank_1_prediction} shows the \isopredict-predicted execution.
As Figure~\ref{fig:smallbank_1_prediction} shows,
a \causal, \unserializable predicted execution exists in which both reads read from the initial state ($t_0$), as demonstrated by the
\pco cycle $t_1 \ltco t_3 \ltco t_2 \ltco t_4 \ltco t_1$.

\subsection{Handling Divergence in the Predicted Execution}
\label{subsec:divergent}

\begin{figure}[tp]
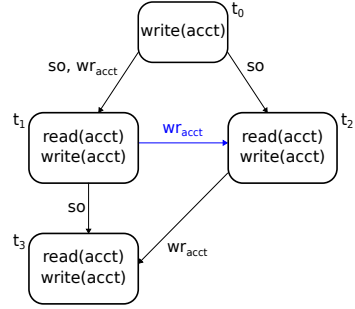
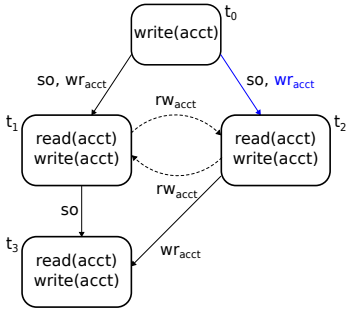
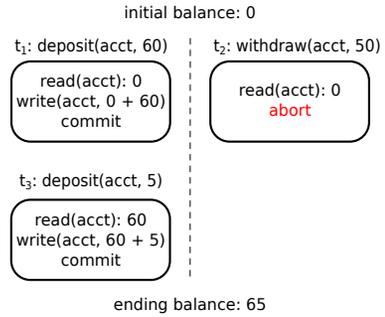
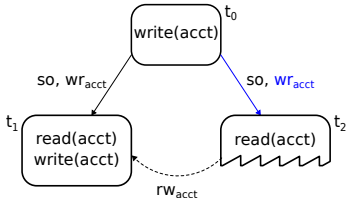
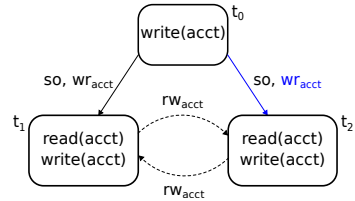

     \centering
     \begin{subfigure}[h]{0.47\textwidth}
         \centering
         \includesvg[inkscapelatex=false,scale=1]{img/validation_val.svg}
         \caption{One session deposits into an account twice, while another session withdraws once.}
         \label{fig:validation_observed}
     \end{subfigure}
     \hfill
     \begin{subfigure}[h]{0.47\textwidth}
         \centering
         \includesvg[inkscapelatex=false,scale=1]{img/validation_obs.svg}
         \caption{The execution history for (\subref{fig:validation_observed}), which is \serializable. The write--read edge from $t_1$ to $t_2$ (shown in blue) is not present in the  predicted execution in (\subref{fig:validation_pred}).}
         \label{fig:validation_hist}
     \end{subfigure}

     % \caption{A deposit transaction and a withdraw transaction execute concurrently over 2 sessions.}
     % \label{fig:validation_val}
% \end{figure}
\bigskip
% \begin{figure}[t]
     \centering
     \begin{subfigure}[h]{0.47\textwidth}
         \centering
         \includesvg[inkscapelatex=false,scale=1]{img/validation_pred.svg}
         \caption{A predicted execution history that is \unserializable. The write--read edge from $t_0$ to $t_2$ (shown in blue) was not present in the observed execution (\subref{fig:validation_hist}).}
         \label{fig:validation_pred}
     \end{subfigure}
     \hfill
     \begin{subfigure}[h]{0.47\textwidth}
         \centering
         \includesvg[inkscapelatex=false,scale=1]{img/validation_pred_val.svg}
         \caption{The \emph{validating} execution based on the predicted execution in (\subref{fig:validation_pred}).
         It diverges because $t_2$ aborts, and the resulting execution is \serializable.}
         % and that will remove the anti-dependency cycle between $t_1$ and $t_2$.}
         \label{fig:validation_failure}
     \end{subfigure}

\bigskip
    \centering
     \begin{subfigure}[h]{0.47\textwidth}
         \centering
         \includesvg[inkscapelatex=false,scale=1]{img/validation_strict.svg}
         \caption{This execution history consisting of the events from the predicted execution in (\subref{fig:validation_pred}) that are within the \emph{strict} prediction boundary is \serializable.}
         %\chujun{Technically our tool only makes \unserializable predictions, so it's a little weird to call this \serializable execution a prediction.}
         %\mike{Tried to address}         
         % The strict prediction boundary will ignore both $t_2$'s write event and $t_3$, and the SMT solver will avoid making the prediction from Figure~\ref{fig:validation_pred} because it's no longer \unserializable.}
         \label{fig:boundary_strict}
     \end{subfigure}
     \hfill
     \begin{subfigure}[h]{0.47\textwidth}
         \centering
         \includesvg[inkscapelatex=false,scale=1]{img/validation_relaxed.svg}
         \caption{This execution history consisting of the events from the predicted execution in (\subref{fig:validation_pred}) that are within the \emph{relaxed} prediction boundary is \unserializable.}
         % \caption{A predicted execution history based on using the \emph{relaxed} prediction boundary.
         % The execution is \unserializable, but it is actually infeasible (as (\subref{fig:validation_failure}) shows).}
         % For some applications, it might be safe to assume a transaction always contains the same set of key--value accesses regardless of its query results. In this case, \Isopredict's relaxed boundary will make a prediction similar to Figure~\ref{fig:validation_pred}'s.}
         \label{fig:boundary_relaxed}
     \end{subfigure}

     %\mike{By the way, I just changed the figure environment from \{figure\}[t] to \{figure\}[tp] to get it to appear here instead of at the end of the paper.}

     % \mike{TODO: Move the transaction descriptions (e.g., ``Transaction 1: deposit(acct, 60)'') \emph{above} the transactions, like in the modified Figure~\ref{fig:motivating_example}.}
     % \chujun{Done. Also shortened the label from ``Transaction 1'' to ``T1''.}

     \caption{Motivation for a prediction boundary (\subref{fig:validation_observed}--\subref{fig:validation_failure}) and illustration of the two kinds of prediction boundaries (\subref{fig:boundary_strict}--\subref{fig:boundary_relaxed}). The target weak isolation model is \causal. Dashed arrows represent \pco edges that are not part of the history.}
     \label{fig:validation}
\end{figure}

% In order for the predicted execution to be \unserializable and the observed execution to be \serializable,
% the predicted execution must have at least one read event that reads from a different writer transaction than in the observed execution (i.e., $\writeread \ne \ogwr$).
Reading from a different write in the predicted execution than in the observed execution,
may lead to different application behaviors.
% in the predicted execution.
Specifically, code in the data store application that is \emph{control dependent} on a read from a different writer transaction may generate different events.
% %
For example, consider the observed execution shown in Figures~\ref{fig:validation_observed} and \ref{fig:validation_hist}, which executes transactions shown in Algorithms~\ref{alg:credit} and \ref{alg:debit}.
% two sessions execute one session executes two deposit transactions (Algorithm~\ref{alg:credit}), while concurrently another session executes one withdrawal transaction (Algorithm~\ref{alg:debit}), as shown .
% Figure~\ref{fig:validation_hist} shows the resulting execution history, and
Figure~\ref{fig:validation_pred} shows an \unserializable predicted history that \isopredict would find using the constraints presented so far. However, the predicted execution is infeasible: $t_2$ aborts if it reads from $t_0$, making it impossible for $t_3$ to read from $t_2$, as Figure~\ref{fig:validation_failure} shows.
% ---and in fact no \unserializable execution exists.
\Isopredict (mostly) avoids make spurious predictions,
by excluding (much of the) potentially divergent behavior.
% As the rest of this section explains,
% \isopredict (mostly) avoids making spurious predictions by excluding events that happen \emph{after} reads that read from different writes in the predicted and observed executions.

\begin{algorithm}[t]
\small
\caption{A procedure in a data store application that withdraws money from an account.}\label{alg:debit}
\begin{algorithmic}
\Procedure{withdraw}{$\mathit{account}$, $\mathit{amount}$}
\State $\mathit{balance} \gets \mathit{DataStore}.\mathit{get}(\mathit{account})$ \Comment{Read balance; implicitly starts transaction}
%\State $balance \gets balance + amount$
\If{$\mathit{balance} < \mathit{amount}$}
\State $\mathit{DataStore}.\mathit{rollback}()$ \Comment{Abort transaction}
\Else
\State $\mathit{DataStore}.\mathit{put}(\mathit{account}, \mathit{balance} - \mathit{amount})$ \Comment{Update balance}
\State $\mathit{DataStore}.\mathit{commit}()$ \Comment{Commit transaction}
\EndIf
\EndProcedure
\end{algorithmic}
\end{algorithm}

% Divergent events are the ones caused by the side--effect of predicted write--read relations when the predicted read result leads to a new set of events that differ from the observed execution.
% This includes new events created by the application and the disappearance of some events from the observed execution.
% Since some of the divergent events are not captured by the observed execution, it is impossible to precisely model their impact on the isolation level of the predicted execution.
% To ensure the soundness of the prediction, which means all predicted executions are possible executions under a weak isolation level and are \unserializable, we exclude all events that \emph{happens after} the predicted read events from our model. \chujun{This includes both divergent events and some of the non--divergent events.}

\subsubsection*{Divergent behavior}
\label{subsec:divergent-behaviors}

To account for divergent behavior, we make a distinction between the \emph{predicted execution}, which is generated by \isopredict based on the observed execution, and what we call the \emph{validating execution}, which is the execution that actually occurs if one tries to produce the predicted execution using the data store application.
Divergent behaviors are behaviors that differ between the predicted and validating executions.
We categorize divergent behaviors into two categories:

\begin{itemize}[leftmargin=*]
  \item The validating execution reads or writes different keys or omits or adds events from the predicted execution, leading to a different execution history with different properties.
  \item A transaction that commits in the predicted execution, aborts in the validating execution (e.g., an application might have logic that aborts if a consistency check fails), as Figures~\ref{fig:validation_pred} and \ref{fig:validation_failure} show.
  % This means that even events ordered \emph{before} reads reading from different writes may be missing from the validating execution.
\end{itemize}
%
% \subsubsection*{How divergent behaviors affect prediction accuracy}
% \label{subsec:divergent_accuracy}
%
% \mike{I think we can define the validating execution as always conforming to the weak isolation model, and thus the only way that divergent events can affect prediction accuracy is by leading to a \serializable execution.}
% \chujun{Agreed.}
%
% Divergent behaviors may affect prediction accuracy, leading to false predictions.
The problem with divergent behavior is that an \unserializable predicted execution can lead to a \emph{\serializable} validating execution. (The validating execution will always be a feasible execution conforming to the weak isolation model because validation ensures these properties; \S\ref{subsec:design-validation}.)

\subsubsection*{Prediction boundary}
\label{subsec:boundary}

% To avoid numerous false predictions,
\Isopredict accounts for divergence by generating \emph{prediction boundary} constraints that exclude events that may be impacted by divergence---specifically, events that \emph{happen-after} (i.e., inverse of \hb) any read event that reads from different writers in the predicted and observed executions.
\Isopredict supports a prediction boundary that is \emph{strict} or \emph{relaxed}, as shown in Table~\ref{tab:prediction-boundaries}. The \emph{strict} boundary excludes events that happen-after \emph{events} that read from a different writer in the predicted execution than in the observed execution. The strict boundary prevents false predictions except when a transaction in the predicted execution aborts in the validating execution. Alternatively, the \emph{relaxed} boundary excludes events that happen-after \emph{transactions} that read from a different writer,
% Table~\ref{tab:prediction-boundaries} compares the two kinds of prediction boundaries.
% The strict boundary is more conservative than the relaxed boundary: The relaxed boundary includes more events in the prediction that may be divergent,
risking more false predictions but increasing the chances of finding an \unserializable predicted execution.
% Predictions under the strict boundary does not suffer from false predictions that are caused by diverged events.
% On one hand, diverged events are excluded from the strict boundary model, so they will not turn an \unserializable prediction into a \serializable one.
% On the other hand, diverged events will not break causal or read committed consistency because the predicted history already satisfies the weak isolation level, and that it's always possible to extend such history while preserving \causal or \rc due to causal extensibility~\cite{Bouajjani2023}.
% \chujun{Added the definition of ``diverged events'' (diverged reads and writes) to \S\ref{subsec:divergent-behaviors}. Not sure if it's clear enough to distinguish ``diverged events'' from ``abort-related divergence'' and ``divergent behaviors''.}
% As the table shows, the strict boundary permits false predictions only due to a transaction aborting in the validating execution that commits in the prediction execution.
% or vice versa.
% In contrast, the relaxed boundary allows false predictions even without abort/commit divergence: If a predicted execution is \unserializable because of events that happen-after reads with different writes, the validating execution may be \serializable as a result of divergence.

\begin{table}
\small
\caption{Comparison of strict and relaxed prediction boundaries.}
\label{tab:prediction-boundaries}

\newcommand\mymrow[1]{\multirow{2}{*}{#1}}
\begin{tabular}{l|p{3.1in}|l}
\bf Prediction & & \bf Divergent behaviors can \\
\bf boundary & \bf Excluded events & \bf cause false predictions \\\hline
Strict & Events that happen-after any \emph{read event} with a different writer & Abort-related only \\\hline
\mymrow{Relaxed} & Events that happen-after any \emph{transaction} containing a read event with a different writer & \mymrow{Any}
\end{tabular}

\end{table}

Figures~\ref{fig:boundary_strict} and \ref{fig:boundary_relaxed} show strict and relaxed boundaries, respectively,
applied to the prediction in Figure~\ref{fig:validation_pred}.
The strict boundary excludes all \emph{events} that happen-after $t_2$'s read (since it has a different writer than in Figure~\ref{fig:validation_hist});
the resulting execution history is \serializable.
The relaxed boundary excludes all \emph{transactions} that happen-after $t_2$'s read; the resulting execution history is \unserializable.
Although the relaxed boundary allows a false prediction in this example,
in our evaluation the relaxed boundary results in few false predictions.
% (\S\ref{subsec:eval-validation}).

% The constraints can use either a strict or relaxed prediction boundary.
% A strict boundary means that events in a transaction after a read with a different last writer are excluded.
% A relaxed boundary means that the events in a transaction after a read with a different last writer are included.

\subsubsection*{Generating prediction boundary constraints}

% \Isopredict generates constraints
% that exclude events or transactions from the predicted execution if they happen-after reads with different last writers in the predicted and observed executions.
Here we present \isopredict's constraints for excluding events using the prediction boundary. We show constraints for the \emph{strict} prediction boundary, but
the constraints for the \emph{relaxed} prediction boundary are similar except they also constrain every session's boundary to be the last event of a transaction.

\iffalse
The strict boundary divides every session into three parts:
\begin{enumerate}
    \item read events that are \emph{before} the boundary, which must read from the same write event as the observed execution;
    \item read events that are \emph{on} the boundary, which can read from any transaction whose last write to the same key is before the boundary,
    \item events that are \emph{after} the boundary, which \isopredict excludes from its constraints.
\end{enumerate}
\fi
% The prediction boundary of each session reflects the total amount of events that should be included in the model.
% For each session, the boundary is either the total number of events or one of its read event's position in the session order.
% When the boundary is equal to total number of events, it means everything from this session is considered inside boundary, and none of its read events should be modified.
% When the boundary is chosen from one of the read event's position, then it means that particular read could read from other possible write events.
The prediction boundary is delimited by a \emph{boundary event} in each session, which
% an event is before the boundary if and only if it is before its session's boundary event. The boundary event for a session
is either (1) a read event, which reads from a different write in the predicted execution than in the observed execution, or (2) the last event in the session (which will always be a commit event).
\Isopredict generates the following constraints on an integer SMT function $\smtboundary(s)$ to ensure that the boundary event for each session is either a read event or the last event (represented by position $\infty$):
\begin{align*}
& \forall s \textnormal{ is a session},
\quad \boxed{\Big(\bigvee_{\substack{t \textnormal{ is a transaction in } s \\ i \in \posk(t)}} \smtboundary(s) = i\Big) \lor \smtboundary(s) = \infty}
\end{align*}
Recall that \posk(s) is the set of positions of reads to $k$ in the transaction $t$.

% \mike{\smtboundary isn't used elsewhere explicitly. I guess it's used informally in the constraints below, e.g., ``is on the boundary'' and ``is not outside the boundary.'' }
% \chujun{True. \smtboundary dictates where the boundary event is for each session, and that determines whether an event is in the boundary or not.}

To ensure that each read that happens-\emph{before} the prediction boundary reads from the same write as in the observed execution, \isopredict generates the following constraints, where $\smtobs(s,i)$  is an integer SMT function that represents the last write of each read in the \emph{observed} execution history (and is thus the analogue of $\smtchoice$ for the observed execution):
\begin{align*}
& \forall t_1, t_2 \in T, t_1 \ne t_2, \forall i \in \posk(t_2) = i, t_2\textnormal{'s read at pos $i$ reads from $t_1$ in } \ogwr,
\quad \boxed{\smtobs(s_2,i) = t_1}
\end{align*}
% where $s_2$ is $t_2$'s session.
\iffalse\footnote{To generate these constraints, \isopredict needs to be able to distinguish writers of reads in transactions that have multiple reads of the same key. We assume this information is part of the observed execution history, but for simplicity we leave it out of the definition of execution history (\S\ref{subsec:background-execution-history}).}\fi
% \begin{align*}
% & \forall k \textnormal{ is a key},
% \forall t_1 \textnormal{ writes } k \textnormal{ in session } s_1,
% \forall t_2 \textnormal{ reads } k \textnormal{ in session } s_2,
% \end{align*}
% \begin{empheq}[box=\fbox]{align*}
% \smtwritereadk(t_1,t_2) = \bigvee_{i \in \posk(t_2)}
% & \Big(i \; = \; \smtboundary(s_2) \land
% \wrposk(t_1) \le \smtboundary(s_1) \land
% \smtchoice(t_2, i) = t_1\Big)
% \\
% \lor \; &
% % \lor & \bigvee_{\substack{i \in \posk(t_2) \textnormal{, $e \in t_1$ and $e$ writes $k$}\\
% \Big(i < \smtboundary(s_2) \land
% \wrposk(t_1) \le \smtboundary(s_1) \land
% \smtobs(t_2, i) = t_1\Big)
% \end{empheq}
% % where $\posk(t)$ is the set of positions of reads to $k$ in the in $t$.
% where \wrposk(t) is the position of transaction $t$'s last write to key $k$.
% A read \emph{before} the prediction boundary must read from the same write as in the observed execution, which the following generated constraints capture:
% %
\begin{align*}
& \forall k \textnormal{ is a key},
\forall t_1 \textnormal{ writes } k,
\forall t_2 \textnormal{ reads } k,
\forall i \in \posk(t_2),
% r is STRICTLY INSIDE boundary ==>  choice(r) = obs(r)
\;\; \boxed{i < \smtboundary(s_2) \;\Rightarrow\,
\smtchoice(s_2, i) = \smtobs(s_2, i)}
\end{align*}
where $s_1$ is $t_1$'s session and $s_2$ is $t_2$'s session.

\newcommand\wrposk{\ensuremath{\mathit{wrpos}_k}}

A read to $k$ \emph{on or before} the prediction boundary
must read from a
% may read from any transaction, as long as the transaction's last
write to $k$ that is \emph{before} the prediction boundary. \Isopredict ensures this property by generating the following constraints:
% %
\begin{align*}
& \forall k \textnormal{ is a key},
\forall t_1 \textnormal{ writes } k,
\forall t_2 \textnormal{ reads } k,
\forall i \in \posk(t_2), \\
% choice(r) = w AND r is INSIDE boundary  ==>  w is INSIDE boundary
& \boxed{\smtchoice(s_2, i) = t_1 \land
i \le \smtboundary(s_2) \implies
\wrposk(t_1) < \smtboundary(s_1)}
\end{align*}
where $s_1$ is $t_1$'s session, $s_2$ is $t_2$'s session, and
\wrposk(t) is the position of $t$'s last write to key $k$.

To exclude events after the prediction boundary,
\isopredict generates modified constraints
% (omitted here for brevity)
for all
% \serializable
arbitration and anti-dependency rules,
% (Equations~\ref{eq:smt_wwquant}--\ref{eq:smt_rwcomin})
% and for \causal and \rc arbitration rules,
% (Equations~\ref{eq:smt_wwcausal} and \ref{eq:smt_wwrc}),
as detailed in
\iftoggle{extended-version}{Appendix~\ref{sec:smt}}{the extended version of this paper~\cite{isopredict-extended-arxiv}}.

\section{Validation}
\label{subsec:design-validation}

Even by using the prediction boundary, \isopredict's predictive analysis may report \unserializable predicted executions for which the corresponding validating execution is \serializable. To rule out such predictions, \isopredict can attempt to \emph{validate} predicted executions, by executing the data store application based on the predicted execution history, and checking whether the resulting \emph{validating execution} is \unserializable.
% Because of divergence, \isopredict may make false predictions. More to the point, based on an observed execution, \isopredict may report an \unserializable predicted execution for an application that has no \unserializable executions. To eliminate such false predictions, \isopredict can optionally attempt to \emph{validate} a predicted execution.
% To validate a predicted execution,
% \Isopredict's validation component attempts to execute the data store application based on the predicted execution history.
% The goal of validation is to produce a weakly isolation (\causal or \rc), \unserializable execution. Achieving this goal is not guaranteed because of divergence: validation may not be able to follow the predicted execution exactly. Validation prioritizes conforming to the weak isolation model (\causal or \rc), and it makes a best-effort attempt to follow the predicted execution history.

\subsubsection*{Validating execution}
\label{subsubsec:validating-execution}

Validation produces the validating execution using a query engine that takes the predicted execution as input.
% Validation controls the value returned by the data store for each read.
At each \readevent{k} event, the query engine checks that (1) the corresponding read in the predicted execution also read from $k$; (2) the writer transaction $t$ from the predicted execution also wrote to $k$ in the validating execution; and (3) reading from $t$ in the validating execution will satisfy the weak isolation model (\causal or \rc).
If any of these conditions is violated, we categorize the execution as having \emph{diverged}, and
the query engine chooses a different, weak isolation model--conforming writer for the read to read from.
Note that it is always possible to keep executing while preserving \causal or \rc~\cite{Bouajjani2023}.
% due to so-called \emph{causal extensibility}
Furthermore, the validating execution may still be \unserializable, as our evaluation shows.

Recall that the predicted execution history contains events only up to the prediction boundary.
To avoid serendipitously introducing \unserializable behaviors that were not part of the predicted execution (which could make it tough to measure the effectiveness of \isopredict's predictive analysis), validation executes each transaction in full that is on the boundary or that happens-before any transaction on the boundary---and then it terminates the execution. This approach is sufficient: If this execution prefix is \unserializable, then so is the full execution.
% Executing each boundary transaction in full is necessary, even when using the strict boundary, to account for the possibility of abort. Cutting off further execution after the boundary is useful because the predicted execution history contains no information for later transactions, which would require validation to choose valid writers for reads to read from, and the choices could serendipitously produce \unserializable behavior that was not part of the predicted execution, making it harder to tease out the effectiveness of \isopredict's prediction approach.

Note that validation must directly control what transaction each read reads from, i.e., the write--read relation (\writeread). Our evaluation extends MonkeyDB~\cite{monkeydb} to allow explicit control of \writeread (\S\ref{sec:impl}).
In settings where MonkeyDB cannot be used, such as production systems, there are other ways to control \writeread. One is using resource locks (e.g., \textsf{sp\_getapplock} in SQL Server) to force specific transaction orders that produce the desired \writeread relation.

\subsubsection*{Checking serializability}

% When the validating execution terminates, it produces a validating execution history.
Validation generates constraints to check whether the validating execution history is \serializable (which can be encoded more efficiently than \unserializable, since \serializable implies a total commit order exists).
If the solver returns ``satisfiable,'' \isopredict reports no prediction. Otherwise (the solver returns ``unsatisfiable''), \isopredict reports the {validating} execution, which is known to be a feasible, \unserializable, weak isolation model--conforming execution.

%% file: 4.implementation.tex
\section{Implementation}
\label{sec:impl}

% \mike{This section needs to be rewritten and reorganized, but I think the content below is basically what it should have. One thing in particular that will help is making it clearer which part of the design each implementation detail is referring to.}
% \chujun{Revised the entire section.}

% \mike{Following comment is probably out of date: This section is pretty good. Try to connect to Design more clearly. For example, it may be unclear to the reader (1) what the ``database consistency checker'' is (how does it fit into what the reader learned in Design?) and (2) what MonkeyDB is used for (basically, we modified MonkeyDB in order to collect traces for the predictive analysis to analyze)?
% Also: This section is out of date w.r.t.\ some aspects of the project.}

This section describes the implementation of \isopredict, which is publicly available~\cite{isopredict-implementation}.

\subsubsection*{Predictive analysis}

We implemented \isopredict's predictive analysis (\S\ref{sec:predictive-analysis}) as a Python program that uses Z3Py, the Python binding of the Z3 SMT solver~\cite{z3}.
Observed and predicted
execution histories are in the form of traces containing read and write events and transaction and session identifiers, including the transaction that each read reads from.
% The predictive analysis takes the observed execution history of an application as its input. The history is in the form of a trace that contains read and write events and transaction and session identifiers, including the transaction each read reads from.
% For every committed transaction, the execution trace includes: a globally unique transaction ID, a globally unique session ID, the data key the current event operates on, and the transaction where the current read event reads its value from.
% The constraint generation module parses the recorded traces and generates predictive SMT constraints (\S\ref{sec:predictive-analysis}) that aim to find an \unserializable execution by letting some of the read events read from alternative writers without breaking the weak isolation level (\causal or \rc).
If Z3 finds a predicted \unserializable execution, it either reports the predicted execution history in both textual and graphical forms, or passes the predicted history to the validation component, depending on how \isopredict is configured.
% Otherwise it reports that no predicted \unserializable execution was found.
% After Z3 solves the constraints, it either  solving the constraints using Z3, the trace encoder will output a set of traces that indicate an erroneous execution that could be further analyzed by the application's developers.

To generate observed execution traces, we extended the implementation of \emph{MonkeyDB}, a transactional key--value data store~\cite{monkeydb}.
MonkeyDB handles relational queries by translating them to key--value queries.
MonkeyDB executes transactions serially,
and we configured it to choose the latest writer at each read,
% (\S\ref{subsec:meth})
so observed executions are always \serializable.
% We chose MonkeyDB as our data store because it supports weak isolation levels including \causal and \rc and it is relatively easy to modify as compared to a production-grade database.

\subsubsection*{Validation}

\Isopredict's validation component replays the client application on a customized query engine that we also built on top of MonkeyDB.
The query engine executes transactions one at a time, in an order dictated by the predicted execution, to ensure that read events always occur after their writers.
At each read, the query engine chooses a last writer that satisfies the weak isolation model and, if possible, matches the predicted execution (\S\ref{subsec:design-validation}).
Validation uses Z3Py to generate and solve SMT constraints to determine if the validating execution history is \unserializable, reporting the validating execution to the user in both textual and graphical forms if so.
% It could also create a graphic representation of the history making it easier for application developers to understand why it is \unserializable.

The customized query engine handles transaction aborts by rewinding the predicted execution trace to the beginning of the current transaction.
% Because observed execution traces include only committed transactions, not aborted transactions, an aborted transaction from the observed execution will not be in the predicted execution trace.
In our experiments, every transaction that aborted during the observed execution also aborts during the validating execution---except in a few cases, when a transaction that aborted in the observed execution and \emph{immediately precedes a committed transaction on the prediction boundary}, actually commits in the validating execution. As for other divergent behavior, the resulting validating execution may or may not be \unserializable.

%% file: 5.experiment.tex
\section{Evaluation}
\label{sec:eval}

This section evaluates how effectively and efficiently \isopredict predicts \unserializable executions under \causal and \rc,
and it compares empirically against prior work MonkeyDB~\cite{monkeydb}.

\subsection{Methodology}
\label{subsec:meth}

\subsubsection*{Prediction strategies}

\begin{table*}
\small
\caption{\label{tab:configs}\isopredict prediction strategies.}

\input{results/config}

\end{table*}

% \Isopredict provides two options for unserializability constraints (\S\ref{subsec:encode_unserial}) and two options for the prediction boundary (\S\ref{subsec:boundary}).
Table~\ref{tab:configs} shows the combinations of unserializability constraints and prediction boundaries that we evaluated, which we call \emph{prediction strategies}.
The \configFull prediction strategy uses precise encoding of unserializability (\S\ref{subsubsec:exact-unser-constraints}), while \configExpress and \configRelaxed encode the sufficient condition for unserializability (\S\ref{subsubsec:sufficient-unser-constraints}).
% that potentially has less coverage.
\configFull and \configExpress encode the strict prediction boundary, while \configRelaxed encode the relaxed prediction boundary.
% \mike{Presumably refer back to \S\ref{sec:design}. Will that section introduce the three prediction strategies (if so, the table probably belongs there), or will it only introduce the encoding precision and prediction boundary options?}
% \chujun{Prediction strategies sound like implementation details to me, so maybe I'll move them to \S\ref{sec:impl}?}
% \mike{I think it's either Design or Methodology.}

\subsubsection*{Benchmarks}

We evaluated \isopredict and MonkeyDB using transactional
workloads from \emph{OLTP-Bench}, a database testing framework that generates various workloads for benchmarking relational databases~\cite{oltp}.
% Since \isopredict is built on the MonkeyDB implementation, we reused the MonkeyDB paper's version of the programs.
Table~\ref{tab:benchmarks_full} shows quantitative characteristics of the evaluated Benchmarks.
\iffalse
\bench{Smallbank} simulates a banking application in which customers can send payments, check their balances, and perform similar actions.
% We configured 30\% of \bench{Smallbank}'s transactions to be read-only.
\bench{Voter} mimics a telephone voting system in which people vote for their favorite TV contestants.
% It has only one type of transaction, which simulates the voting process.
% This transaction inserts a record to the votes table if the voter has not voted; otherwise it is read-only. \mike{Don't need this here at least.}
\bench{TPC-C} simulates activities from a wholesale supplier including order processing, payment processing, and delivery management.
% We configured \bench{TPC-C} so that 8\% of the transactions are read-only.
\bench{Wikipedia} emulates its real-world counterpart with operations such as updating pages, editing the watch list, and retrieving pages.
% 91\% of the transactions was configured to be read-only.
% \mike{What's the rationale for the read-only percentages chosen above? Did MonkeyDB and/or any other work do the same? Are these the defaults of OLTP-Bench? Need to explain how we picked these numbers / why they're not cherrypicked.}
% \chujun{The workload configs are identical to MonkeyDB paper's so that we could have a fair comparison.}
% \mike{Addressed above}
\fi

Our experiments used versions of the OLTP-Bench programs that the MonkeyDB authors ported to use simplified SQL queries recognized by MonkeyDB~\cite{monkeydb}.
In these versions, each benchmark runs a nondeterministic number of transactions based on a specified time limit.
For the purposes of our evaluation, we modified the benchmarks to be more deterministic for two reasons.
First, determinism provides a more stable comparison among \isopredict's prediction strategies. Second, determinism helps with validation, since the validating execution can run the benchmark with the same RNG seed that the observed execution used.
(To use validation in a production setting, one should record and replay the application~\cite{replay,r3}.)
% Each transaction has a randomly chosen ``type'' and consists of operations on randomly chosen keys.
We modified the benchmarks to be more deterministic by
% \spyros{We could point out that commercial DBs come with such capabilities built-in: see Oracle Database Replay and https://dl.acm.org/doi/10.1145/1376616.1376732}
(1) fixing the number of sessions and transactions per session and
% and configured each benchmark to run the same fraction of read-only transactions as in the MonkeyDB evaluation~\cite{monkeydb} for a fair comparison.
(2) adding a random number generator (RNG) seed as a parameter to each benchmark.
% These modifications achieve two purposes. First, they provide a more stable comparison among \isopredict's prediction strategies and MonkeyDB. Second, they help with validation as the validating execution runs the benchmark with the same RNG seed that the observed execution used.
%\yang{My understanding is that, in a real setting, validation should be done by record-and-replay. Making the benchmark deterministic is a convenient way
%to do replay without recording, but is not necessary in a real setting. My concern here is that this sentence may give a false impression that
%the workload must be deterministic to some extent for validation to work.}
%Alternatively, one could record and replay the benchmark's nondeterministic behaviors.
%\chujun{Added a sentence above.}
% \mike{I think we should say ``validate/validation'' instead of ``replay'' since the predicted execution is different from the original execution.}
%\mike{Reorder paragraph flow?}
Although these modifications increase determinism, the benchmarks still execute nondeterministically
% even when using \isopredict's serial execution:
because the \emph{interleaving} of transactions is timing dependent. This source of nondeterminism does not hinder validation, which executes transactions in an order consistent with the predicted execution's \hb relation.
% \mike{Previously the text said the order should be consistent with commit order (\co), but that doesn't make sense (e.g., \co is cyclic), right?}
% \chujun{Agreed. \hb order is definitely acyclic while \co is cyclic.}
% which in general differs from the \emph{observed} execution's serial execution order.

% Throughout the evaluation, we use the same 10 RNG seeds, and use each one for every experiment, leading to 10 trials for every experiment.
% \mike{This is discussed later.}

\begin{table*}[t]
\small
\caption{\label{tab:benchmarks_full} Average number of events and committed transactions across 10 trials of each OLTP-Bench program.}

\input{results/benchmarks_full}
% Each execution runs 3 sessions, each of which attempts 4 transactions (small workload) or 8 transactions (large workload).}
\end{table*}

\begin{algorithm}[t]
\small
\caption{Code executed by each of \bench{Voter}'s transactions.}\label{alg:voter}
%\hrule
%\smallskip
\begin{algorithmic}
\Procedure{Vote}{id}
\State $\mathit{votes} \gets \mathit{DataStore}.\mathit{get}(\mathit{id})$
\If{$\mathit{votes} < 1$}
    \State $\mathit{DataStore}.\mathit{put}(\mathit{id}, 1)$
\EndIf
\State $\mathit{DataStore}.\mathit{commit}()$
\EndProcedure
\end{algorithmic}
\end{algorithm}

\iffalse
\begin{table*}
\newcounter{tmp}%
\begin{minipage}{0.5\linewidth}
\small
\input{results/benchmarks}
\caption{\label{tab:benchmarks} Average number of events and committed transactions across 10 trials of each OLTP-Bench program. Each execution runs 3 sessions, each of which attempts 4 transactions.}
\end{minipage}%
\hfill
\begin{minipage}{0.4\linewidth}
\renewcommand\tablename{\bf Algorithm}
\setcounter{tmp}{\value{table}}
\setcounter{table}{\value{algorithm}}
\small
\vspace*{-1.5em}
\hrule
\smallskip
\caption{Code executed by each of \bench{Voter}'s transactions.}%\label{alg:voter}
\vspace*{-1em}
\hrule
\smallskip
\begin{algorithmic}
\Procedure{Vote}{id}
\State $\mathit{votes} \gets \mathit{DataStore}.\mathit{get}(\mathit{id})$
\If{$\mathit{votes} < 1$}
    \State $\mathit{DataStore}.\mathit{put}(\mathit{id}, 1)$
\EndIf
\State $\mathit{DataStore}.\mathit{commit}()$
\EndProcedure
\end{algorithmic}
\smallskip
\hrule
\end{minipage}
\setcounter{table}{\value{tmp}}
\end{table*}
\fi

We configured each benchmark with both \emph{small} and \emph{large} workloads, in which three sessions each execute four or eight transactions, resulting in 12 or 24 \emph{attempted} transactions, respectively.
% four or eight transactions, providing ``small'' and ``large'' workload sizes to help measure \isopredict's scalability. Thus the number of \emph{attempted} transactions is 12 and 24, respectively, in the small and large workloads.
The number of \emph{committed} transactions is somewhat fewer because all programs except \bench{Voter} occasionally abort a transaction based on application-specific logic.
% even when executing transactions serially.
% Prior work MonkeyDB used the same number of sessions and a similar number of transactions~\cite{monkeydb}.\footnote{MonkeyDB's evaluation ran the benchmarks with a time limit, so the number of transactions executed varied, but on average was 10--20 transactions according to our measurements.}

\iffalse
We also implemented 10 microbenchmarks that cover various corner cases. We used the microbenchmarks to check the correctness of our implementation (results not shown).
\fi

% \textcolor{red}{We verified the correctness of our modification by running MonkeyDB's TPC-C benchmark under Serial Mode.
% As a result, only 2 assertions out of the 12 are violated, and that is supposedly caused by a bug in OLTP Benchmark~\cite{monkeydb,oltpissue}.}

\subsubsection*{Platform}

All experiments ran on an Intel Xeon server at 2.3 GHz with 16 cores, hyperthreading enabled, and 187 GB of RAM, running Linux.

\subsection{\IsoPredict's Effectiveness and Performance}
\label{subsec:isopredict-results}

% \sout{For each one of the four OLTP Benchmarks, we first ran it 100 times on MonkeyDB's Serial Mode with 3 sessions and a scale factor of 1.
% The time limit for each iteration was set to 1 second because a small workload is enough to evaluate the effectiveness of our predictive analysis.
% Then we ran our predictive analysis with all three prediction strategies on the execution traces and we compared their performance.
% To prevent the SMT solver from running forever on undecidable problems, we have a time limit on the constraint solving process.}

Tables~\ref{tab:results_full_causal} and \ref{tab:results_full_rc} show \isopredict's effectiveness and performance at predicting \unserializable executions under \causal and \rc, respectively.
For each benchmark and each of \isopredict's three prediction strategies, we ran \isopredict on 10 executions, each of which used one of 10 RNG seeds, which we kept consistent across prediction strategies and isolation levels.
% %
% Overall, the results show that \isopredict frequently predicts and validates feasible \unserializable executions, as shown in the \emph{Validated} column of each table.
% from observed serializable executions.
\iffalse
Under \causal, \isopredict's \configRelaxed prediction strategy predicts the most feasible executions  because the relaxed boundary enables more \unserializable predictions.
% without degradation in performance or validation failures.
Under \rc, all of the prediction strategies provide the same results;
\textcolor{red}{however, on executions with twice as many transactions, \configExpress has a slight advantage over \configRelaxed because the latter has a few validation failures.} \mike{TODO: Check text for consistency with full results being in this section.}
\fi

\begin{table*}
\footnotesize
\caption{\Isopredict effectiveness and performance \textbf{under \causal}.
``T/O'' means the solver did not finish within 24 hours.
``Unk'' means the solver returned ``unknown'' without reaching the timeout.}
\label{tab:results_full_causal}

\begin{subtable}[t]{\textwidth}
\centering

\input{results/oltp_cc}
\input{results/valid_cc}
\newcommand\unkOrTO{Unk}
\input{summary}
\caption{Small workload}
\end{subtable}

\begin{subtable}[t]{\textwidth}
\centering

\input{results/oltp_cc_8}
\input{results/valid_cc_8}
\newcommand\unkOrTO{T/O}
\input{summary}
\caption{Large workload}
\end{subtable}

\end{table*}

\begin{table*}
\footnotesize
\caption{\Isopredict effectiveness and performance \textbf{under \rc}.
``T/O'' means the solver did not finish within 24 hours.
``Unk'' means the solver returned ``unknown'' without reaching the timeout.}
\label{tab:results_full_rc}

\begin{subtable}[t]{\textwidth}
\centering

\input{results/oltp_rc}
\input{results/valid_rc}
\newcommand\unkOrTO{Unk}
\input{summary}
\caption{Small workload}
\end{subtable}

\begin{subtable}[t]{\textwidth}
\centering

\input{results/oltp_rc_8}
\input{results/valid_rc_8}
\newcommand\unkOrTO{T/O}
\input{summary}
\caption{Large workload}
\end{subtable}

\end{table*}

\subsubsection*{Predictive analysis}
\label{subsec:prediction-rates}

The \emph{Sat} column under \emph{Prediction} reports the number of \unserializable executions (out of 10) that \isopredict found.
% The prediction rate varies across the benchmarks and workload sizes, the target isolation models, and the prediction strategies.
The \configRelaxed prediction strategy generally predicts more than the other strategies because it uses the relaxed boundary.
% \mike{It's not just changing multiple events in a transaction, but also being able to keep all events in the rest of the transaction, right?}
% \chujun{That is true.}
% \mike{Reworded}
Although \configFull can theoretically predict more executions than \configExpress, this never happened in our experiments.

\Isopredict consistently predicts more \unserializable executions under \rc than under \causal, which makes sense because \rc is strictly weaker than \causal. \bench{Voter} has the biggest difference---there were no successful predictions under \causal.
% This is because unserializable behaviors such as reading an older write, which in \bench{Voter}'s case indicates reading from the initial state because there is only one writing (i.e., non-read-only) transaction, are allowed in \rc but not in \causal \textcolor{red}{for transactions that follow the newer writes in session order}.
% \mike{Unclear, especially the highlighted part.}
% \chujun{Replaced with the following text.}
% \mike{Looks good. Pushed details to footnote.}
This is because every observed execution of \bench{Voter} has only one writing (i.e., non-read-only) transaction (see Algorithm~\ref{alg:voter}), which is not sufficient to predict an \unserializable execution under \causal.\footnote{More specifically, the initial state transaction $t_0$ and the writing transaction $t_w$ constitute the only pair of conflicting writes.
% and it is trivial that any reading transaction is only allowed to read from either the initial state or the writing transaction under \causal.
If a transaction $t_r$ reads from the initial state, then a commit order with $t_r$ preceding $t_w$ is acyclic.
On the other hand, if $t_r$ reads from another transaction, a commit order $t_r$ following $t_w$ is acyclic.}
% Therefore, no matter where these read-only transactions read from, the predicted history will always be \serializable.}
Similarly, \isopredict has low prediction rates for \bench{Wikipedia}, which has few writing transactions.
In contrast, under \rc, a transaction may legally read both the initial state and the writing transaction, which
% This violates both causal consistency and serializability, and this
is why \isopredict has higher prediction rates for \bench{Voter} and \bench{Wikipedia} under \rc than under \causal.
% %
% In contrast, \bench{TPC-C} and \bench{Smallbank} generally have more writes for reads to choose from, leading to more predicted \unserializable behaviors.
\S\ref{subsec:prediction-examples} and \iftoggle{extended-version}{Appendix~\ref{sec:prediction_patterns}}{the extended version of this paper~\cite{isopredict-extended-arxiv}} present some observed and predicted executions from the evaluated benchmarks.

\subsubsection*{Validation}
\label{subsec:eval-validation}

% When the SMT solver detects a predicted execution, \isopredict validates it by trying to make the program execute the predicted execution.
We configured \isopredict to validate every predicted \unserializable execution. The \emph{Validated} column reports the number of validating executions that were \unserializable.
% As the results show,
Across all experiments, \emph{all but three predicted executions were successfully validated} as \unserializable.

The \emph{Diverged} column shows that, in many cases, the validating execution diverged, i.e., it could not match the predicted execution history
(\S\ref{subsec:design-validation}).
Unsurprisingly, the relaxed boundary experienced significantly more divergence than the strict boundary.
However, divergence rarely resulted in failed validation:
% almost all divergeces did \emph{not} affect the ability to produce an \unserializable execution.
% \Isopredict's best-effort validation still produced an \unserializable execution based on the prediction.
Among the 81 divergent executions across Tables~\ref{tab:results_full_causal} and \ref{tab:results_full_rc}, only three
failed validation (i.e., produced \serializable executions).
% and the rest produced \unserializable executions (validation successes).
% In theory, a validation failure occurs because a previously aborted transaction commits in the validating execution, leading to a \serializable execution; a previously committed transaction aborts in the validating execution, leading to a \serializable execution; or, for \configRelaxed only, divergent transaction operations lead to a \serializable execution (\S\ref{subsec:design-validation}).
One validation failure
% (\bench{Wikipedia}'s large workload under \rc)
was caused by divergent behavior unrelated to aborts (\S\ref{subsec:design-validation}), and
the other two failures
% (\bench{Smallbank}'s small workload under \causal and large workload under \rc)
were caused by previously aborted transactions being committed (an implementation issue discussed in \S\ref{sec:impl}).
% \yang{Is there an explanation why divergence usually leads to a different unserializable execution?}
% \chujun{Did you mean why divergent executions are unserializable despite diverging? If that's the case, then we don't know.}
% \mike{Intuitively, the divergent behavior doesn't affect the validating execution enough (compared with the predicted execution) to yield a \serializable execution. But we don't have further insights into what's going on.}

% \mike{Wrote the rest of the paragraph by piecing together various statements plus some guesswork:}
% All benchmarks issue queries that are affected by the results of previous queries, so there is always possibility of divergence when using \configRelaxed configuration.
% In comparison, divergence under \configFull and \configExpress configurations are affected by the aborted transactions, which rarely happened during our experiment.
% As shown in the tables, there were way less divergence under \configFull and \configExpress than using \configRelaxed configuration.
% However, the majority of diverged executions were still unserializable executions.
% \mike{Maybe that's not true or knowable, depending on the exact definition of divergence? All we can say for sure is that best-effort execution of the predicted execution yielded a valid unserializable execution.}

\subsubsection*{Performance}

The four rightmost columns of each table report the performance of \isopredict's predictive analysis, which consists of two components: (1) the time the analysis takes to generate SMT constraints (\emph{Constraint gen.\/})
% based on the execution log and prediction model,
and (2) SMT solving time (\emph{Solving time}).
Each table also reports the size of the generated constraints (\emph{\# Literals}),\footnote{The \configExpress and \configRelaxed prediction strategies
generate different constraints, but they have the same size.}
which correlates with constraint generation time.
%  which is the time Z3 takes to determine whether the SMT constraints are satisfiable.
SMT solving is significantly faster for successful prediction (\emph{Sat}) than for failed prediction (\emph{Unsat}),\footnote{It makes sense that successful prediction, which finds a single satisfying solution, is faster than failed prediction, which requires the solver to prove that no satisfying solution exists.} so the table reports the two average solving times separately.

Compared to the other prediction strategies, \configFull, which generates a single quantified constraint, spends less time generating constraints but more time solving constraints because its constraints are inherently harder to solve.
% than \configExpress and \configRelaxed's constraints.
% The table shows that \isopredict's \configFull prediction strategy spends less time on generating constraints but takes longer time in constraint solving.
% \configExpress takes more time than \configFull to generate constraints because it has more constraints to approximate the one-liner quantifier from \configFull, but \configExpress's constraint solving is significantly faster because it simplifies the unserializability definition with an approximation that is a sufficient but not necessary condition.
\configRelaxed and \configExpress have performance similar to each other,
which makes sense since they share the same approximation techniques.

Generating constraints can take a long time---often longer than constraint-solving time.
% After all, \emph{solving} time is exponential in the number of constraints in the worst case!
% We have not attempted to optimize our Python code that generates constraints, which sometimes uses deeply nested loops over the execution history to generate a few constraints. A more optimized approach could use efficient lookups in specialized data structures.
We investigated this issue by using the \emph{perf}~\cite{perf} and \emph{py-spy}~\cite{py-spy} performance analysis tools
% \footnote{\url{https://github.com/benfred/py-spy}} \yang{PL's convention is to put url in a footnote? System papers usually just use a reference.}
on the slowest instance of constraint generation: the large workload of \bench{TPC-C} under \rc using the \configRelaxed strategy (Table~\ref{tab:results_full_rc}). To the best of our understanding, 97\% of time is spent in Python code (\isopredict and Z3Py), and 3\% is spent in C code (Z3). Of the time spent in Python, 81\% is spent in Z3Py functions, with most time spent in the following Z3PY API functions and their callees: \textsf{__call__()}, \textsf{And()}, and \textsf{Or()}.
The \textsf{__call__()} function is part of Z3Py's implementation of SMT functions, which act as callable objects in Python. The \textsf{And()} and \textsf{Or()} functions create conjunction and disjunction clauses, respectively.
Z3Py functions call into Z3 code written in C; an unknown fraction of the time spent in Z3Py is due to making cross-language calls from Z3Py to Z3.
%mike{TODO: Investigate Z3Py's calls into Z3 code.} \yang{I remember py-spy does not sample C stacks? If so, we should say we first tried perf and find very small amount of time is spent in C code. Then we try py-spy.} \mike{Revised.}

% \begin{figure*}
% \centering
%      \begin{subfigure}[b]{0.475\textwidth}
%          \centering
%          \includesvg[width=\textwidth]{results/smallbank.svg}
%          \caption{Smallbank}
%          \label{fig:smallbank}
%      \end{subfigure}
%      \hfill
%     \begin{subfigure}[b]{0.475\textwidth}
%          \centering
%          \includesvg[width=\textwidth]{results/voter.svg}
%          \caption{Voter}
%          \label{fig:voter}
%      \end{subfigure}
%     \vskip\baselineskip
%     \begin{subfigure}[b]{0.475\textwidth}
%          \centering
%          \includesvg[width=\textwidth]{results/tpcc.svg}
%          \caption{TPC-C}
%          \label{fig:tpcc}
%      \end{subfigure}
%      \hfill
%      \begin{subfigure}[b]{0.475\textwidth}
%          \centering
%          \includesvg[width=\textwidth]{results/wiki.svg}
%          \caption{Wikipedia}
%          \label{fig:wiki}
%      \end{subfigure}
% \caption{Serial Mode OLTP Benchmark Prediction Results}
% \label{fig:oltp_barchart}
% \end{figure*}

\subsection{Comparison with MonkeyDB}
\label{subsec:cmp_monkeydb}

% We also tested our predictive analysis on \causal and \rc executions from OLTP Benchmarks and compared its effectiveness with MonkeyDB's assertion-checking strategy (\S\ref{subsec:cmp_monkeydb}).
% MonkeyDB is a transactional key--value data store that simulates weak consistency levels by randomly picking write values for database queries while respecting the weak consistency level at the same time~\cite{monkeydb}.
% It also supports SQL queries by translating them into key--value operations.
% MonkeyDB evaluates its effectiveness at generating abnormal behaviors under weak consistency levels by checking the final state of the database against several invariants that would not be violated under \serializable.
% We compared our work with MonkeyDB in terms of the percentage of OLTP Benchmark runs that are unserializable.

MonkeyDB is a transactional key--value data store that aims to produce unusual executions
% \savespace{of transactional data store applications }
that are legal under a target isolation level~\cite{monkeydb}.
MonkeyDB handles each read to a key by returning a randomly chosen value
% \mike{Does MonkeyDB use a ``pick a random value'' strategy for every read, or does it do something else?}
% \chujun{it randomly picks one legal value for every read event.}
among the set of legal values under the target isolation level.

MonkeyDB and \isopredict both aim to find erroneous executions under weak isolation, but they use completely different approaches. MonkeyDB relies on a customized query engine that produces a single execution, while \isopredict uses predictive analysis to analyze an equivalence class of many executions at once.
They also differ in how they define and expose \unserializable behavior:
\Isopredict tries to find an \unserializable execution, while
MonkeyDB uses programmer-crafted assertions to detect \unserializable behaviors.
% The MonkeyDB authors wrote assertions for each evaluated benchmark
% that should fail only when the execution is \unserializable.

% The OLTP Benchmark on MonkeyDB works in 3 stages:
% \begin{enumerate}
%     \item Initial loading of data. No commit or rollback operations are issued by the benchmark at this stage, and MonkeyDB runs in Serial Mode.
%     \item Querying and updating the database at a certain rate for a set period of time. This is where the majority of benchmarking takes place, and MonkeyDB would switch to a weak isolation level at this point.
%     \item Querying the final state of the database. MonkeyDB switches to Serial Mode again, and there are no commit or rollback operations issued. The query results are then compared with a set of hand-picked constraints for anomaly detection.
% \end{enumerate}

% One thing worth noticing is that MonkeyDB's assertions aim to find anomalies from a single execution, while our approach analyzes a group of executions that share the same $\mathit{History\langle T, wr, so \rangle}$ at the same time.
% Violating the constraints in one execution does not prevent other executions from the same group from adhering to the \serializability constraints.
% Hence there is no correlation between violating MonkeyDB's assertion and breaking our \serializability constraints.

Tables~\ref{tab:comparison_monkeydb_full:cc} and \ref{tab:comparison_monkeydb_full:rc} compare MonkeyDB and \isopredict's effectiveness at predicting \unserializable executions. To account for MonkeyDB's randomized approach, we ran it 100 times for each configuration: 10 times for each of the 10 RNG seeds used as benchmark input (\S\ref{subsec:meth}). The percentage of these executions with an assertion failure is reported in the \emph{Fail} column.
% The assertion failure rate varies from 0 to 100\% depending on the program, workload size, and isolation level.

\begin{table*}[t]
\newcommand\mynum[1]{\ifthenelse{\equal{#1}{0}}{0\%}{#1{}0\%}}
\small
\caption{\label{tab:comparison_monkeydb_full:cc}Comparison between MonkeyDB~\cite{monkeydb} and \isopredict (\configRelaxed strategy) \textbf{under \causal}. The numbers report how often a benchmark assertion failed (\textnormal{\em Fail}) or the history was \unserializable (\textnormal{\em Unser}).}

\begin{subtable}[h]{0.47\textwidth}
\centering

\input{results/oltp_cc}
\input{results/comparison}
\caption{Small workload}
\end{subtable}%
\hfill
\begin{subtable}[h]{0.47\textwidth}
\centering

\input{results/oltp_cc_8}
\input{results/comparison_8}
\caption{Large workload}
\end{subtable}

%\end{table*}
\bigskip
%\begin{table*}

\setcounter{subtable}{0}
\caption{\label{tab:comparison_monkeydb_full:rc}Comparison between MonkeyDB~\cite{monkeydb}, \isopredict (\configExpress strategy), and regular execution using MySQL \textbf{under \rc}. Each number is the percentage of runs in which a benchmark assertion failed (\textnormal{\em Fail}) or the history was \unserializable (\textnormal{\em Unser}).}

\setlength\tabcolsep{3.4pt} % default value: 6pt

\begin{subtable}[h]{0.45\textwidth}
\centering

\input{results/oltp_rc}
\input{results/comparison_rc}
\caption{Small workload}
\end{subtable}%
\hfill
\begin{subtable}[h]{0.45\textwidth}
\centering

\input{results/oltp_rc_8}
\input{results/comparison_rc_8}
\caption{Large workload}
\end{subtable}

\end{table*}

To compare MonkeyDB and \isopredict directly, we computed whether each execution produced by MonkeyDB was \unserializable,
% Assertion failures are detected simply by MonkeyDB's assertion checking.
by generating and solving constraints corresponding to the definition of \serializable.
% \footnote{The MonkeyDB-generated execution histories include a final assertion-checking transaction, which can lead to a history that is \unserializable even when the history \emph{without} the transaction is \serializable, but there was little difference in practice.}
An assertion failure is a sufficient but unnecessary condition for an \unserializable execution; hence, for MonkeyDB, the number of executions failing assertions (\emph{Fail}) never exceeds the number of \unserializable executions (\emph{Unser}).

% In contrast to MonkeyDB, \isopredict uses predictive analysis on an observed \serializable execution.
The \emph{\IsoPredict} column shows the percentage of executions that led to \unserializable predictions that were successfully validated (i.e., same results as the \emph{Validation} columns in Tables~\ref{tab:results_full_causal} and \ref{tab:results_full_rc}).
The tables use the best-performing prediction strategy for each isolation level.
% (\S\ref{subsec:isopredict-results}).
% The table includes two \isopredict prediction strategies: \configRelaxed (which works best under \causal) and \configExpress (which works best under \rc). The results in the \isopredict columns match the results in Tables~\ref{tab:oltpinfo-causal} and \ref{tab:oltpinfo-rc}.
% Overall, the quantitative comparison between is basically a wash:

\emph{Quantitatively}, MonkeyDB and \isopredict are comparable, finding erroneous executions at similar rates, except for two cases.
% In about half of the configurations, \isopredict has a higher prediction rate;
% in the other half, MonkeyDB has the higher rate.
% \subsubsection*{Detailed differences between MonkeyDB and \isopredict}
% Here we explain the two starkest prediction differences between MonkeyDB and \isopredict.
In one case---\bench{Voter} under \causal---MonkeyDB produces \unserializable executions, but \isopredict never predicts any.
%As \S\ref{subsec:prediction-rates} and Algorithm~\ref{alg:voter} explained,
\bench{Voter} issues only one write transaction under \serializable (Algorithm~\ref{alg:voter}),
from which it is impossible
to predict an \unserializable execution under \causal, because \isopredict cannot predict
% future
events that did not happen in the observed execution.
% events that happen after reads with different writers in the observed and predicted executions.
% This is due to the fact that the \bench{Voter} program issues only one write transaction if the execution is \serializable.
% \Isopredict predicts anomalies based on an observed execution and it is incapable of making assumptions that read-only transactions could turn into write transactions under certain conditions.
% As a result, when the observed execution contains exactly one write transaction like in the \bench{Voter} program, \isopredict is unable to predict any anomalies under \causal.
% unless they are executed under a weak isolation level.
In contrast, since MonkeyDB chooses values on the fly, its choices of reads can lead \bench{Voter} to perform additional writes, leading to \unserializable behavior.
% \chujun{Do we need to prove why 1 write transaction is not enough for serializability violation under causal consistency and/or why it's enough for serializability violation under \rc?}
% \mike{Don't need a proof, but need to make it clearer what's going on (with a diagram?). In particular, why is \isopredict able to predict unserializable executions under \rc?}
In another case---\bench{Wikipedia} under \causal---%
% \isopredict produces \unserializable executions at a higher rate than MonkeyDB under both \causal and \rc. (Under \causal,
\isopredict is able to predict several \unserializable executions while MonkeyDB never has assertion failures, since its
% final-state
assertions are not sensitive enough to detect \unserializable behaviors.

\emph{Qualitatively}, the approaches differ in two significant ways.
First, \isopredict does not require programmers to write assertions. Second and more significantly, \isopredict predicts \unserializable executions from observed executions, which in theory could be produced by any data store. In contrast, MonkeyDB's approach requires its specialized query engine.

\subsubsection*{Comparison with regular execution}

Both MonkeyDB and \isopredict routinely produce \unserializable executions for the evaluated programs, but a natural question is whether executing these programs normally on a real-world data store yields \unserializable executions. To evaluate this question, we executed the programs using MySQL~\cite{mysql} in \rc mode (MySQL does not support \causal). As for the MonkeyDB runs, we executed each program 100 times---10 times for each of the 10 RNG seeds used as input to the program---and evaluated the assertions used by MonkeyDB.

Table~\ref{tab:comparison_monkeydb_full:rc}'s \emph{MySQL} columns show the percentage of runs in which an assertion failed, a sufficient condition for an \unserializable history.
% \mike{TODO: Add \emph{MySQL} numbers for \rc with large workload and update text.}
% \chujun{Updated numbers and text.}
The results show that \bench{Smallbank}, \bench{Voter}, and \bench{Wikipedia} never experienced an assertion failure under regular execution.\footnote{It is an open question whether MySQL in \rc mode can actually produce \unserializable executions for these programs. Data store implementations may preclude behaviors that are theoretically possible under the target isolation level.}
\bench{TPC-C} experienced an assertion failure half of the time on the small workload and 70\% of the time on the large workload.
In contrast, MonkeyDB and \isopredict often produce assertion-failing, \unserializable executions.
% for these programs under \rc.
% \yang{This result can be interpreted in two opposite ways: 1) Running concurrency tests blindly has a low chance to find an unserializable execution, so we need tools like MonkeyDB and \isopredict;
% 2) MonkeyDB and \isopredict generate unserializable executions that are not feasible on a real database, maybe due to MySQL's implementation. Then it becomes questionable how useful
% these tools are. To determine which one is right, can we somehow reproduce a SmallBank/Voter/Wikipedia unserializable execution on MySQL?}
% \chujun{I don't think we can validate our predictions on MySQL since that requires modifying MySQL in the same way we modified MonkeyDB. The validation process seems to be reasonable enough about showing the predictions are possible, and I think it's just uncommon to happen when running on MySQL.}
% \mike{Actually, in future work we would like to validate predicted executions using unmodified MySQL. This paper already mentions a few things that might be relevant: \textsf{sp\_getapplock} in SQL Server, R3~\cite{r3}, Oracle Database Replay~\cite{replay}.
% In any case, I don't think we should say anything about that here.}
% \mike{Added a footnote above to try to address the issue we discussed. Can probably be improved.}

\subsubsection*{Differences between our MonkeyDB results and the MonkeyDB paper's results}

In our experiments, MonkeyDB triggered fewer assertion failures than reported in the MonkeyDB paper~\cite{monkeydb}.
% For example, the MonkeyDB paper reports assertion failures for \bench{Wikipedia} under \causal. In our results, the assertion failure rate is 0 after our bug fixes, although some of the executions are still \unserializable.
These differences exist because we found and fixed a few bugs in the ported benchmarks and their assertions, which eliminated a few spurious failures.
We confirmed all of the bugs and fixes with the MonkeyDB authors~\cite{monkeydb-personal-communication}.
% likely because the assertions do not provide complete coverage.
To be clear, the differences do not impact the MonkeyDB paper's takeaway: MonkeyDB often produces \unserializable, erroneous executions for the evaluated programs.

%% file: results/config.tex
\begin{tabular}{l|lll}
\bf Pred.\ strategy & \bf Encoding precision & \bf Pred.\ boundary & \bf Divergence $\Rightarrow$ false predictions? \\
\hline
\configFull & Exact encoding & Strict & Only because of aborts \\
\configExpress & Approximate encoding & Strict & Only because of aborts \\
\configRelaxed & Approximate encoding & Relaxed & Yes \\
\end{tabular}

%% file: results/benchmarks_full.tex
\newcommand\mytx[2]{#2 & (#1)}

% \begin{subtable}[t]{\textwidth}
\centering
\begin{tabular}{l|rr|rr|rr|rr}
& \mc{4}{c|}{\bf Small workload} & \mc{4}{c}{\bf Large workload} \\
& \mc{2}{c|}{\bf KV accesses} & \mc{2}{c|}{\bf Committed txns} & \mc{2}{c|}{\bf KV accesses} & \mc{2}{c}{\bf Committed txns} \\
\bf Program & Reads & Writes & Total & (Read-only) & Reads & Writes & Total & (Read-only) \\
\hline
\bench{Smallbank} & 669.7 & 14.7 & \mytx{3.5}{11.0} & 1271.3 & 30.5 & \mytx{6.6}{20.3} \\
\bench{Voter} & 763.0 & 6.0 & \mytx{11.0}{12.0} & 919.0 & 6.0 & \mytx{23.0}{24.0} \\
\bench{TPC-C} & 3297.3 & 763.0 & \mytx{0.9}{11.9} & 7025.6 & 1502.4 & \mytx{1.7}{23.8} \\
\bench{Wikipedia} & 1067.7 & 55.1 & \mytx{8.8}{9.9} &  2677.1 & 111.1 & \mytx{20.6}{22.8} \\
\end{tabular}

% \caption{Small workload}
% \label{subtab:bench-small}
% \end{subtable}

% \begin{subtable}[t]{\textwidth}
% \centering
% \begin{tabular}{l|rr|rr|r}
%             & \mc{2}{c|}{\bf Key--value accesses} & \mc{2}{c|}{\bf Committed transactions} & \\
% \bf Program & Reads & Writes & Total & (Read-only) & \bf Sessions \\
%   \hline
% \bench{Smallbank} & 1271.3 & 30.5 & \mytx{6.6}{20.3} & 3 \\
% \bench{Voter} & 919.0 & 6.0 & \mytx{23.0}{24.0} & 3 \\
% \bench{TPC-C} & 7025.6 & 1502.4 & \mytx{1.7}{23.8} & 3 \\
% \bench{Wikipedia} & 2677.1 & 111.1 & \mytx{20.6}{22.8} & 3 \\
% \end{tabular}
% \caption{Large workload}
% \label{subtab:bench-large}
% \end{subtable}

%% file: results/oltp_cc.tex
% smallbank_express_causal
\newcommand{\bankexpgencon}{22.9}
\newcommand{\bankexpliteral}{366339}
\newcommand{\bankexpz}{??}
\newcommand{\bankexpcntsat}{4}
\newcommand{\bankexpeventsat}{719.0}
\newcommand{\bankexptxsat}{11.5}
\newcommand{\bankexpgenconsat}{25.5}
\newcommand{\bankexpsolvesat}{1.0}
\newcommand{\bankexpcntunsat}{6}
\newcommand{\bankexpeventunsat}{661.3}
\newcommand{\bankexptxunsat}{10.7}
\newcommand{\bankexpgenconunsat}{21.2}
\newcommand{\bankexpsolveunsat}{3.2}
\newcommand{\bankexpcntunknown}{0}
\newcommand{\bankexpeventunknown}{}
\newcommand{\bankexptxunknown}{}
\newcommand{\bankexpgenconunknown}{}
\newcommand{\bankexpsolveunknown}{}
% smallbank_full_causal
\newcommand{\bankfullgencon}{8.8}
\newcommand{\bankfullliteral}{140278}
\newcommand{\bankfullz}{??}
\newcommand{\bankfullcntsat}{4}
\newcommand{\bankfulleventsat}{719.0}
\newcommand{\bankfulltxsat}{11.5}
\newcommand{\bankfullgenconsat}{10.0}
\newcommand{\bankfullsolvesat}{13.9}
\newcommand{\bankfullcntunsat}{6}
\newcommand{\bankfulleventunsat}{661.3}
\newcommand{\bankfulltxunsat}{10.7}
\newcommand{\bankfullgenconunsat}{7.9}
\newcommand{\bankfullsolveunsat}{11.3}
\newcommand{\bankfullcntunknown}{0}
\newcommand{\bankfulleventunknown}{}
\newcommand{\bankfulltxunknown}{}
\newcommand{\bankfullgenconunknown}{}
\newcommand{\bankfullsolveunknown}{}
% smallbank_relaxed_causal
\newcommand{\bankrelgencon}{22.9}
\newcommand{\bankrelliteral}{366339}
\newcommand{\bankrelz}{??}
\newcommand{\bankrelcntsat}{10}
\newcommand{\bankreleventsat}{684.4}
\newcommand{\bankreltxsat}{11.0}
\newcommand{\bankrelgenconsat}{22.9}
\newcommand{\bankrelsolvesat}{0.6}
\newcommand{\bankrelcntunsat}{0}
\newcommand{\bankreleventunsat}{}
\newcommand{\bankreltxunsat}{}
\newcommand{\bankrelgenconunsat}{}
\newcommand{\bankrelsolveunsat}{}
\newcommand{\bankrelcntunknown}{0}
\newcommand{\bankreleventunknown}{}
\newcommand{\bankreltxunknown}{}
\newcommand{\bankrelgenconunknown}{}
\newcommand{\bankrelsolveunknown}{}
% tpcc_express_causal
\newcommand{\tpccexpgencon}{425.8}
\newcommand{\tpccexpliteral}{6508254}
\newcommand{\tpccexpz}{??}
\newcommand{\tpccexpcntsat}{9}
\newcommand{\tpccexpeventsat}{4143.6}
\newcommand{\tpccexptxsat}{11.9}
\newcommand{\tpccexpgenconsat}{429.1}
\newcommand{\tpccexpsolvesat}{35.1}
\newcommand{\tpccexpcntunsat}{1}
\newcommand{\tpccexpeventunsat}{3311.0}
\newcommand{\tpccexptxunsat}{12.0}
\newcommand{\tpccexpgenconunsat}{396.0}
\newcommand{\tpccexpsolveunsat}{105.2}
\newcommand{\tpccexpcntunknown}{0}
\newcommand{\tpccexpeventunknown}{}
\newcommand{\tpccexptxunknown}{}
\newcommand{\tpccexpgenconunknown}{}
\newcommand{\tpccexpsolveunknown}{}
% tpcc_full_causal
\newcommand{\tpccfullgencon}{220.4}
\newcommand{\tpccfullliteral}{3493396}
\newcommand{\tpccfullz}{??}
\newcommand{\tpccfullcntsat}{9}
\newcommand{\tpccfulleventsat}{4143.6}
\newcommand{\tpccfulltxsat}{11.9}
\newcommand{\tpccfullgenconsat}{222.7}
\newcommand{\tpccfullsolvesat}{230.4}
\newcommand{\tpccfullcntunsat}{1}
\newcommand{\tpccfulleventunsat}{3311.0}
\newcommand{\tpccfulltxunsat}{12.0}
\newcommand{\tpccfullgenconunsat}{199.7}
\newcommand{\tpccfullsolveunsat}{752.3}
\newcommand{\tpccfullcntunknown}{0}
\newcommand{\tpccfulleventunknown}{}
\newcommand{\tpccfulltxunknown}{}
\newcommand{\tpccfullgenconunknown}{}
\newcommand{\tpccfullsolveunknown}{}
% tpcc_relaxed_causal
\newcommand{\tpccrelgencon}{425.5}
\newcommand{\tpccrelliteral}{6508254}
\newcommand{\tpccrelz}{??}
\newcommand{\tpccrelcntsat}{10}
\newcommand{\tpccreleventsat}{4060.3}
\newcommand{\tpccreltxsat}{11.9}
\newcommand{\tpccrelgenconsat}{425.5}
\newcommand{\tpccrelsolvesat}{22.7}
\newcommand{\tpccrelcntunsat}{0}
\newcommand{\tpccreleventunsat}{}
\newcommand{\tpccreltxunsat}{}
\newcommand{\tpccrelgenconunsat}{}
\newcommand{\tpccrelsolveunsat}{}
\newcommand{\tpccrelcntunknown}{0}
\newcommand{\tpccreleventunknown}{}
\newcommand{\tpccreltxunknown}{}
\newcommand{\tpccrelgenconunknown}{}
\newcommand{\tpccrelsolveunknown}{}
% voter_express_causal
\newcommand{\voteexpgencon}{131.7}
\newcommand{\voteexpliteral}{1526284}
\newcommand{\voteexpz}{??}
\newcommand{\voteexpcntsat}{0}
\newcommand{\voteexpeventsat}{}
\newcommand{\voteexptxsat}{}
\newcommand{\voteexpgenconsat}{}
\newcommand{\voteexpsolvesat}{}
\newcommand{\voteexpcntunsat}{10}
\newcommand{\voteexpeventunsat}{769.0}
\newcommand{\voteexptxunsat}{12.0}
\newcommand{\voteexpgenconunsat}{131.7}
\newcommand{\voteexpsolveunsat}{10.4}
\newcommand{\voteexpcntunknown}{0}
\newcommand{\voteexpeventunknown}{}
\newcommand{\voteexptxunknown}{}
\newcommand{\voteexpgenconunknown}{}
\newcommand{\voteexpsolveunknown}{}
% voter_full_causal
\newcommand{\votefullgencon}{61.7}
\newcommand{\votefullliteral}{686679}
\newcommand{\votefullz}{??}
\newcommand{\votefullcntsat}{0}
\newcommand{\votefulleventsat}{}
\newcommand{\votefulltxsat}{}
\newcommand{\votefullgenconsat}{}
\newcommand{\votefullsolvesat}{}
\newcommand{\votefullcntunsat}{10}
\newcommand{\votefulleventunsat}{769.0}
\newcommand{\votefulltxunsat}{12.0}
\newcommand{\votefullgenconunsat}{61.7}
\newcommand{\votefullsolveunsat}{64.5}
\newcommand{\votefullcntunknown}{0}
\newcommand{\votefulleventunknown}{}
\newcommand{\votefulltxunknown}{}
\newcommand{\votefullgenconunknown}{}
\newcommand{\votefullsolveunknown}{}
% voter_relaxed_causal
\newcommand{\voterelgencon}{132.1}
\newcommand{\voterelliteral}{1526284}
\newcommand{\voterelz}{??}
\newcommand{\voterelcntsat}{0}
\newcommand{\votereleventsat}{}
\newcommand{\votereltxsat}{}
\newcommand{\voterelgenconsat}{}
\newcommand{\voterelsolvesat}{}
\newcommand{\voterelcntunsat}{10}
\newcommand{\votereleventunsat}{769.0}
\newcommand{\votereltxunsat}{12.0}
\newcommand{\voterelgenconunsat}{132.1}
\newcommand{\voterelsolveunsat}{10.0}
\newcommand{\voterelcntunknown}{0}
\newcommand{\votereleventunknown}{}
\newcommand{\votereltxunknown}{}
\newcommand{\voterelgenconunknown}{}
\newcommand{\voterelsolveunknown}{}
% wikipedia_express_causal
\newcommand{\wikiexpgencon}{36.3}
\newcommand{\wikiexpliteral}{529313}
\newcommand{\wikiexpz}{??}
\newcommand{\wikiexpcntsat}{0}
\newcommand{\wikiexpeventsat}{}
\newcommand{\wikiexptxsat}{}
\newcommand{\wikiexpgenconsat}{}
\newcommand{\wikiexpsolvesat}{}
\newcommand{\wikiexpcntunsat}{10}
\newcommand{\wikiexpeventunsat}{1122.8}
\newcommand{\wikiexptxunsat}{9.9}
\newcommand{\wikiexpgenconunsat}{36.3}
\newcommand{\wikiexpsolveunsat}{1.3}
\newcommand{\wikiexpcntunknown}{0}
\newcommand{\wikiexpeventunknown}{}
\newcommand{\wikiexptxunknown}{}
\newcommand{\wikiexpgenconunknown}{}
\newcommand{\wikiexpsolveunknown}{}
% wikipedia_full_causal
\newcommand{\wikifullgencon}{13.9}
\newcommand{\wikifullliteral}{179967}
\newcommand{\wikifullz}{??}
\newcommand{\wikifullcntsat}{0}
\newcommand{\wikifulleventsat}{}
\newcommand{\wikifulltxsat}{}
\newcommand{\wikifullgenconsat}{}
\newcommand{\wikifullsolvesat}{}
\newcommand{\wikifullcntunsat}{9}
\newcommand{\wikifulleventunsat}{1132.6}
\newcommand{\wikifulltxunsat}{10.0}
\newcommand{\wikifullgenconunsat}{14.7}
\newcommand{\wikifullsolveunsat}{24.0}
\newcommand{\wikifullcntunknown}{1}
\newcommand{\wikifulleventunknown}{1035.0}
\newcommand{\wikifulltxunknown}{9.0}
\newcommand{\wikifullgenconunknown}{6.6}
\newcommand{\wikifullsolveunknown}{91.7}
% wikipedia_relaxed_causal
\newcommand{\wikirelgencon}{36.3}
\newcommand{\wikirelliteral}{529313}
\newcommand{\wikirelz}{??}
\newcommand{\wikirelcntsat}{2}
\newcommand{\wikireleventsat}{1171.5}
\newcommand{\wikireltxsat}{11.0}
\newcommand{\wikirelgenconsat}{53.4}
\newcommand{\wikirelsolvesat}{2.5}
\newcommand{\wikirelcntunsat}{8}
\newcommand{\wikireleventunsat}{1110.6}
\newcommand{\wikireltxunsat}{9.6}
\newcommand{\wikirelgenconunsat}{32.0}
\newcommand{\wikirelsolveunsat}{1.0}
\newcommand{\wikirelcntunknown}{0}
\newcommand{\wikireleventunknown}{}
\newcommand{\wikireltxunknown}{}
\newcommand{\wikirelgenconunknown}{}
\newcommand{\wikirelsolveunknown}{}

%% file: results/valid_cc.tex
\newcommand\bankfullvalid{\valid{4}{\bankfullcntsat}{0}}
\newcommand\bankexpvalid{\valid{4}{\bankexpcntsat}{1}}
\newcommand\bankrelvalid{\valid{9}{\bankrelcntsat}{1}}

\newcommand\votefullvalid{\valid{0}{0}{0}}
\newcommand\voteexpvalid{\valid{0}{0}{0}}
\newcommand\voterelvalid{\valid{0}{0}{0}}

\newcommand\tpccfullvalid{\valid{9}{\tpccfullcntsat}{0}}
\newcommand\tpccexpvalid{\valid{9}{\tpccexpcntsat}{0}}
\newcommand\tpccrelvalid{\valid{10}{\tpccrelcntsat}{0}}

\newcommand\wikifullvalid{\valid{0}{\wikifullcntsat}{0}}
\newcommand\wikiexpvalid{\valid{0}{\wikiexpcntsat}{0}}
\newcommand\wikirelvalid{\valid{2}{\wikirelcntsat}{1}}

%% file: summary.tex
\newcommand{\pred}[2]{#2 & \number\numexpr 10-#1-#2\relax & #1}
\newcommand{\predunk}[2]{#2*\hspace*{-2pt} & \number\numexpr 10-#1-#2\relax & #1}
\newcommand{\valid}[3]{\bf #1 & (#3)}
\newcommand{\tm}[1]{\ifdefempty{#1}{--}{\num[group-separator={,},group-minimum-digits={4}]{#1}\,s}}
\newcommand{\nc}[1]{\def\abc{\the\numexpr #1 / 1000\relax}\num[group-separator={,},group-minimum-digits={4}]{\abc}\,K}
\begin{tabular}{@{}l@{\;\;\;}l|r@{\;\;\;}r@{\;\;\;}r|r@{\;\;\;}r|r@{\;\;\;}r@{\;\;\;}|r@{\;\;\;}r@{}}
  %\toprule
& \bf Prediction & \mc{3}{c|}{\bf Prediction} & \mc{2}{c|}{\bf Validation} & \mc{2}{c|}{\bf Constraint gen.} & \mc{2}{c}{\bf Solving time} \\
% & & & & \multicolumn{3}{c}{Solve time (s)} \\
\bf Program & \bf strategy & \unkOrTO & Unsat & Sat & Validated & (Diverged) & \# Literals & Time & Sat & Unsat \\
\hline
\multirow{3}{*}{\bench{Smallbank}}
  & \configFull    & \pred{\bankfullcntsat}{\bankfullcntunknown} & \bankfullvalid & \nc{\bankfullliteral} & \tm{\bankfullgencon} & \tm{\bankfullsolvesat} & \tm{\bankfullsolveunsat} \\
  & \configExpress & \pred{\bankexpcntsat}{\bankexpcntunknown}   & \bankexpvalid  & \nc{\bankexpliteral}  & \tm{\bankexpgencon}  & \tm{\bankexpsolvesat}  & \tm{\bankexpsolveunsat} \\
  & \configRelaxed & \pred{\bankrelcntsat}{\bankrelcntunknown}   & \bankrelvalid  & \nc{\bankrelliteral}  & \tm{\bankrelgencon}  & \tm{\bankrelsolvesat}  & \tm{\bankrelsolveunsat} \\
\hline
\multirow{3}{*}{\bench{Voter}}
  & \configFull    & \pred{\votefullcntsat}{\votefullcntunknown} & \votefullvalid & \nc{\votefullliteral} & \tm{\votefullgencon} & \tm{\votefullsolvesat} & \tm{\votefullsolveunsat} \\
  & \configExpress & \pred{\voteexpcntsat}{\voteexpcntunknown}   & \voteexpvalid  & \nc{\voteexpliteral}  & \tm{\voteexpgencon}  & \tm{\voteexpsolvesat}  & \tm{\voteexpsolveunsat} \\
  & \configRelaxed & \pred{\voterelcntsat}{\voterelcntunknown}   & \voterelvalid  & \nc{\voterelliteral}  & \tm{\voterelgencon}  & \tm{\voterelsolvesat}  & \tm{\voterelsolveunsat} \\
\hline
\multirow{3}{*}{\bench{TPC-C}}
  & \configFull    & \pred{\tpccfullcntsat}{\tpccfullcntunknown} & \tpccfullvalid & \nc{\tpccfullliteral} & \tm{\tpccfullgencon} & \tm{\tpccfullsolvesat} & \tm{\tpccfullsolveunsat} \\
  & \configExpress & \pred{\tpccexpcntsat}{\tpccexpcntunknown}   & \tpccexpvalid  & \nc{\tpccexpliteral}  & \tm{\tpccexpgencon}  & \tm{\tpccexpsolvesat}  & \tm{\tpccexpsolveunsat} \\
  & \configRelaxed & \pred{\tpccrelcntsat}{\tpccrelcntunknown}   & \tpccrelvalid  & \nc{\tpccrelliteral}  & \tm{\tpccrelgencon}  & \tm{\tpccrelsolvesat}  & \tm{\tpccrelsolveunsat} \\
\hline
\multirow{3}{*}{\bench{Wikipedia}}
  & \configFull    & \pred{\wikifullcntsat}{\wikifullcntunknown} & \wikifullvalid & \nc{\wikifullliteral} & \tm{\wikifullgencon} & \tm{\wikifullsolvesat} & \tm{\wikifullsolveunsat} \\
  & \configExpress & \pred{\wikiexpcntsat}{\wikiexpcntunknown}   & \wikiexpvalid  & \nc{\wikiexpliteral}  & \tm{\wikiexpgencon}  & \tm{\wikiexpsolvesat}  & \tm{\wikiexpsolveunsat} \\
  & \configRelaxed & \pred{\wikirelcntsat}{\wikirelcntunknown}   & \wikirelvalid  & \nc{\wikirelliteral}  & \tm{\wikirelgencon}  & \tm{\wikirelsolvesat}  & \tm{\wikirelsolveunsat} \\
\end{tabular}

%% file: results/oltp_cc_8.tex
% smallbank_express_causal
\newcommand{\bankexpgencon}{121.0}
\newcommand{\bankexpliteral}{2174895}
\newcommand{\bankexpcntsat}{9}
\newcommand{\bankexpeventsat}{1298.0}
\newcommand{\bankexptxsat}{20.3}
\newcommand{\bankexpgenconsat}{119.9}
\newcommand{\bankexpsolvesat}{332.5}
\newcommand{\bankexpcntunsat}{0}
\newcommand{\bankexpeventunsat}{}
\newcommand{\bankexptxunsat}{}
\newcommand{\bankexpgenconunsat}{}
\newcommand{\bankexpsolveunsat}{}
\newcommand{\bankexpcntunknown}{1}
\newcommand{\bankexpeventunknown}{1336.0}
\newcommand{\bankexptxunknown}{20.0}
\newcommand{\bankexpgenconunknown}{130.2}
\newcommand{\bankexpsolveunknown}{7200.1}
% smallbank_full_causal
\newcommand{\bankfullgencon}{55.6}
\newcommand{\bankfullliteral}{1072874}
\newcommand{\bankfullcntsat}{5}
\newcommand{\bankfulleventsat}{1308.0}
\newcommand{\bankfulltxsat}{20.4}
\newcommand{\bankfullgenconsat}{56.3}
\newcommand{\bankfullsolvesat}{8618.9}
\newcommand{\bankfullcntunsat}{1}
\newcommand{\bankfulleventunsat}{1126.0}
\newcommand{\bankfulltxunsat}{17.0}
\newcommand{\bankfullgenconunsat}{32.7}
\newcommand{\bankfullsolveunsat}{2366.2}
\newcommand{\bankfullcntunknown}{4}
\newcommand{\bankfulleventunknown}{1338.0}
\newcommand{\bankfulltxunknown}{21.0}
\newcommand{\bankfullgenconunknown}{60.5}
\newcommand{\bankfullsolveunknown}{86401.5}
% smallbank_relaxed_causal
\newcommand{\bankrelgencon}{118.8}
\newcommand{\bankrelliteral}{1072874}
\newcommand{\bankrelcntsat}{10}
\newcommand{\bankreleventsat}{1301.8}
\newcommand{\bankreltxsat}{20.3}
\newcommand{\bankrelgenconsat}{118.8}
\newcommand{\bankrelsolvesat}{19.3}
\newcommand{\bankrelcntunsat}{0}
\newcommand{\bankreleventunsat}{}
\newcommand{\bankreltxunsat}{}
\newcommand{\bankrelgenconunsat}{}
\newcommand{\bankrelsolveunsat}{}
\newcommand{\bankrelcntunknown}{0}
\newcommand{\bankreleventunknown}{}
\newcommand{\bankreltxunknown}{}
\newcommand{\bankrelgenconunknown}{}
\newcommand{\bankrelsolveunknown}{}
% tpcc_express_causal
\newcommand{\tpccexpgencon}{3416.1}
\newcommand{\tpccexpliteral}{60833606}
\newcommand{\tpccexpcntsat}{8}
\newcommand{\tpccexpeventsat}{8576.0}
\newcommand{\tpccexptxsat}{23.8}
\newcommand{\tpccexpgenconsat}{3645.1}
\newcommand{\tpccexpsolvesat}{1210.3}
\newcommand{\tpccexpcntunsat}{0}
\newcommand{\tpccexpeventunsat}{}
\newcommand{\tpccexptxunsat}{}
\newcommand{\tpccexpgenconunsat}{}
\newcommand{\tpccexpsolveunsat}{}
\newcommand{\tpccexpcntunknown}{2}
\newcommand{\tpccexpeventunknown}{8336.0}
\newcommand{\tpccexptxunknown}{24.0}
\newcommand{\tpccexpgenconunknown}{2500.0}
\newcommand{\tpccexpsolveunknown}{7202.4}
% tpcc_full_causal
\newcommand{\tpccfullgencon}{1914.6}
\newcommand{\tpccfullliteral}{36433544}
\newcommand{\tpccfullcntsat}{3}
\newcommand{\tpccfulleventsat}{8205.3}
\newcommand{\tpccfulltxsat}{23.7}
\newcommand{\tpccfullgenconsat}{1646.5}
\newcommand{\tpccfullsolvesat}{30413.1}
\newcommand{\tpccfullcntunsat}{3}
\newcommand{\tpccfulleventunsat}{9039.0}
\newcommand{\tpccfulltxunsat}{23.7}
\newcommand{\tpccfullgenconunsat}{1873.5}
\newcommand{\tpccfullsolveunsat}{24281.2}
\newcommand{\tpccfullcntunknown}{4}
\newcommand{\tpccfulleventunknown}{8386.8}
\newcommand{\tpccfulltxunknown}{24.0}
\newcommand{\tpccfullgenconunknown}{2146.6}
\newcommand{\tpccfullsolveunknown}{86406.0}
% tpcc_relaxed_causal
\newcommand{\tpccrelgencon}{3332.3}
\newcommand{\tpccrelliteral}{60833606}
\newcommand{\tpccrelcntsat}{10}
\newcommand{\tpccreleventsat}{8528.0}
\newcommand{\tpccreltxsat}{23.8}
\newcommand{\tpccrelgenconsat}{3332.3}
\newcommand{\tpccrelsolvesat}{186.2}
\newcommand{\tpccrelcntunsat}{0}
\newcommand{\tpccreleventunsat}{}
\newcommand{\tpccreltxunsat}{}
\newcommand{\tpccrelgenconunsat}{}
\newcommand{\tpccrelsolveunsat}{}
\newcommand{\tpccrelcntunknown}{0}
\newcommand{\tpccreleventunknown}{}
\newcommand{\tpccreltxunknown}{}
\newcommand{\tpccrelgenconunknown}{}
\newcommand{\tpccrelsolveunknown}{}
% voter_express_causal
\newcommand{\voteexpgencon}{490.5}
\newcommand{\voteexpliteral}{5623468}
\newcommand{\voteexpcntsat}{0}
\newcommand{\voteexpeventsat}{}
\newcommand{\voteexptxsat}{}
\newcommand{\voteexpgenconsat}{}
\newcommand{\voteexpsolvesat}{}
\newcommand{\voteexpcntunsat}{10}
\newcommand{\voteexpeventunsat}{925.0}
\newcommand{\voteexptxunsat}{24.0}
\newcommand{\voteexpgenconunsat}{490.5}
\newcommand{\voteexpsolveunsat}{47.2}
\newcommand{\voteexpcntunknown}{0}
\newcommand{\voteexpeventunknown}{}
\newcommand{\voteexptxunknown}{}
\newcommand{\voteexpgenconunknown}{}
\newcommand{\voteexpsolveunknown}{}
% voter_full_causal
\newcommand{\votefullgencon}{235.1}
\newcommand{\votefullliteral}{2622819}
\newcommand{\votefullcntsat}{0}
\newcommand{\votefulleventsat}{}
\newcommand{\votefulltxsat}{}
\newcommand{\votefullgenconsat}{}
\newcommand{\votefullsolvesat}{}
\newcommand{\votefullcntunsat}{1}
\newcommand{\votefulleventunsat}{925.0}
\newcommand{\votefulltxunsat}{24.0}
\newcommand{\votefullgenconunsat}{234.4}
\newcommand{\votefullsolveunsat}{5708.7}
\newcommand{\votefullcntunknown}{9}
\newcommand{\votefulleventunknown}{925.0}
\newcommand{\votefulltxunknown}{24.0}
\newcommand{\votefullgenconunknown}{235.2}
\newcommand{\votefullsolveunknown}{7201.1}
% voter_relaxed_causal
\newcommand{\voterelgencon}{496.1}
\newcommand{\voterelliteral}{5623468}
\newcommand{\voterelcntsat}{0}
\newcommand{\votereleventsat}{}
\newcommand{\votereltxsat}{}
\newcommand{\voterelgenconsat}{}
\newcommand{\voterelsolvesat}{}
\newcommand{\voterelcntunsat}{10}
\newcommand{\votereleventunsat}{925.0}
\newcommand{\votereltxunsat}{24.0}
\newcommand{\voterelgenconunsat}{496.1}
\newcommand{\voterelsolveunsat}{47.1}
\newcommand{\voterelcntunknown}{0}
\newcommand{\votereleventunknown}{}
\newcommand{\votereltxunknown}{}
\newcommand{\voterelgenconunknown}{}
\newcommand{\voterelsolveunknown}{}
% wikipedia_express_causal
\newcommand{\wikiexpgencon}{263.7}
\newcommand{\wikiexpliteral}{4315973}
\newcommand{\wikiexpcntsat}{1}
\newcommand{\wikiexpeventsat}{2483.0}
\newcommand{\wikiexptxsat}{22.0}
\newcommand{\wikiexpgenconsat}{171.0}
\newcommand{\wikiexpsolvesat}{15.6}
\newcommand{\wikiexpcntunsat}{9}
\newcommand{\wikiexpeventunsat}{2822.1}
\newcommand{\wikiexptxunsat}{22.9}
\newcommand{\wikiexpgenconunsat}{274.0}
\newcommand{\wikiexpsolveunsat}{30.1}
\newcommand{\wikiexpcntunknown}{0}
\newcommand{\wikiexpeventunknown}{}
\newcommand{\wikiexptxunknown}{}
\newcommand{\wikiexpgenconunknown}{}
\newcommand{\wikiexpsolveunknown}{}
% wikipedia_full_causal
\newcommand{\wikifullgencon}{111.9}
\newcommand{\wikifullliteral}{1773203}
\newcommand{\wikifullcntsat}{1}
\newcommand{\wikifulleventsat}{2483.0}
\newcommand{\wikifulltxsat}{22.0}
\newcommand{\wikifullgenconsat}{63.3}
\newcommand{\wikifullsolvesat}{910.2}
\newcommand{\wikifullcntunsat}{1}
\newcommand{\wikifulleventunsat}{2727.0}
\newcommand{\wikifulltxunsat}{22.0}
\newcommand{\wikifullgenconunsat}{79.9}
\newcommand{\wikifullsolveunsat}{1876.8}
\newcommand{\wikifullcntunknown}{8}
\newcommand{\wikifulleventunknown}{2834.0}
\newcommand{\wikifulltxunknown}{23.0}
\newcommand{\wikifullgenconunknown}{122.0}
\newcommand{\wikifullsolveunknown}{7200.6}
% wikipedia_relaxed_causal
\newcommand{\wikirelgencon}{258.3}
\newcommand{\wikirelliteral}{4315973}
\newcommand{\wikirelcntsat}{2}
\newcommand{\wikireleventsat}{2614.0}
\newcommand{\wikireltxsat}{22.5}
\newcommand{\wikirelgenconsat}{221.9}
\newcommand{\wikirelsolvesat}{20.3}
\newcommand{\wikirelcntunsat}{8}
\newcommand{\wikireleventunsat}{2831.8}
\newcommand{\wikireltxunsat}{22.9}
\newcommand{\wikirelgenconunsat}{267.4}
\newcommand{\wikirelsolveunsat}{25.3}
\newcommand{\wikirelcntunknown}{0}
\newcommand{\wikireleventunknown}{}
\newcommand{\wikireltxunknown}{}
\newcommand{\wikirelgenconunknown}{}
\newcommand{\wikirelsolveunknown}{}

%% file: results/valid_cc_8.tex
\newcommand\bankfullvalid{\valid{5}{\bankfullcntsat}{1}}
\newcommand\bankexpvalid{\valid{9}{\bankexpcntsat}{0}}
\newcommand\bankrelvalid{\valid{10}{\bankrelcntsat}{0}}

\newcommand\votefullvalid{\valid{0}{0}{0}}
\newcommand\voteexpvalid{\valid{0}{0}{0}}
\newcommand\voterelvalid{\valid{0}{0}{0}}

\newcommand\tpccfullvalid{\valid{3}{\tpccfullcntsat}{0}}
\newcommand\tpccexpvalid{\valid{8}{\tpccexpcntsat}{0}}
\newcommand\tpccrelvalid{\valid{10}{\tpccrelcntsat}{2}}

\newcommand\wikifullvalid{\valid{1}{\wikifullcntsat}{0}}
\newcommand\wikiexpvalid{\valid{1}{\wikiexpcntsat}{0}}
\newcommand\wikirelvalid{\valid{2}{\wikirelcntsat}{2}}

%% file: results/oltp_rc.tex
% smallbank_express_readcommitted
\newcommand{\bankexpgencon}{24.3}
\newcommand{\bankexpliteral}{369597}
\newcommand{\bankexpz}{??}
\newcommand{\bankexpcntsat}{10}
\newcommand{\bankexpeventsat}{684.4}
\newcommand{\bankexptxsat}{11.0}
\newcommand{\bankexpgenconsat}{24.3}
\newcommand{\bankexpsolvesat}{0.8}
\newcommand{\bankexpcntunsat}{0}
\newcommand{\bankexpeventunsat}{}
\newcommand{\bankexptxunsat}{}
\newcommand{\bankexpgenconunsat}{}
\newcommand{\bankexpsolveunsat}{}
\newcommand{\bankexpcntunknown}{0}
\newcommand{\bankexpeventunknown}{}
\newcommand{\bankexptxunknown}{}
\newcommand{\bankexpgenconunknown}{}
\newcommand{\bankexpsolveunknown}{}
% smallbank_full_readcommitted
\newcommand{\bankfullgencon}{10.0}
\newcommand{\bankfullliteral}{143537}
\newcommand{\bankfullz}{??}
\newcommand{\bankfullcntsat}{10}
\newcommand{\bankfulleventsat}{684.4}
\newcommand{\bankfulltxsat}{11.0}
\newcommand{\bankfullgenconsat}{10.0}
\newcommand{\bankfullsolvesat}{2.3}
\newcommand{\bankfullcntunsat}{0}
\newcommand{\bankfulleventunsat}{}
\newcommand{\bankfulltxunsat}{}
\newcommand{\bankfullgenconunsat}{}
\newcommand{\bankfullsolveunsat}{}
\newcommand{\bankfullcntunknown}{0}
\newcommand{\bankfulleventunknown}{}
\newcommand{\bankfulltxunknown}{}
\newcommand{\bankfullgenconunknown}{}
\newcommand{\bankfullsolveunknown}{}
% smallbank_relaxed_readcommitted
\newcommand{\bankrelgencon}{24.4}
\newcommand{\bankrelliteral}{369597}
\newcommand{\bankrelz}{??}
\newcommand{\bankrelcntsat}{10}
\newcommand{\bankreleventsat}{684.4}
\newcommand{\bankreltxsat}{11.0}
\newcommand{\bankrelgenconsat}{24.4}
\newcommand{\bankrelsolvesat}{0.6}
\newcommand{\bankrelcntunsat}{0}
\newcommand{\bankreleventunsat}{}
\newcommand{\bankreltxunsat}{}
\newcommand{\bankrelgenconunsat}{}
\newcommand{\bankrelsolveunsat}{}
\newcommand{\bankrelcntunknown}{0}
\newcommand{\bankreleventunknown}{}
\newcommand{\bankreltxunknown}{}
\newcommand{\bankrelgenconunknown}{}
\newcommand{\bankrelsolveunknown}{}
% tpcc_express_readcommitted
\newcommand{\tpccexpgencon}{569.2}
\newcommand{\tpccexpliteral}{6868917}
\newcommand{\tpccexpz}{??}
\newcommand{\tpccexpcntsat}{10}
\newcommand{\tpccexpeventsat}{4060.3}
\newcommand{\tpccexptxsat}{11.9}
\newcommand{\tpccexpgenconsat}{569.2}
\newcommand{\tpccexpsolvesat}{27.2}
\newcommand{\tpccexpcntunsat}{0}
\newcommand{\tpccexpeventunsat}{}
\newcommand{\tpccexptxunsat}{}
\newcommand{\tpccexpgenconunsat}{}
\newcommand{\tpccexpsolveunsat}{}
\newcommand{\tpccexpcntunknown}{0}
\newcommand{\tpccexpeventunknown}{}
\newcommand{\tpccexptxunknown}{}
\newcommand{\tpccexpgenconunknown}{}
\newcommand{\tpccexpsolveunknown}{}
% tpcc_full_readcommitted
\newcommand{\tpccfullgencon}{359.0}
\newcommand{\tpccfullliteral}{3855159}
\newcommand{\tpccfullz}{??}
\newcommand{\tpccfullcntsat}{10}
\newcommand{\tpccfulleventsat}{4060.3}
\newcommand{\tpccfulltxsat}{11.9}
\newcommand{\tpccfullgenconsat}{359.0}
\newcommand{\tpccfullsolvesat}{52.0}
\newcommand{\tpccfullcntunsat}{0}
\newcommand{\tpccfulleventunsat}{}
\newcommand{\tpccfulltxunsat}{}
\newcommand{\tpccfullgenconunsat}{}
\newcommand{\tpccfullsolveunsat}{}
\newcommand{\tpccfullcntunknown}{0}
\newcommand{\tpccfulleventunknown}{}
\newcommand{\tpccfulltxunknown}{}
\newcommand{\tpccfullgenconunknown}{}
\newcommand{\tpccfullsolveunknown}{}
% tpcc_relaxed_readcommitted
\newcommand{\tpccrelgencon}{588.9}
\newcommand{\tpccrelliteral}{6868917}
\newcommand{\tpccrelz}{??}
\newcommand{\tpccrelcntsat}{10}
\newcommand{\tpccreleventsat}{4060.3}
\newcommand{\tpccreltxsat}{11.9}
\newcommand{\tpccrelgenconsat}{588.9}
\newcommand{\tpccrelsolvesat}{22.8}
\newcommand{\tpccrelcntunsat}{0}
\newcommand{\tpccreleventunsat}{}
\newcommand{\tpccreltxunsat}{}
\newcommand{\tpccrelgenconunsat}{}
\newcommand{\tpccrelsolveunsat}{}
\newcommand{\tpccrelcntunknown}{0}
\newcommand{\tpccreleventunknown}{}
\newcommand{\tpccreltxunknown}{}
\newcommand{\tpccrelgenconunknown}{}
\newcommand{\tpccrelsolveunknown}{}
% voter_express_readcommitted
\newcommand{\voteexpgencon}{133.0}
\newcommand{\voteexpliteral}{1527349}
\newcommand{\voteexpz}{??}
\newcommand{\voteexpcntsat}{10}
\newcommand{\voteexpeventsat}{769.0}
\newcommand{\voteexptxsat}{12.0}
\newcommand{\voteexpgenconsat}{133.0}
\newcommand{\voteexpsolvesat}{12.3}
\newcommand{\voteexpcntunsat}{0}
\newcommand{\voteexpeventunsat}{}
\newcommand{\voteexptxunsat}{}
\newcommand{\voteexpgenconunsat}{}
\newcommand{\voteexpsolveunsat}{}
\newcommand{\voteexpcntunknown}{0}
\newcommand{\voteexpeventunknown}{}
\newcommand{\voteexptxunknown}{}
\newcommand{\voteexpgenconunknown}{}
\newcommand{\voteexpsolveunknown}{}
% voter_full_readcommitted
\newcommand{\votefullgencon}{62.5}
\newcommand{\votefullliteral}{687744}
\newcommand{\votefullz}{??}
\newcommand{\votefullcntsat}{10}
\newcommand{\votefulleventsat}{769.0}
\newcommand{\votefulltxsat}{12.0}
\newcommand{\votefullgenconsat}{62.5}
\newcommand{\votefullsolvesat}{12.9}
\newcommand{\votefullcntunsat}{0}
\newcommand{\votefulleventunsat}{}
\newcommand{\votefulltxunsat}{}
\newcommand{\votefullgenconunsat}{}
\newcommand{\votefullsolveunsat}{}
\newcommand{\votefullcntunknown}{0}
\newcommand{\votefulleventunknown}{}
\newcommand{\votefulltxunknown}{}
\newcommand{\votefullgenconunknown}{}
\newcommand{\votefullsolveunknown}{}
% voter_relaxed_readcommitted
\newcommand{\voterelgencon}{132.7}
\newcommand{\voterelliteral}{1527349}
\newcommand{\voterelz}{??}
\newcommand{\voterelcntsat}{10}
\newcommand{\votereleventsat}{769.0}
\newcommand{\votereltxsat}{12.0}
\newcommand{\voterelgenconsat}{132.7}
\newcommand{\voterelsolvesat}{12.7}
\newcommand{\voterelcntunsat}{0}
\newcommand{\votereleventunsat}{}
\newcommand{\votereltxunsat}{}
\newcommand{\voterelgenconunsat}{}
\newcommand{\voterelsolveunsat}{}
\newcommand{\voterelcntunknown}{0}
\newcommand{\votereleventunknown}{}
\newcommand{\votereltxunknown}{}
\newcommand{\voterelgenconunknown}{}
\newcommand{\voterelsolveunknown}{}
% wikipedia_express_readcommitted
\newcommand{\wikiexpgencon}{38.3}
\newcommand{\wikiexpliteral}{533451}
\newcommand{\wikiexpz}{??}
\newcommand{\wikiexpcntsat}{7}
\newcommand{\wikiexpeventsat}{1160.4}
\newcommand{\wikiexptxsat}{10.3}
\newcommand{\wikiexpgenconsat}{44.9}
\newcommand{\wikiexpsolvesat}{2.1}
\newcommand{\wikiexpcntunsat}{3}
\newcommand{\wikiexpeventunsat}{1035.0}
\newcommand{\wikiexptxunsat}{9.0}
\newcommand{\wikiexpgenconunsat}{22.9}
\newcommand{\wikiexpsolveunsat}{0.5}
\newcommand{\wikiexpcntunknown}{0}
\newcommand{\wikiexpeventunknown}{}
\newcommand{\wikiexptxunknown}{}
\newcommand{\wikiexpgenconunknown}{}
\newcommand{\wikiexpsolveunknown}{}
% wikipedia_full_readcommitted
\newcommand{\wikifullgencon}{15.4}
\newcommand{\wikifullliteral}{184105}
\newcommand{\wikifullz}{??}
\newcommand{\wikifullcntsat}{7}
\newcommand{\wikifulleventsat}{1160.4}
\newcommand{\wikifulltxsat}{10.3}
\newcommand{\wikifullgenconsat}{19.1}
\newcommand{\wikifullsolvesat}{3.9}
\newcommand{\wikifullcntunsat}{1}
\newcommand{\wikifulleventunsat}{1035.0}
\newcommand{\wikifulltxunsat}{9.0}
\newcommand{\wikifullgenconunsat}{6.7}
\newcommand{\wikifullsolveunsat}{8.0}
\newcommand{\wikifullcntunknown}{2}
\newcommand{\wikifulleventunknown}{1035.0}
\newcommand{\wikifulltxunknown}{9.0}
\newcommand{\wikifullgenconunknown}{6.7}
\newcommand{\wikifullsolveunknown}{106.3}
% wikipedia_relaxed_readcommitted
\newcommand{\wikirelgencon}{38.1}
\newcommand{\wikirelliteral}{533451}
\newcommand{\wikirelz}{??}
\newcommand{\wikirelcntsat}{7}
\newcommand{\wikireleventsat}{1160.4}
\newcommand{\wikireltxsat}{10.3}
\newcommand{\wikirelgenconsat}{44.9}
\newcommand{\wikirelsolvesat}{1.7}
\newcommand{\wikirelcntunsat}{3}
\newcommand{\wikireleventunsat}{1035.0}
\newcommand{\wikireltxunsat}{9.0}
\newcommand{\wikirelgenconunsat}{22.2}
\newcommand{\wikirelsolveunsat}{0.5}
\newcommand{\wikirelcntunknown}{0}
\newcommand{\wikireleventunknown}{}
\newcommand{\wikireltxunknown}{}
\newcommand{\wikirelgenconunknown}{}
\newcommand{\wikirelsolveunknown}{}

%% file: results/valid_rc.tex
\newcommand\bankfullvalid{\valid{10}{\bankfullcntsat}{0}}
\newcommand\bankexpvalid{\valid{10}{\bankexpcntsat}{0}}
\newcommand\bankrelvalid{\valid{10}{\bankrelcntsat}{0}}

\newcommand\votefullvalid{\valid{10}{\votefullcntsat}{2}}
\newcommand\voteexpvalid{\valid{10}{\voteexpcntsat}{7}}
\newcommand\voterelvalid{\valid{10}{\voterelcntsat}{10}}

\newcommand\tpccfullvalid{\valid{10}{\tpccfullcntsat}{0}}
\newcommand\tpccexpvalid{\valid{10}{\tpccexpcntsat}{0}}
\newcommand\tpccrelvalid{\valid{10}{\tpccrelcntsat}{3}}

\newcommand\wikifullvalid{\valid{7}{\wikifullcntsat}{2}}
\newcommand\wikiexpvalid{\valid{7}{\wikiexpcntsat}{1}}
\newcommand\wikirelvalid{\valid{7}{\wikirelcntsat}{7}}

%% file: results/oltp_rc_8.tex
% smallbank_express_readcommitted
\newcommand{\bankexpgencon}{124.5}
\newcommand{\bankexpliteral}{2187328}
\newcommand{\bankexpcntsat}{10}
\newcommand{\bankexpeventsat}{1301.8}
\newcommand{\bankexptxsat}{20.3}
\newcommand{\bankexpgenconsat}{124.5}
\newcommand{\bankexpsolvesat}{19.0}
\newcommand{\bankexpcntunsat}{0}
\newcommand{\bankexpeventunsat}{}
\newcommand{\bankexptxunsat}{}
\newcommand{\bankexpgenconunsat}{}
\newcommand{\bankexpsolveunsat}{}
\newcommand{\bankexpcntunknown}{0}
\newcommand{\bankexpeventunknown}{}
\newcommand{\bankexptxunknown}{}
\newcommand{\bankexpgenconunknown}{}
\newcommand{\bankexpsolveunknown}{}
% smallbank_full_readcommitted
\newcommand{\bankfullgencon}{60.3}
\newcommand{\bankfullliteral}{1085308}
\newcommand{\bankfullcntsat}{10}
\newcommand{\bankfulleventsat}{1301.8}
\newcommand{\bankfulltxsat}{20.3}
\newcommand{\bankfullgenconsat}{60.3}
\newcommand{\bankfullsolvesat}{624.6}
\newcommand{\bankfullcntunsat}{0}
\newcommand{\bankfulleventunsat}{}
\newcommand{\bankfulltxunsat}{}
\newcommand{\bankfullgenconunsat}{}
\newcommand{\bankfullsolveunsat}{}
\newcommand{\bankfullcntunknown}{0}
\newcommand{\bankfulleventunknown}{}
\newcommand{\bankfulltxunknown}{}
\newcommand{\bankfullgenconunknown}{}
\newcommand{\bankfullsolveunknown}{}
% smallbank_relaxed_readcommitted
\newcommand{\bankrelgencon}{128.7}
\newcommand{\bankrelliteral}{2187328}
\newcommand{\bankrelcntsat}{10}
\newcommand{\bankreleventsat}{1361.1}
\newcommand{\bankreltxsat}{21.3}
\newcommand{\bankrelgenconsat}{128.7}
\newcommand{\bankrelsolvesat}{51.7}
\newcommand{\bankrelcntunsat}{0}
\newcommand{\bankreleventunsat}{}
\newcommand{\bankreltxunsat}{}
\newcommand{\bankrelgenconunsat}{}
\newcommand{\bankrelsolveunsat}{}
\newcommand{\bankrelcntunknown}{0}
\newcommand{\bankreleventunknown}{}
\newcommand{\bankreltxunknown}{}
\newcommand{\bankrelgenconunknown}{}
\newcommand{\bankrelsolveunknown}{}
% tpcc_express_readcommitted
\newcommand{\tpccexpgencon}{3981.9}
\newcommand{\tpccexpliteral}{62462272}
\newcommand{\tpccexpcntsat}{10}
\newcommand{\tpccexpeventsat}{8528.0}
\newcommand{\tpccexptxsat}{23.8}
\newcommand{\tpccexpgenconsat}{3981.9}
\newcommand{\tpccexpsolvesat}{279.7}
\newcommand{\tpccexpcntunsat}{0}
\newcommand{\tpccexpeventunsat}{}
\newcommand{\tpccexptxunsat}{}
\newcommand{\tpccexpgenconunsat}{}
\newcommand{\tpccexpsolveunsat}{}
\newcommand{\tpccexpcntunknown}{0}
\newcommand{\tpccexpeventunknown}{}
\newcommand{\tpccexptxunknown}{}
\newcommand{\tpccexpgenconunknown}{}
\newcommand{\tpccexpsolveunknown}{}
% tpcc_full_readcommitted
\newcommand{\tpccfullgencon}{2571.0}
\newcommand{\tpccfullliteral}{38062210}
\newcommand{\tpccfullcntsat}{10}
\newcommand{\tpccfulleventsat}{8528.0}
\newcommand{\tpccfulltxsat}{23.8}
\newcommand{\tpccfullgenconsat}{2571.0}
\newcommand{\tpccfullsolvesat}{898.6}
\newcommand{\tpccfullcntunsat}{0}
\newcommand{\tpccfulleventunsat}{}
\newcommand{\tpccfulltxunsat}{}
\newcommand{\tpccfullgenconunsat}{}
\newcommand{\tpccfullsolveunsat}{}
\newcommand{\tpccfullcntunknown}{0}
\newcommand{\tpccfulleventunknown}{}
\newcommand{\tpccfulltxunknown}{}
\newcommand{\tpccfullgenconunknown}{}
\newcommand{\tpccfullsolveunknown}{}
% tpcc_relaxed_readcommitted
\newcommand{\tpccrelgencon}{4040.4}
\newcommand{\tpccrelliteral}{62462272}
\newcommand{\tpccrelcntsat}{10}
\newcommand{\tpccreleventsat}{8528.0}
\newcommand{\tpccreltxsat}{23.8}
\newcommand{\tpccrelgenconsat}{4040.4}
\newcommand{\tpccrelsolvesat}{201.0}
\newcommand{\tpccrelcntunsat}{0}
\newcommand{\tpccreleventunsat}{}
\newcommand{\tpccreltxunsat}{}
\newcommand{\tpccrelgenconunsat}{}
\newcommand{\tpccrelsolveunsat}{}
\newcommand{\tpccrelcntunknown}{0}
\newcommand{\tpccreleventunknown}{}
\newcommand{\tpccreltxunknown}{}
\newcommand{\tpccrelgenconunknown}{}
\newcommand{\tpccrelsolveunknown}{}
% voter_express_readcommitted
\newcommand{\voteexpgencon}{491.7}
\newcommand{\voteexpliteral}{5625421}
\newcommand{\voteexpcntsat}{10}
\newcommand{\voteexpeventsat}{925.0}
\newcommand{\voteexptxsat}{24.0}
\newcommand{\voteexpgenconsat}{491.7}
\newcommand{\voteexpsolvesat}{76.2}
\newcommand{\voteexpcntunsat}{0}
\newcommand{\voteexpeventunsat}{}
\newcommand{\voteexptxunsat}{}
\newcommand{\voteexpgenconunsat}{}
\newcommand{\voteexpsolveunsat}{}
\newcommand{\voteexpcntunknown}{0}
\newcommand{\voteexpeventunknown}{}
\newcommand{\voteexptxunknown}{}
\newcommand{\voteexpgenconunknown}{}
\newcommand{\voteexpsolveunknown}{}
% voter_full_readcommitted
\newcommand{\votefullgencon}{255.7}
\newcommand{\votefullliteral}{2624772}
\newcommand{\votefullcntsat}{10}
\newcommand{\votefulleventsat}{1858.0}
\newcommand{\votefulltxsat}{25.0}
\newcommand{\votefullgenconsat}{255.7}
\newcommand{\votefullsolvesat}{212.0}
\newcommand{\votefullcntunsat}{0}
\newcommand{\votefulleventunsat}{}
\newcommand{\votefulltxunsat}{}
\newcommand{\votefullgenconunsat}{}
\newcommand{\votefullsolveunsat}{}
\newcommand{\votefullcntunknown}{0}
\newcommand{\votefulleventunknown}{}
\newcommand{\votefulltxunknown}{}
\newcommand{\votefullgenconunknown}{}
\newcommand{\votefullsolveunknown}{}
% voter_relaxed_readcommitted
\newcommand{\voterelgencon}{495.2}
\newcommand{\voterelliteral}{5625421}
\newcommand{\voterelcntsat}{10}
\newcommand{\votereleventsat}{925.0}
\newcommand{\votereltxsat}{24.0}
\newcommand{\voterelgenconsat}{495.2}
\newcommand{\voterelsolvesat}{75.4}
\newcommand{\voterelcntunsat}{0}
\newcommand{\votereleventunsat}{}
\newcommand{\votereltxunsat}{}
\newcommand{\voterelgenconunsat}{}
\newcommand{\voterelsolveunsat}{}
\newcommand{\voterelcntunknown}{0}
\newcommand{\votereleventunknown}{}
\newcommand{\votereltxunknown}{}
\newcommand{\voterelgenconunknown}{}
\newcommand{\voterelsolveunknown}{}
% wikipedia_express_readcommitted
\newcommand{\wikiexpgencon}{272.8}
\newcommand{\wikiexpliteral}{4349712}
\newcommand{\wikiexpcntsat}{10}
\newcommand{\wikiexpeventsat}{2788.2}
\newcommand{\wikiexptxsat}{22.8}
\newcommand{\wikiexpgenconsat}{272.8}
\newcommand{\wikiexpsolvesat}{29.2}
\newcommand{\wikiexpcntunsat}{0}
\newcommand{\wikiexpeventunsat}{}
\newcommand{\wikiexptxunsat}{}
\newcommand{\wikiexpgenconunsat}{}
\newcommand{\wikiexpsolveunsat}{}
\newcommand{\wikiexpcntunknown}{0}
\newcommand{\wikiexpeventunknown}{}
\newcommand{\wikiexptxunknown}{}
\newcommand{\wikiexpgenconunknown}{}
\newcommand{\wikiexpsolveunknown}{}
% wikipedia_full_readcommitted
\newcommand{\wikifullgencon}{124.6}
\newcommand{\wikifullliteral}{1806942}
\newcommand{\wikifullcntsat}{10}
\newcommand{\wikifulleventsat}{2788.2}
\newcommand{\wikifulltxsat}{22.8}
\newcommand{\wikifullgenconsat}{124.6}
\newcommand{\wikifullsolvesat}{81.4}
\newcommand{\wikifullcntunsat}{0}
\newcommand{\wikifulleventunsat}{}
\newcommand{\wikifulltxunsat}{}
\newcommand{\wikifullgenconunsat}{}
\newcommand{\wikifullsolveunsat}{}
\newcommand{\wikifullcntunknown}{0}
\newcommand{\wikifulleventunknown}{}
\newcommand{\wikifulltxunknown}{}
\newcommand{\wikifullgenconunknown}{}
\newcommand{\wikifullsolveunknown}{}
% wikipedia_relaxed_readcommitted
\newcommand{\wikirelgencon}{272.9}
\newcommand{\wikirelliteral}{4349712}
\newcommand{\wikirelcntsat}{10}
\newcommand{\wikireleventsat}{2788.2}
\newcommand{\wikireltxsat}{22.8}
\newcommand{\wikirelgenconsat}{272.9}
\newcommand{\wikirelsolvesat}{16.9}
\newcommand{\wikirelcntunsat}{0}
\newcommand{\wikireleventunsat}{}
\newcommand{\wikireltxunsat}{}
\newcommand{\wikirelgenconunsat}{}
\newcommand{\wikirelsolveunsat}{}
\newcommand{\wikirelcntunknown}{0}
\newcommand{\wikireleventunknown}{}
\newcommand{\wikireltxunknown}{}
\newcommand{\wikirelgenconunknown}{}
\newcommand{\wikirelsolveunknown}{}

%% file: results/valid_rc_8.tex
\newcommand\bankfullvalid{\valid{9}{\bankfullcntsat}{1}}
\newcommand\bankexpvalid{\valid{10}{\bankexpcntsat}{1}}
\newcommand\bankrelvalid{\valid{10}{\bankrelcntsat}{1}}

\newcommand\votefullvalid{\valid{10}{\votefullcntsat}{2}}
\newcommand\voteexpvalid{\valid{10}{\voteexpcntsat}{6}}
\newcommand\voterelvalid{\valid{10}{\voterelcntsat}{10}}

\newcommand\tpccfullvalid{\valid{10}{\tpccfullcntsat}{2}}
\newcommand\tpccexpvalid{\valid{10}{\tpccexpcntsat}{2}}
\newcommand\tpccrelvalid{\valid{10}{\tpccrelcntsat}{4}}

\newcommand\wikifullvalid{\valid{10}{\wikifullcntsat}{1}}
\newcommand\wikiexpvalid{\valid{10}{\wikiexpcntsat}{1}}
\newcommand\wikirelvalid{\valid{9}{\wikirelcntsat}{10}}

%% file: results/comparison.tex
\begin{tabular}{@{}l|rr|r@{}}
        & \mc{2}{c|}{\bf MonkeyDB} & \mc{1}{c@{}}{\bf \isopredict}  \\
\bf Program & Fail & Unser & Unser \\\hline
\bench{Smallbank} &  70\%   &   98\%   & \mynum{9} \\
\bench{Voter}     &  70\%   &   80\%   & \mynum{0} \\
\bench{TPC-C}     &  98\%   &   100\%   & \mynum{10} \\
\bench{Wikipedia} &  0\%   &    11\%   & \mynum{2}  \\
\end{tabular}

%% file: results/comparison_8.tex
\begin{tabular}{@{}l|rr|r@{}}
        & \mc{2}{c|}{\bf MonkeyDB} & \mc{1}{c@{}}{\bf \isopredict}  \\
\bf Program & Fail & Unser & Unser \\\hline
\bench{Smallbank} &  84\%   &   100\%   & \mynum{10} \\
\bench{Voter}     &  56\%   &   80\%   & \mynum{0} \\
\bench{TPC-C}     &  100\%   &   100\%   & \mynum{10} \\
\bench{Wikipedia} &  0\%   &   19\%   & \mynum{2} \\
\end{tabular}

%% file: results/comparison_rc.tex
\begin{tabular}{@{}l|rr|r|r@{}}
        & \mc{2}{c|}{\bf MonkeyDB} & \mc{1}{c|}{\bf \isopredict} & \mc{1}{c@{}}{\bf MySQL} \\
\bf Program & Fail & Unser & Unser & Fail \\\hline
\bench{Smallbank} & 100\%   &   100\%  & \mynum{10} &  0\% \\
\bench{Voter}     &  89\%   &   100\%  & \mynum{10} &  0\% \\
\bench{TPC-C}     & 100\%   &   100\%  & \mynum{10} & 50\% \\
\bench{Wikipedia} &  54\%   &   54\%   & \mynum{7}  &  0\% \\
\end{tabular}

%% file: results/comparison_rc_8.tex
\begin{tabular}{@{}l|rr|r|r@{}}
        & \mc{2}{c|}{\bf MonkeyDB} & \mc{1}{c|}{\bf \Isopredict} & \mc{1}{c@{}}{\bf MySQL}  \\
\bf Program & Fail & Unser & Unser & Fail \\\hline
\bench{Smallbank} &  100\%   &   100\%  & \mynum{10} & 0\% \\
\bench{Voter}     &  95\%   &   100\%   & \mynum{10} & 0\% \\
\bench{TPC-C}     &  100\%   &   100\%  & \mynum{10} & 70\% \\
\bench{Wikipedia} &  89\%   &   89\%    & \mynum{10} & 0\% \\
\end{tabular}

%% file: 6.related.tex
\section{Related Work}
\label{sec:related}

% \mike{See previously submitted NSF proposal (\url{https://www.overleaf.com/project/618d79fa2c92b8207d32626d})}
% \mike{Done}

% \mike{Some related work from my notes that may not have been added to our paper yet:
% \begin{itemize}
% \item Model checking database applications: \url{https://ieeexplore.ieee.org/abstract/document/7515467} and \url{https://www.cs.purdue.edu/homes/suresh/papers/concur18.pdf}
% \item Article about three program analyses for database applications: \url{http://sites.computer.org/debull/A14mar/p48.pdf}
% \item PLDI'23 paper on DPOR model checking
% \item Paper possibly mentioned by Constantin: \url{https://dl.acm.org/doi/10.1145/3498711} (although I think we probably don't need to cite it since it's about model checking non-DB programs)
% \end{itemize}}

% To the best of our knowledge, no prior work predicts unserializable executions soundly (i.e., without reporting infeasible unserializable behaviors) from an observed serializable execution.
The closest existing approaches to \isopredict are arguably MonkeyDB~\cite{monkeydb}, IsoDiff~\cite{isodiff}, 2AD~\cite{Warszawski2017ACIDRain}, and Sinha et al.'s predictive analysis~\cite{sinha2012}.
% %
As \S\ref{subsec:cmp_monkeydb} explained, MonkeyDB produces a single execution, which may or may not be unserializable, while
% MonkeyDB is a database designed for testing concurrent execution under weak isolation levels. It handles read requests by choosing a random value from the legal values under the specified weak isolation level~\cite{monkeydb}. MonkeyDB is not a predictive analysis: It inherently produces a single execution, which may or may not be unserializable.
% In contrast,
\isopredict predicts unserializable executions from an observed execution.
% essentially analyzing many executions at once using constraint solving.
\Isopredict can in theory work with any data store that can generate execution traces, while MonkeyDB requires its specialized query engine.
% with the writer transaction of each read.
% (Exploring other data stores is outside of our work's scope.)
% In contrast, MonkeyDB requires its specialized query engine.

\iffalse
\mike{Can we say more about the potential benefits of sound predictive analysis in this context?
One key potential benefit is that we can predict using observed production executions.}
\yang{I am a little concerned about this because production traces usually don't record wr relationship. Supporting this
will need the database to record a "last writer" for every record, which will introduce both memory and CPU overhead.}
\mike{MonkeyDB~\cite{monkeydb} or Biswas and Enea~\cite{biswas2019} points out that \writeread can be reconstructed offline
from values as long as values written to each key are unique. If values aren't unique, then unique IDs / versions can be used.
In a real data store implementation, would unique IDs / versions likely already be available?
Alternatively, it seems like we could translate everything to be in terms of values rather than last writers.}
\yang{I think they are available in some databases, but probably not all (Spyros probably knows more). OK I am fine with
listing applicability to production executions as a major benefits here, but probably need some explanation in implementation section.}
\mike{Yeah, we need to be careful not to oversell. Here seems too early to try to motivate this potential benefit of predictive analysis. Maybe in a discussion section later (which the results summary of Intro can forward reference).}
\fi

IsoDiff and 2AD detect unserializable behaviors based on an observed execution~\cite{isodiff,Warszawski2017ACIDRain}. They build an abstract graph that does not take into account potential dependencies between read values. As the 2AD paper acknowledges, ``2AD's abstract histories are value-agnostic and do not account for control flow within a program; in effect, 2AD's abstract history construction process assumes that each variable read and written can assume arbitrary values. However, there are often dependencies (e.g., $y = x + 1$) between the values that variables assume''~\cite{Warszawski2017ACIDRain}.
% [Sec 3.1.4].
As a result, 2AD incurs high false positive rates even after using programmer-guided refinement: 37 reported ``witnesses'' on average per application,
% [Sec 4.2.3],
but only 22 bugs across 12 applications, or 2 bugs on average per application~\cite{Warszawski2017ACIDRain}.
% [Sec 4.2.5].
% because it does nothing to account for  divergent behavior caused by a predicted application reading from a different write.
% As such, IsoDiff can arguably be categorized as an unsound predictive analysis.
% IsoDiff uses techniques to reduce false positives, which in turn can eliminate some true positives.
% \yang{This sentence does not feel right. I am fine with just deleting it. It's true that IsoDiff does not account for divergent behavior.}
% \mike{Removed it}

In contrast, \isopredict accounts for dependencies among read values through its axiomatic encoding of constraints, which permits encoding of potential dependencies using the prediction boundary. \IsoPredict may still report false positives, but for narrower reasons: divergent aborts or (only when using the relaxed boundary) intra-transaction dependencies.
% In contrast, \isopredict accounts for divergent behavior with a few caveats (Table~\ref{tab:configs}), leading to 99\% of reported unserializable executions being feasible.
% IsoDiff predicts unserializable executions under read committed (\rc) and snapshot isolation, while our approach targets \causal and \rc, so they are not directly empirically comparable.
% \mike{Oops, they're empirically comparable now that \isopredict supports \rc.}

% While \isopredict is (to the best of our knowledge) the first sound analysis for predicting unserializable behaviors on database applications,
Sinha et al.'s analysis predicts atomicity violations in shared-memory multithreaded programs by encoding the conditions for unserializability as SMT constraints~\cite{sinha2012}. A key difference with \isopredict is that
Sinha et al.'s work deals with execution histories of shared-memory programs, in which all pairs of conflicting accesses are fully ordered, while \isopredict deals with execution histories of distributed data store applications, in which conflicting accesses are \emph{unordered} in general. As a result, Sinha et al.'s work only needs to encode graph cyclicity, while \isopredict must encode that \emph{every} potential commit order is acyclic. Addressing this unique challenge led us to develop \isopredict's approximate encoding (\S\ref{subsubsec:sufficient-unser-constraints}).
% %
Other differences include the different prediction spaces: Sinha et al.'s analysis predicts different orderings of critical sections on the same lock, while \isopredict predicts different write--read orders.
% Sinha et al.'s work does not consider weak ordering since data-race-free shared-memory executions are sequentially consistent~\cite{memory-models-cacm-2010,adve90weakordering}.
% Unlike \isopredict, Sinha et al.'s work does not need to consider divergence due to transaction aborts.
% While no prior work besides Sinha et al.'s predicts atomicity violations in shared-memory program to our knowledge,

% Less related to \isopredict are dynamic predictive analyses that predict data races and deadlocks from an observed shared-memory program execution~\cite{said-nfm-2011,wcp,smarttrack,rvpredict-pldi-2014,predict-deadlocks}.
% However, solving ``reordering and concurrent events'' problems is different and generally simpler from solving ``reachability and cyclicity'' problems.

\subsubsection*{Dynamic analysis}

Non-predictive dynamic analysis can check if an observed execution satisfies an isolation level.
ECRacer
% is a dynamic program analysis that
checks whether an observed
% eventually consistent
execution is serializable, using a relaxed definition of serializability that accounts for commutative operations~\cite{serializability-for-eventual-consistency-2017}.
In contrast, \isopredict finds \emph{new} executions that violate serializability.

Prior work uses run-time testing and constraint solving to check if a data store provides a stated weak isolation level~\cite{biswas2019,elle,cobra,Zhang2023Viper,checking-causal}.
In contrast, \isopredict assumes the data store provides the target weak isolation level and predicts feasible unserializable executions.

Model checking explores multiple executions, avoiding exhaustively exploring all possible executions by using techniques such as dynamic partial order reduction (DPOR)~\cite{abdulla2023,Bouajjani2023,Ghafoor2016}.
Conschecker uses a DPOR-based stateless model checking algorithm to verify distributed shared-memory programs under causal consistency~\cite{abdulla2023}.
Bouajjani et al.'s work adapts DPOR-based algorithms to transactional database applications to check them under a range of isolation levels~\cite{Bouajjani2023}.

% , works, which focus on the correctness of a single execution, \isopredict predicts incorrectness by finding \unserializable executions that are legal under a certain weak isolation level.

\subsubsection*{Static analysis}

Static analysis can find unserializable behavior, but precision and performance scale poorly with program size.
% Static analysis does not execute the program, but relies on the program's source code to infer possible executions.
C\textsuperscript{4} and Nagar and Jagannathan's analysis detect serializability violations under causal consistency, eventual consistency, and snapshot isolation~\cite{serializability-for-causal-consistency-2018,nagar2018automated}. Clotho uses static analysis, model checking, and test generation to detect unserializable executions; it
% The problem with static analysis is that its execution time and false positives (infeasible \unserializable executions) scale poorly with problem size.
avoids false positives by verifying the feasibility of unserializable behaviors~\cite{clotho}. In contrast, \isopredict detects unserializable behaviors with high precision by basing it on a single observed execution.
% and is more effective in avoiding false predictions due to its conservative handling of divergence.

\subsubsection*{Isolation levels}

\Isopredict generates constraints based on isolation levels encoded in Biswas and Enea's axiomatic framework~\cite{biswas2019}.
Other prior work besides Biswas and Enea's has introduced axiomatic encodings of weak isolation levels~\cite{Bouajjani2017,Perrin2016,cerone15concur,kaki18oopsla}.

Adya et al.\ define various isolation levels with dependency graphs where each level allows certain types of cycles~\cite{adya2000}. Their approach encompasses ``classical'' database isolation levels such as read committed and snapshot isolation, but not isolation levels typically used in distributed data stores such as causal consistency~\cite{Alglave2014,Bouajjani2017,burckhardt2014,hamza2015algorithmic,Perrin2016} and eventual consistency~\cite{burckhardt2014}.
% In a distributed setting, strong isolation levels such as serializability~\cite{biswas2019,Szekeres2018} are hard to implement, and applications usually run on weakly isolated data stores, which makes them more error prone.
% \yang{Please also check this paper: ``Making Consistency More Consistent: A Unified Model for Coherence, Consistency and Isolation.''. You may not use it, but it may worth a comparison.}
% \chujun{Added.}
% \yang{These two papers~\cite{Crooks2017SBC,Szekeres2018} provide another flavor of definitions. The later claims to support coherence, consistency, and isolation. Not sure how well they can be encoded into SMT though.}
% \mike{Are they axiomatic encodings?} \yang{I don't think so. I am mentioning them just for completeness.}

\Isopredict currently supports only \causal and \rc, by encoding axioms from Biswas and Enea's framework~\cite{biswas2019}. We expect that extending \isopredict to more isolation levels from their framework---\textsc{read atomic} (a.k.a.\ \textsc{repeated reads}) and \textsc{snapshot isolation}---to be straightforward. We do not know how difficult it would be to encode other isolation levels (e.g., \textsc{eventual consistency} and \textsc{monotonic atomic view}) into Biswas and Enea's framework or into \isopredict.

%% file: 7.conclusion.tex
\section{Conclusion}

\Isopredict is the first predictive analysis for detecting unserializable behaviors of applications backed by weakly isolated data stores. \Isopredict's design introduces novel approaches to address challenges involving constraint complexity, constraint encoding, and divergent behaviors.
% This work introduces an SMT-based approach that detects potentially unserializable behaviors by performing dynamic predictive analysis on executions of applications that are backed by weakly isolated data stores.
% We have implemented it in \isopredict, which is a tool that can perform the predictive analysis and can validate the feasibility of its predictions.
An evaluation shows that, based on observed executions of data store applications, \isopredict effectively, precisely, and efficiently predicts feasible, unserializable behaviors.
% The evaluation on four weakly isolated applications shows \isopredict is very effective in predicting unserializable behaviors both under causal consistency and under read committed.
% Due to its conservative handling of divergence, which are unforeseen behaviors in predicted executions that differ from the observed ones, \isopredict's predictions are mostly feasible executions that exhibit undesirable outcomes.

%% file: proof.tex
\section{Proof that Anti-Dependency Implies Commit Order}
\label{sec:proof}

Here we prove the following claim from \S\ref{subsec:anti-dependency}:
Anti-dependency order must imply commit order, i.e., $\antidependency \subseteq \co$ for every valid \co. The proof proceeds by showing that violating anti-dependency order violates arbitration order:
% %
\begin{proof}
Suppose there exist $t_1, t_2$ such that $\antidependency(t_1,t_2)$, but $\neg\co(t_1,t_2$). By the definition of anti-dependency, let $k$ be a key and $t_w$ be a transaction such that 
$t_2$ writes $k$, $\writeread_k(t_w, t_1)$, and $\co(t_w, t_2)$.
Because $\neg\co(t_1,t_2)$ and \co is a total order, therefore $\co(t_2, t_1)$.
Then $\co(t_2, t_w)$ according to the arbitration rule (Equation~\ref{eq:serial_ar}).
However, $\co(t_2, t_w)$ contradicts
$\co(t_w, t_2)$ since \co is a total order.
% Therefore, $\co(t_1, t_2)$ for all \serializable \co.
\end{proof}

%% file: smt.tex
\section{\IsoPredict's Full Constraints using the Prediction Boundary}
\label{sec:smt}

This section shows the constraints generated by \isopredict's predictive analysis using the strict prediction boundary. For completeness we show all constraints generated by \isopredict, including those that are unchanged compared with \S\ref{sec:predictive-analysis}.

% \mike{We discussed how a few more constraints might need to exclude events after the boundary?}
% \chujun{I forgot to copy the \smtobs constraints here during the previous edit, and I think that's causing some of the confusions.}

\subsection{Encoding of Feasible Execution}

% \paragraph{Session Order.}
\begin{align*}
\multirow{2}{*}{$\forall t_1, t_2 \in T, t_1 \ne t_2, \quad$}
%\begin{cases}
\quad \boxed{\smtso(t_1,t_2)} & \quad \textnormal{if } \so(t_1,t_2) \\
\boxed{\neg\smtso(t_1,t_2)} & \quad \textnormal{otherwise}
%\end{cases}
\end{align*}

% \paragraph{Write--read order.}
% To ensure that each read that happens \emph{before} the prediction boundary reads from the same write as in the observed execution, \isopredict generates the following constraints, where $\smtobs(s,i)$  is an integer SMT function that represents the last write of each read in the \emph{observed} execution history (and is thus the analogue of $\smtchoice$ for the observed execution):
\begin{align*}
& \forall t_1, t_2 \in T, t_1 \ne t_2, \forall i \in \posk(t_2) = i, t_2\textnormal{'s read at pos $i$ reads from $t_1$ in } \ogwr,
\quad \boxed{\smtobs(s_2,i) = t_1}
\end{align*}

\begin{align*}
& \forall k \textnormal{ is a key},
\forall t_1 \textnormal{ writes } k,
\forall t_2 \textnormal{ reads } k,
\forall i \in \posk(t_2),
% r is STRICTLY INSIDE boundary ==>  choice(r) = obs(r)
\;\; \boxed{i < \smtboundary(s_2) \implies
\smtchoice(s_2, i) = \smtobs(s_2, i)}
\end{align*}
where $s_1$ is $t_1$'s session and $s_2$ is $t_2$'s session.
% \chujun{This constraint says if $t_2$'s read event $i$ is within the boundary, then $\smtchoice(s_2, i)$ must be the same as $\smtobs(s_2, i)$. If the writer is outside the boundary, the constraints below won't be satisfiable. This means \smtwriteread only includes transactions that are in the boundary and serializability/weak isolation rules don't need additional constraints to exclude transactions beyond the boundary.}
% \mike{I don't think the constraint below does that. In any case, we don't want to restrict the writer transaction of a read after the boundary; $\smtchoice(s_2, i)$ needs to be \emph{something}. Instead, we want to make sure that such write--read edges don't get included as \co edges.}
% \chujun{Added a new $\mathit{IsIncluded}$ function that tells whether a transaction should be included for serializability and weak isolation.}

\begin{align*}
& \forall k \textnormal{ is a key},
\forall t_1 \textnormal{ writes } k,
\forall t_2 \ne t_1 \textnormal{ reads } k,
\forall i \in \posk(t_2), \\
%\end{align*}
%\begin{align*}
& \boxed{\smtchoice(s_2, i) = t_1 \land
i \le \smtboundary(s_2) \implies
\wrposk(t_1) < \smtboundary(s_1)}
\end{align*}
where $s_1$ is $t_1$'s session and $s_2$ is $t_2$'s session, and
\wrposk(t) is the position of $t$'s last write to key $k$.

\begin{align*}
& \forall s \textnormal{ is a session},
\quad \boxed{\Big(\bigvee_{\substack{t \textnormal{ is a transaction in } s \\ i \in \posk(t)}} \smtboundary(s) = i\Big) \lor \smtboundary(s) = \infty}
\end{align*}
Recall that \posk(s) is the set of positions of reads to $k$ in the transaction $t$.

\begin{align*}
& \forall k \textnormal{ is a key},
\forall t_1 \textnormal{ writes } k,
\forall t_2 \textnormal{ reads } k,
t_1 \ne t_2, \\
& \boxed{\smtwritereadk(t_1,t_2) = \bigvee_{i \in \posk(t_2)} \smtchoice(s_2, i) = t_1 \land i \le \smtboundary(s_2)}
\end{align*}
where $s_2$ is $t_2$'s session.

\begin{align*}
& \forall t_1,t_2 \in \tx, t_1 \ne t_2,
\quad \boxed{\smtwriteread(t_1, t_2) = \bigvee_{k \textnormal{ is a key}} \smtwritereadk(t_1, t_2)} \\
\end{align*}

% \chujun{Added a few new constraints below:}

\iffalse
Boundary transactions are transactions that prediction boundary events belong to.
\begin{align*}
& \boxed{\mathit{BoundaryTxn}(t) \coloneq \smtboundary(s) \in \pos(t)} \\
\end{align*}
where $s$ is the session of transaction $t$.

Whether a transaction should be included in the analysis depends on whether none of the boundary transactions happen before it:
\begin{align*}
& \boxed{\mathit{IsIncluded}(t) \coloneq \bigwedge_{\forall t' \in \tx} \neg \bigl(\mathit{BoundaryTxn}(t') \land \smthb(t', t) \bigr)} \\
\end{align*}

\mike{Don't the constraints need to reason about which \emph{events}, not which transactions, should be included?}

\mike{We already have a simpler way to define whether an event $i \in \posk(t)$ should be included:
$i \le \smtboundary(s)$ where $s$ is $t$'s session.}
\fi

\subsection{Encoding of Unserializability}

\subsubsection{Precise encoding}

% \paragraph{Unserializability}
\begin{align*}
\boxed{\forall \smtco, \neg\mathit{IsSerializable}(\smtco)}
\end{align*}
where $\mathit{IsSerializable}$ is defined as follows:
\begin{empheq}[box=\fbox]{align*}
\mathit{IsSerializable}(\smtco) \coloneq \; & \mathit{Distinct}(\smtco(t_1),\dots,\smtco(t_n)) \; \land \\
& \bigwedge_{\forall t_1, t_2 \in T, t_1 \ne t_2}
(\smtso(t_1, t_2) \lor \smtwriteread(t_1, t_2) \lor \mathit{Arbitration}(t_1, t_2)) \Rightarrow \smtco(t_1) < \smtco(t_2)
% \land & \bigwedge_{\forall t_1, t_2 \in T, t_1 \ne t_2} \smtwriteread(t_1, t_2) \Rightarrow \smtco(t_1) < \smtco(t_2)\\
% \land & \bigwedge_{\forall t_1, t_2 \in T, t_1 \ne t_2} \smtso(t_1, t_2) \Rightarrow \smtco(t_1) < \smtco(t_2)\\
% \land & \bigwedge_{\forall t_1, t_2 \in T, t_1 \ne t_2} \mathit{Arbitration}(t_1, t_2) \Rightarrow \smtco(t_1) < \smtco(t_2)
\end{empheq}
where $t_1, \dots, t_n$ are all transactions in \tx,
and $\mathit{Distinct}(v_1,\dots,v_k)$ is a built-in SMT function that requires all input values to be distinct from each other.

% \paragraph{Arbitration}
\begin{empheq}[box=\fbox]{align*}
\mathit{Arbitration}(t_1, t_2) \coloneq \!\! \bigvee_{\substack{\forall k, t_1 \textnormal{ and } t_2 \textnormal{ write } k \\ \forall t_3 \in \tx \setminus \{t_1,t_2\}, t_3 \textnormal{ reads } k}}
\!\! \smtwritereadk(t_2, t_3) \land \smtco(t_1) < \smtco(t_3) \land \wrposk(t_1) < \smtboundary(s_1)
% \begin{subarray}{l}
% \displaystyle \smtwritereadk(t_2, t_3) \land \smtco(t_1) < \smtco(t_3) \;
% \end{subarray}
\end{empheq}
% \mike{\posk(t) is a set, so its use in the above constraint and other constraints doesn't work.}
% \chujun{Replaced it with the new $IsIncluded$ function (in the serializability constraints, not sure if it's still needed here).}

\subsubsection{Approximate encoding}

% \paragraph{Arbitration}
\begin{align*}
& \forall t_1, t_2 \in T, t_1 \ne t_2,
\end{align*}
\begin{empheq}[box=\fbox]{align*}
\smtarserial(t_1, t_2) =
\bigvee_{\substack{\forall k, t_1 \textnormal{ and } t_2 \textnormal{ write } k \\ \forall t_3 \in \tx \setminus \{t_1,t_2\}, t_3 \textnormal{ reads } k}}
\;\;
\begin{subarray}{l}
\displaystyle \smtwritereadk(t_2, t_3) \land \smtcomin(t_1, t_3) \land \rank(t_1,t_2) > \rank(t_1,t_3) \; \land \\ \displaystyle \wrposk(t_1) < \smtboundary(s_1)
\end{subarray}
% \begin{subarray}{l}
% \displaystyle \smtwritereadk(t_2, t_3) \land \smtcomin(t_1, t_3) \land \rank(t_1,t_2) > \rank(t_1,t_3) \;
% \end{subarray}
\end{empheq}
% \mike{Should $\posk(t_3) < \smtboundary(s_3)$ be $\posk(t_3) \le \smtboundary(s_3)$? (The writes can do the same, but it doesn't amtter since a write can't be \emph{on} the boundary.) Similarly for other constraints that use the boundary.}
% \chujun{That makes sense.}

% \paragraph{Anti-dependency}
\begin{empheq}[box=\fbox]{align*}
\smtantidependency(t_1, t_2) = \bigvee_{\substack{\forall k, 
 t_1 \textnormal{ reads } k \land t_2 \textnormal{ writes } k \\ \forall t_3 \in \tx \setminus \{t_1,t_2\}, t_3 \textnormal{ writes } k}}
\;\;
\begin{subarray}{l}
\displaystyle \smtwritereadk(t_3, t_1) \land \smtcomin(t_3, t_2) \land \rank(t_1,t_2) > \rank(t_3,t_2) \land \\ \displaystyle \wrposk(t_2) < \smtboundary(s_2)
\end{subarray}
% \begin{subarray}{l}
% \displaystyle \smtwritereadk(t_3, t_1) \land \smtcomin(t_3, t_2) \; \land \\ \displaystyle \rank(t_1,t_2) > \rank(t_3,t_2) 
% \end{subarray}
\end{empheq}

% \mike{The above constraints have \smtpco taking one parameter, while the below constraint has it taking two parameters.}
% \chujun{Changed them to using two parameters.}

% \paragraph{Unserializability}
% \begin{align*}
% \forall t_1, t_2 \in T, t_1 \ne t_2,
% \end{align*}
\begin{empheq}[box=\fbox]{align*}
\smtpco(t_1, t_2) =\;&
\smtso(t_1,t_2) \lor \smtwriteread(t_1,t_2) \lor \smtarserial(t_1,t_2) \lor \smtantidependency(t_1,t_2) \; \lor \\
&
% \Big(
\bigvee_{t \in \tx \setminus \{t_1,t_2\}} \!\! \smtpco(t_1,t) \land \smtpco(t,t_2) \land \rank(t_1,t_2) > \rank(t_1,t) \land \rank(t_1,t_2) > \rank(t,t_2)
% \Big)
\end{empheq}
% \mike{Is that right? Is rank correctly encoded / is anything missing?}
% \chujun{That's correct}

% To ensure that \pco is cyclic, the analysis generates the following constraint:
% %
\begin{align*}
\boxed{\bigvee_{\forall t_1,t_2 \in \tx, t_1 \ne t_2} \smtpco(t_1,t_2) \land \smtpco(t_2,t_1)}
\end{align*}

\subsection{Encoding of Weak Isolation}

\subsubsection{Causal consistency}

\begin{align*}
& \forall t_1,t_2 \in \tx, t_1 \ne t_2,
\end{align*}
\begin{align*}
& \boxed{\smthb(t_1, t_2) = \smtso(t_1, t_2)
\lor \smtwriteread(t_1, t_2)
\lor \bigvee_{\forall t \in T\setminus\{t_1, t_2\}}\smthb(t_1, t) \land \smthb(t, t_2)}
\end{align*}

% \paragraph{Arbitration}
\begin{empheq}[box=\fbox]{align*}
\smtarcausal(t_1, t_2) = \bigvee_{\substack{\forall k, t_1 \textnormal{ and } t_2 \textnormal{ write } k \\
\forall t_3 \in \tx \setminus \{t_1, t_2\}, t_3 \textnormal{ reads } k}}
\begin{subarray}{l}
\displaystyle \smtwritereadk(t_2, t_3) \land \smthb(t_1, t_3)
\land \wrposk(t_1) < \smtboundary(s_1)
\end{subarray}
% \begin{subarray}{l}
% \displaystyle \smtwritereadk(t_2, t_3) \land \smthb(t_1, t_3)
% \end{subarray}
\end{empheq}

% \paragraph{Causal Consistency}
\begin{align*}
% 0 \leq \cocausal(t) < \textnormal{total number of transactions} \\
% \cocausal(t_1) \ne \cocausal(t_2) \\
\quad \boxed{\smthb(t_1, t_2) \lor \smtarcausal(t_1, t_2) \; \Rightarrow \; \smtcocausal(t_1) < \smtcocausal(t_2)}
\end{align*}
% \mike{This requires $t_1$ and $t_2$ to be within the boundary, but what if $\smthb(t_1, t_2)$ or $\smtarcausal(t_1, t_2)$ only because of some ordering involving transaction $t$ that isn't within the boundary?}

\subsubsection{Read committed}

% \paragraph{Arbitration}
\begin{align*}
\forall t_1, t_2 \in T, t_1 \ne t_2,
\end{align*}
\begin{empheq}[box=\fbox]{align*}
\smtarrc(t_1, t_2) = \bigvee_{\substack{\forall k, \; t_1 \textnormal{ and } t_2 \textnormal{ write } k \\
\forall t_3 \in \tx \setminus \{t_1, t_2\}, \; t_3 \textnormal{ reads } k \\
\forall i \in \pos(t_3), \forall j \in \posk(t_3), \;  i < j}}
\begin{subarray}{l}
\displaystyle \smtchoice(s_3, i) = t_1 \land \smtchoice(s_3, j) = t_2
\land j \le \smtboundary(s_3)
\end{subarray}
% \begin{subarray}{l}
% \displaystyle \smtchoice(s_3, i) = t_1 \land \smtchoice(s_3, j) = t_2
% \end{subarray}
\end{empheq}
where $\pos(t)$ is the set of positions of read events in transaction $t$, $\posk(t)$ is the set of reads to $k$ in transaction $t$, and $s_3$ is $t_3$'s session.

% \paragraph{Read Committed}
\begin{align*}
% 0 \leq \corc(t) < \textnormal{total number of transactions} \\
% \cocausal(t_1) \ne \cocausal(t_2) \\
\quad \boxed{\smthb(t_1, t_2) \lor \smtarrc(t_1, t_2) \; \Rightarrow \; \smtcorc(t_1) < \smtcorc(t_2)}
\end{align*}

%% file: predictions.tex
\FloatBarrier

\section{Patterns of Observed and Predicted Executions}
\label{sec:prediction_patterns}

\begin{figure}[H]
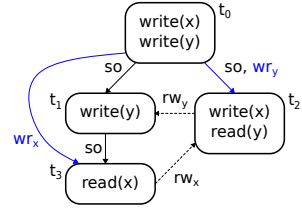
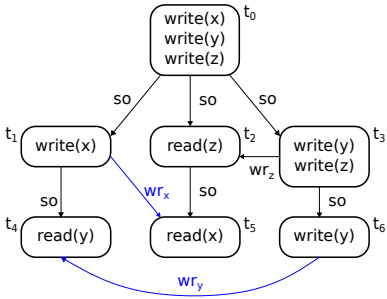
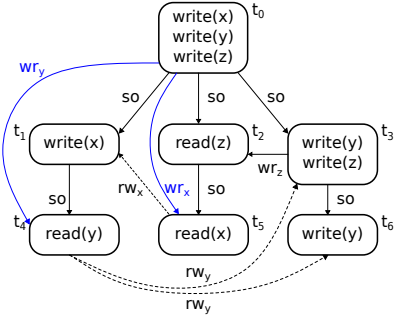
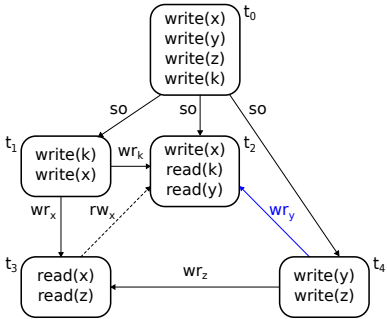
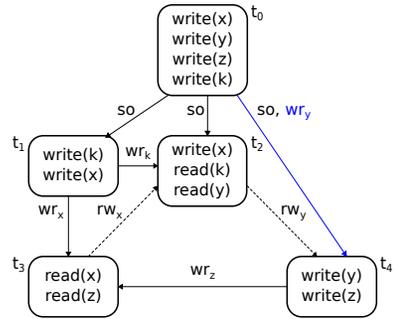

     \centering
     \begin{subfigure}[h]{0.48\textwidth}
         \centering
         \includesvg[inkscapelatex=false,scale=1]{img/smallbank_2_obs.svg}
         \caption{An observed execution of \bench{Smallbank}.}
         \label{fig:pattern_1_observed}
     \end{subfigure}
     \hfill
     \begin{subfigure}[h]{0.48\textwidth}
         \centering
         \includesvg[inkscapelatex=false,scale=1]{img/smallbank_2_pred.svg}
         \caption{A predicted execution based on (\subref{fig:pattern_1_observed}).}
         \label{fig:pattern_1_prediction}
     \end{subfigure}

    \begin{subfigure}[h]{0.48\textwidth}
         \centering
         \includesvg[inkscapelatex=false,scale=1]{img/smallbank_3_obs.svg}
         \caption{An observed execution of \bench{Smallbank}.}
         \label{fig:pattern_2_observed}
     \end{subfigure}
     \hfill
     \begin{subfigure}[h]{0.48\textwidth}
         \centering
         \includesvg[inkscapelatex=false,scale=1]{img/smallbank_3_pred.svg}
         \caption{A predicted execution based on (\subref{fig:pattern_2_observed}).}
         \label{fig:pattern_2_prediction}
     \end{subfigure}

    \begin{subfigure}[h]{0.48\textwidth}
         \centering
         \includesvg[inkscapelatex=false,scale=1]{img/tpcc_1_obs.svg}
         \caption{An observed execution of \bench{TPC-C}.}
         \label{fig:pattern_3_observed}
     \end{subfigure}
     \hfill
     \begin{subfigure}[h]{0.48\textwidth}
         \centering
         \includesvg[inkscapelatex=false,scale=1]{img/tpcc_1_pred.svg}
         \caption{A predicted execution based on (\subref{fig:pattern_3_observed}).}
         \label{fig:pattern_3_prediction}
     \end{subfigure}

     \begin{subfigure}[h]{0.48\textwidth}
         \centering
         \includesvg[inkscapelatex=false,scale=1]{img/tpcc_2_obs.svg}
         \caption{An observed execution of \bench{TPC-C}.}
         \label{fig:pattern_4_observed}
     \end{subfigure}
     \hfill
     \begin{subfigure}[h]{0.48\textwidth}
         \centering
         \includesvg[inkscapelatex=false,scale=1]{img/tpcc_2_pred.svg}
         \caption{A predicted execution based on (\subref{fig:pattern_4_observed}).}
         \label{fig:pattern_4_prediction}
     \end{subfigure}

     \caption{Observed executions that resulted in \causal (and thus \rc), \unserializable predicted executions.}
     %Each history shows a subset of executed transactions, and each transaction shows a subset of executed events.}
     \label{fig:prediction-patterns}
\end{figure}

Figure~\ref{fig:prediction-patterns} shows several observed executions and their \unserializable predictions from our experiments.
% and these predictions satisfy both \causal and \rc.
% \mike{Which program is each observed execution from (see ``??'' in captions)?}
The actual executions consist of dozens of transactions and thousands of events, but the figures show only the transactions and events relevant to predicting \unserializable behavior.

% \begin{figure}[t]
%      \centering

%      \caption{Another pattern of predictable anomaly found in \bench{TPC-C}}
%      \chujun{Need to revise the captions of these figures.}
% \end{figure}